\documentclass[structabstract]{aa}
\usepackage{supertabular}
\usepackage{natbib}
\usepackage{graphicx}
\usepackage{longtable,lscape}
\def\aap{A\&A}                
\def\apjl{ApJ}                
\def\apjs{ApJS}               
\def\amm{$\mathrm{NH_3}$}

\def\h13cop{$\mathrm{H^{13}CO^+}$}

\usepackage{txfonts}
\begin{document}
   \title{Initial phases of massive star formation in high infrared
     extinction clouds.\thanks{Tables \ref{ta:ext} and \ref{ta:nh3} are only
       available in electronic form at http://www.aanda.org. Table \ref{ta:ext}
       and fits images associated with the extinction maps of
       Fig.~\ref{fig:galext} and the 1.2\,mm continuum maps appearing in Fig.~\ref{fig:mips} can be
       queried from the CDS via anonymous ftp to cdsarc.u-strasbg.fr
       (130.79.128.5) or via http://cdswed.u-strasbg.fr/cgi-bin/gcat?J/A+A/.}}

   \subtitle{I. Physical parameters}

   \author{K.L.J. Rygl
          \inst{1}\thanks{Member of the International Max Planck Research School
(IMPRS) for Astronomy and Astrophysics at the Universities
of Bonn and Cologne},
          F. Wyrowski\inst{1},
          F. Schuller\inst{1} \and K.M. Menten\inst{1}
          }

   \institute{Max-Planck-Institut f\"ur Radioastronomie (MPIfR),
              Auf dem H\"ugel 69, 53121 Bonn, Germany\\
              \email{[kazi;wyrowski;schuller;kmenten]@mpifr-bonn.mpg.de}
                     }

  \date{}

 
  \abstract
   {}
   {The earliest phases of massive star formation are found in cold and dense infrared dark
   clouds (IRDCs). Since the detection method of IRDCs is very sensitive to
   the local properties of the background emission, we present here an
   alternative method to search for high column density in the Galactic plane by using infrared extinction maps. Using this method we find clouds between 1 and 5\,kpc, of which many were missed by previous surveys. By studying the physical conditions of a subsample of these clouds, we aim at a better understanding of the initial conditions of massive star formation. } 
   {We have made extinction maps of the Galactic plane based on the $3.6-4.5$~$\mu$m color excess between the two shortest wavelength Spitzer IRAC bands, reaching to visual extinctions of $\sim$100\,mag and column densities of $9\times10^{22}~\mathrm{cm^{-2}}$. From this we compiled a new sample of cold and compact high extinction clouds. 
We used the MAMBO array at the IRAM 30m telescope to study
the morphology, masses and densities of the clouds and the dense clumps within
them. The latter were followed up by pointed ammonia observations with the 100m Effelsberg telescope, to determine rotational temperatures and kinematic distances.}
   { Extinction maps of the Galactic plane trace large scale structures such as the spiral arms. The extinction method probes lower column densities, $N_{\mathrm{H_2}} \sim 4\times 10^{22}\,\mathrm{cm^{-2}}$, than the 1.2 mm continuum, which reaches up to $N_{\mathrm{H_2}} \sim 3\times 10^{23}\,\mathrm{cm^{-2}}$ but is less sensitive to large scale structures. The 1.2 mm emission maps reveal that the high extinction clouds contain extended cold dust emission, from filamentary structures to still diffuse clouds. Most of the clouds are dark in 24\,$\mu$m, but several show already signs of star formation via maser emission or bright infrared sources, suggesting that the high extinction clouds contain a variety of evolutionary stages. The observations suggest an evolutionary scheme from dark, cold and diffuse clouds, to clouds with a stronger 1.2\,mm peak and to finally clouds with many strong 1.2 mm peaks, which are also warmer, more turbulent and already have some star formation signposts.}
   {}
   \keywords{dust, extinction -- ISM: clouds -- ISM: structure -- Stars: formation -- Radio lines: ISM -- Submillimeter}
\authorrunning{Rygl et al.}
\titlerunning{Initial phases of massive star formation in high infrared extinction clouds}
   \maketitle
%
\section{Introduction}
Massive stars play a fundamental role in the evolution of galaxies through their
strong UV radiation, stellar winds and supernovae explosions, which contribute to the chemical enrichment of the interstellar medium.  
Massive stars are rare, hence usually found at large distances. They form very
rapidly while still deeply embedded in their natal molecular clouds. These characteristics impose several
observational obstacles, like the necessity of high resolution and sensitivity
in an un-absorbed frequency range, to study their formation.

Currently, the earliest stage of massive star formation is thought to take place in
the very dense clumps found in Infrared Dark Clouds (IRDCs). The properties of
IRDCs are shown by \citet{carey:1998} to be dense ($n_{\mathrm{gas}}>10^5\,\mathrm{cm^{-3}}$)
and cool ($T<20\,\mathrm{K}$) aggregations of gas and dust in the
Galaxy. They contain clumps with typical masses of $\geq100\,\mathrm{M_\odot}$ \citep{rathborne:2006,pillai:2006,simon:2006b}. 
From IRDC clumps to the next stage, the high-mass protostellar objects \citep[HMPOs,][]{beuther:2002,sridharan:2002}, the temperatures
increase ($30\,\mathrm{K}<\mathrm{T}<60\,\mathrm{K}$), the line widths
increase, densities and masses rise \citep{motte:2007}.  
HMPOs are usually found prior to the formation of ultra compact H{\sc ii}
(UCH{\sc ii}) regions, before the newly formed star begins to ionize its surrounding medium. 

\citet{motte:2007} demonstrated the difficulty of finding massive objects in
an early evolutionary phase: in their survey of 3 deg$^2$ in Cygnus X, they found little evidence for
dense clumps without any trace of star formation, however dense clumps
with already ongoing star formation were found to be abundantly present. Based on these
results, the statistical life time of the high-mass protostars and prestellar cores was estimated $\leq3\times10^4$\,yr \citep{motte:2007}, which
is much shorter than what is found in nearby low-mass star-forming regions: $\sim$$3\times10^5$\,yr \citep{kirk:2005}. 
It is these very early stages, which provide important clues to construct a theoretical model
of massive star formation, since the initial fragmentation of the gas and dust in a clump
will be different in the case of monolithical collapse \citep{krumholz:2009} compared to the
competitive accretion model \citep{bonnell:2001,clark:2008}.

Massive stars generally form in clusters \citep{lada:2003}, of which the
precursors are massive clumps or the so-called precluster forming clumps,
hereafter just clumps, of a $\sim$1\,pc size. For many massive star-forming
regions, we do not yet have the capacity to resolve the clumps into
prestellar cores and study the fragmentation \citep{beuther:2007a,rathborne:2007,rathborne:2008,zhang:2009,swift:2009}. In this paper, we report on the
physical parameters of the clumps, such as their morphology, density and temperature. 
Based on this, we hypothesize on an evolutionary sequence of cluster
formation.

Our understanding of the clumps increased considerably with the discovery of IRDCs. IRDCs are detected by a
local absence of infrared (IR) emission against the diffuse mid IR emission of the
Galactic plane \citep{perault:1996,egan:1998} and are observed numerously
throughout the Milky Way \citep{simon:2006b}. At the typical low temperatures of
IRDCs ($\sim20$\,K), the dust emission peaks in the far-infrared and is
optically thin at mm/submm wavelengths. For a
majority of IRDCs the mm dust emission coincides with the morphology of the IR
absorption \citep{rathborne:2006,pillai:2006}. 
Many clumps in IRDCs show signs of star formation via infrared emission at 24\,$\mu$m, or SiO emission from shocks driven by outflows
\citep{motte:2007,beuther:2007b,chambers:2009}. Observations tell that clumps in IRDCs span a very wide range of masses, indicating that not all will form clusters with massive stars \citep{rathborne:2006,pillai:2006}.

The detection method of IRDCs is very sensitive to the local properties of the background emission.
Also, not all massive dust condensations will be infrared dark if there is enough foreground emission.
Hence, to find the high mass end of molecular clouds in an unbiased fashion,
new, complementary approaches are needed. We have developed such a new method, well known from studies of low mass star-forming regions, to target more efficiently the most massive clouds: \citet{lada:1994} pioneered the method of measuring high amounts of extinction
through stellar color excess in the infrared. Applied to the 2\,$\mu$m data of
the 2MASS survey, they covered the range up to 40\,magnitudes in visual extinction, $A_V$. However, this is
not sufficient to probe the dense birthplaces of massive stars. Here the results of the Spitzer Space Telescope GLIMPSE survey \citep{benjamin:2003} came to help: by applying the
extinction curve of \citet{indebetouw:2005} we have extended the color excess
method to reach up to  peaks in $A_V$ of $\sim$100 magnitudes (or column
  densities $N_\mathrm{H_2}$ of $9\times10^{22}$\,cm$^{-2}$), thus entering the realm where massive star
formation becomes possible. The extinction method, however, is limited by the number of available background stars, and will therefore detect mainly nearby clouds (discussed in Sect.~\ref{sec:hecmap}).
In the meanwhile, complementary, unbiased dust continuum surveys were carried out:
the ATLASGAL survey of the complete inner Galactic plane at 870\,$\mu$m by
\citet{schuller:2009} and the 1.1\,mm BOLOCAM survey \citep{rosolowsky:2009} of the Galactic plane accessible from the northern hemisphere. 

We selected the more compact and high extinction (mean $A_{\mathrm{V}}>20\,\mathrm{mag}$ 
  or $N_\mathrm{H_2}>2\times10^{22}$\,cm$^{-2}$) sources
from large scale extinction maps of the inner Galactic plane
($-60\degr < l < 60\degr$, $0.9\degr<b<-0.9\degr$). These
{\em high extinction clouds} (HECs) were studied in the millimeter dust continuum and
the rotational transitions of ammonia (\amm). \amm\ has proven to be a
reliable tracer of dense gas in dark clouds: not only does the \amm\ emission
match the submillimeter dust emission peaks \citep{pillai:2006}, but also do observations show
that, unlike other molecules, \amm\ does not deplete from the gas phase for typical IRDC densities -- \citet{pagani:2005} observed that \amm\ depletes at densities of $10^{6}\,\mathrm{cm^{-3}}$ in agreement with the prediction of \citet{bergin:1997}. Moreover, throughout the evolutionary stages of massive star formation, \amm\ shows an increasing trend in averaged line widths and temperatures from less to more evolved sources \citep{pillai:2006}, which indicates that the \amm\ molecule is also a tracer of evolutionary phase. 

This study presents an overview from the high extinction complexes on Galactic size-scales, covering several
tens of parsecs, to the clumps found in the 1.2\,mm continuum of $0.1-0.7$\,pc in size.
The connection between the largest and smallest scale is important for a
comprehensive view of cluster formation in giant molecular clouds. In section 2 , we present the method of extinction mapping. Observations
and data reduction are described in section 3, and the results follow in
section 4. The analysis and discussion of
the physical parameters are given in section 5, and are compared to previous studies and theoretical predictions in section 6.


\section{High extinction clouds}

\subsection{Method: extinction mapping}
\label{sec:hecmap}
The overall distribution of dust in a cloud can be traced by the
extinction of background starlight at visual and near-infrared
wavelengths as it passes through a cloud \citep{lada:1994}. Since extinction decreases with wavelength,
observations at longer wavelengths probe deeper into the cloud and trace
denser regions. Additionally the number of detectable background stars
increases at these wavelengths. With the advance of infrared cameras it became
possible to detect several hundreds of background stars through a cloud, allowing
to convert the infrared images covering them in extinction maps of useful resolution. 

The GLIMPSE survey employed the Infrared Array Cameras \citep{fazio:2004} onboard the Spitzer Space telescope, operating at 3.6, 4.5, 5.8 and 8.0\,$\mu$m. We used the data provided by the GLIMPSE\,I survey (release  April 2005), which covered longitudes of $l=10-65\degr,~295-350\degr$ with $b=\pm1\degr$. The calibration of the data is described in \citet{reach:2005}. The inner 20\degr\ of the Galactic Plane, except for the innermost $\pm1\degr$, was taken from the GLIMPSE\,II survey (2007). 

To construct the extinction maps we used the averaged (3.6\,$\mu$m--4.5\,$\mu$m) color excess, because the
extinction law determination for these wavelengths is the most accurate of all the Spitzer bands.
The averaged color excess, $<E(3.6\,\mu\mathrm{m}-4.5\,\mu\mathrm{m})>$, was calculated from the
color excess in a large scale field (a box of size $108\arcsec\times108\arcsec$):
\begin{eqnarray}
<E(3.6\,\mu\mathrm{m}-4.5\,\mu\mathrm{m})> = \hspace{3.9cm}\nonumber\\<(3.6\,\mu\mathrm{m}-4.5\,\mu\mathrm{m})-(3.6\,\mu\mathrm{m}-4.5\,\mu\mathrm{m})_0>\,,
\end{eqnarray}
where the background stars are taken to be common-type K giants.
Measurements of $(3.6\,\mu\mathrm{m}-4.5\,\mu\mathrm{m})_0$ in such control
fields showed that K giants have an average color of $\sim$0\,mag with a dispersion of 0.2\,mag. Starting from the reddening law for
$<E(H-K)>$ in \citet{lada:1994}, one can extend it following \citet{indebetouw:2005}, and get the relation between the averaged color excess to the averaged visual extinction, $<A_V>$:
\begin{equation}
<A_V> = 81.8 <E(3.6\,\mu\mathrm{m}-4.5\,\mu\mathrm{m})>\hspace{0.5cm}[\mathrm{mag}].
\label{eq:ce}
\end{equation}

The color excess map can be contaminated by embedded stars in the cloud itself or by foreground stars. The latter will increase in number as the cloud is located at a farther distance. Since the foreground stars will not be reddened, they decrease the average color excess of the field. For example, if the number of foreground stars equals the number of background stars the color excess will be halved. It also means that for far away clouds the color excess will be underestimated.
\citet{rathborne:2006} and \citet{chambers:2009} find signs of active massive star formation in one of our clouds. Such young red objects will contribute to the measured color excess, which will lead to an
overestimation of the derived extinction. However, the selection of clouds
associated with very early phases of star formation will not be affected.

In general, the limits of extinction mapping are set by the number of available background stars; their number has to be sufficient for a statistically
meaningful color excess determination. Thus, the reach of the extinction method will change with Galactic latitude, because at higher latitudes the number of stars decreases. In the Galactic plane, there will be ``horizon'' to which one can measure a sufficient color excess, however this horizon will be far from uniform; it depends for every direction on the number of K giants in front and behind the clouds, which will differ when crossing a spiral arm or moving in toward the Galactic center.

Recently, \citet{chapman:2009} studied the changes in the mid-infrared
extinction law within a large region with high resolution. They find that
while in regions with a K-band extinction of $A_{\mathrm{K}}\leq 0.5$\,mag the
extinction law is well fitted by an extinction factor of $R_{\mathrm{V}}=3.1$ \citep{wein:2001}, the regions with $A_{\mathrm{K}}\geq 1.0$\,mag are more consistent with the \citet{wein:2001} model of $R_{\mathrm{V}}=5.5$, which uses larger maximum dust grain sizes. The high extinction clouds are by definition very dense regions for which $A_{\mathrm{K}} > 1.0$\,mag. This means that the visual extinctions and column densities reached by the extinction mapping are a factor $\sim$1.8 higher than estimated from Eq.~\ref{eq:ce}, where we used $R_{\mathrm{V}}=3.1$. 
The near and mid-infrared extinction law is of great interest, and recently
many publications appeared on the mid-infrared \citep{flaherty:2007,nishiyama:2009,zasowski:2009} and the near-infrared \citep{moore:2005,froebrich:2006,stead:2009} extinction law. Within the range of 3.6 and 4.5\,$\mu$m, the range we used for our extinction maps, there is a reasonable agreement between the results of \citet{indebetouw:2005} and most recent studies.

\subsection{Catalog of high extinction clouds}

\begin{figure*}[!htb]
\centering
\includegraphics[width=0.8\textwidth]{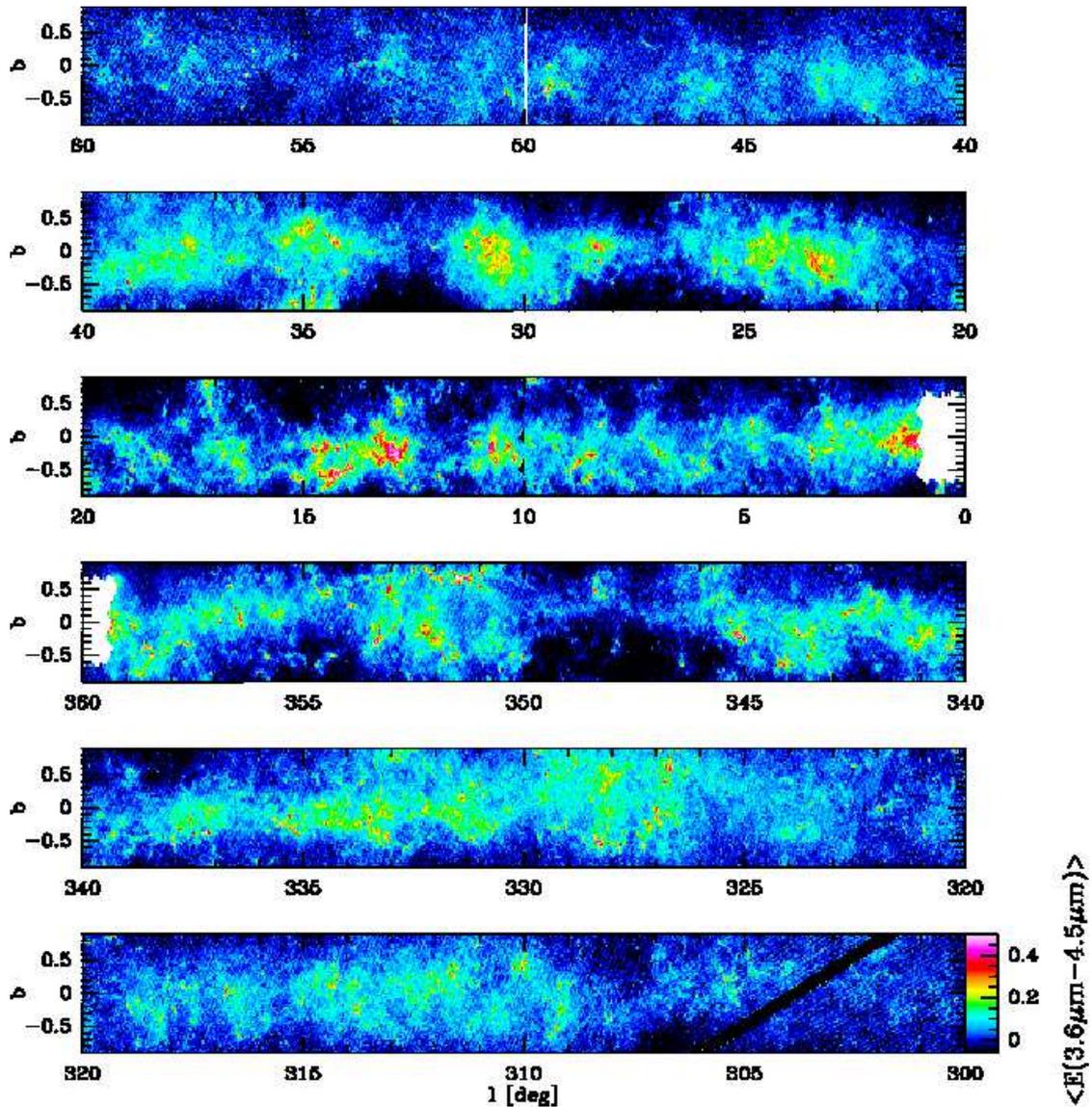}
\caption{\label{fig:galext} The color excess $\mathrm{<E(3.6\mu\mathrm{m}-4.5\mu\mathrm{m})>}$ for the
  first (top three graphs) and fourth (bottom three graphs) quadrant of the Galaxy. The extinction maps are available in fits format on the CDS.}
\end{figure*}

We made extinction maps for the complete inner part of the Galaxy with a
resolution of 108\arcsec, plotted in Fig.~\ref{fig:galext}. The
  mid-infrared extinction is changing with longitude and latitude, because it is
  sensitive to large-scale structures. Fig.~\ref{fig:ll} shows histograms of the average color excess in
  longitude and latitude. Most of the peaks in the color excess can be associated to Galactic spiral arms.
The large-scale structure of the extinction was excluded by selecting only compact high extinction
regions with an color excess above 0.25\,mag (equivalent to a hydrogen column density of $2\times10^{22}$\,cm$^{-2}$).
In a second step, smaller extinction maps with a higher resolution
($54\arcsec$) were computed to obtain more accurate positions of the highest extinction peaks.
These peaks were selected by eye. Regions with known sources, such as H{\sc ii} regions and
HMPOs, were discarded, leaving a sample of unknown and possibly cold and massive
clouds. In this paper we studied 25 high extinction clouds (HECs) in the first
Galactic quadrant. These were the clouds, visible from the northern
  hemisphere, with the highest extinction peaks which were not associated with
  H{\sc ii} regions from the \citet{becker:1994} survey and which had no available
mm/submm maps from the literature. A complete catalog of all the high extinction clouds is
given in Table \ref{ta:ext}, where for each cloud the center position in
  J2000 coordinates is listed with its corresponding peak color excess.
 
\onltab{1}{
\longtab{1}{
\begin{longtable}{l c c c|l c c c}
\caption{\label{ta:ext}Catalog of high extinction clouds (HECs) with the peak color excess, $E<(3.6\,\mu\mathrm{m}-4.5\,\mu\mathrm{m})>$}\\
\hline
\hline
HEC name & R.A. & Dec. & Color Excess & HEC name & R.A. & Dec. & Color Excess\\
     & (h:m:s) & (\degr:\,\arcmin:\,\arcsec) & (mag) &      & (h:m:s) & (\degr:\,\arcmin:\,\arcsec) & (mag)\\
     & (J2000) & (J2000) & &     & (J2000) & (J2000) & \\
\noalign{\smallskip}
\hline
\noalign{\smallskip}
\endfirsthead
\caption{continued.}\\
\hline\hline
HEC name & R.A. & Dec. & Color Excess & HEC name & R.A. & Dec. & Color Excess\\
     & (h:m:s) & (\degr:\,\arcmin:\,\arcsec) & (mag)&      & (h:m:s) & (\degr:\,\arcmin:\,\arcsec) & (mag)\\
     & (J2000) & (J2000) & &     & (J2000) & (J2000) & \\
\noalign{\smallskip}
\hline
\noalign{\smallskip}
\multicolumn{4}{c|}{{\it 1st quadrant clouds}} & \multicolumn{4}{c}{{\it 4th quadrant clouds}}\\
\endhead
\multicolumn{4}{c|}{{\it 1st quadrant clouds}} & \multicolumn{4}{c}{{\it 4th quadrant clouds}}\\
G011.09--00.54  & 18:12:01.2  & --19:36:07  &  0.40 & G313.26--00.72 &  14:20:34.2 &  --61:47:42  &  0.48  \\  
G012.54--00.35  & 18:14:15.8  & --18:14:11  &  0.29 & G313.72--00.29 &  14:22:57.1 &  --61:13:49  &  0.69  \\  
G012.65--00.17  & 18:13:49.6  & --18:02:53  &  0.45 & G314.21+00.21 &  14:25:18.8 &  --60:35:22  &  0.70   \\  
G012.73--00.58  & 18:15:33.2  & --18:08:45  &  0.29 & G314.27+00.09 &  14:26:07.4 &  --60:40:48  &  0.67   \\  
G012.89+00.48  & 18:11:52.7  & --17:31:27  &  0.55  & G314.31+00.12 &  14:26:21.3 &  --60:38:31  &  0.58   \\  
G013.28--00.34  & 18:15:40.8  & --17:34:32  &  0.57 & G316.45--00.63 &  14:44:46.4 &  --60:29:47  &  0.47  \\  
G013.38--00.27  & 18:15:38.8  & --17:27:26  &  0.47 & G316.73+00.06 &  14:44:28.6 &  --59:45:12  &  0.49   \\  
G013.91--00.51  & 18:17:34.8  & --17:06:07  &  0.58 & G316.77--00.02 &  14:45:02.6 &  --59:48:25  &  0.89  \\  
G013.97--00.45  & 18:17:27.6  & --17:01:34  &  0.51 & G317.70+00.11 &  14:51:11.0 &  --59:16:52  &  0.53   \\  
G014.33--00.65  & 18:18:55.4  & --16:48:12  &  0.53 & G317.88--00.26 &  14:53:46.2 &  --59:31:40  &  0.89  \\  
G014.39--00.75  & 18:19:24.1  & --16:47:34  &  0.42 & G318.05+00.09 &  14:53:43.9 &  --59:08:36  &  0.90   \\  
G014.45--00.09  & 18:17:06.1  & --16:25:37  &  0.50 & G318.78--00.17 &  14:59:40.2 &  --59:02:11  &  0.84  \\  
G014.63--00.57  & 18:19:13.6  & --16:29:56  &  0.57 & G320.19+00.85 &  15:05:20.9 &  --57:28:03  &  0.76   \\  
G014.72--00.21  & 18:18:04.7  & --16:14:47  &  0.46 & G321.93--00.01 &  15:19:42.4 &  --57:18:40  &  1.06  \\  
G015.09--00.60  & 18:20:14.0  & --16:06:38  &  0.33 & G322.16+00.64 &  15:18:33.9 &  --56:38:03  &  0.67   \\  
G015.21--00.62  & 18:20:33.2  & --16:00:37  &  0.40 & G323.19+00.15 &  15:26:47.7 &  --56:29:15  &  0.64   \\  
G015.51--00.42  & 18:20:23.8  & --15:39:22  &  0.31 & G323.72--00.28 &  15:31:44.6 &  --56:32:20  &  0.52  \\ 
G016.33--00.55  & 18:22:27.7  & --14:59:20  &  0.31 & G325.51+00.42 &  15:39:08.4 &  --54:55:47  &  0.57   \\  
G016.37--00.21  & 18:21:18.7  & --14:48:01  &  0.36 & G326.40+00.93 &  15:41:58.5 &  --53:59:04  &  0.86   \\  
G016.54--00.39  & 18:22:19.1  & --14:43:50  &  0.28 & G326.47+00.70 &  15:43:16.2 &  --54:07:35  &  0.80   \\  
G016.81--00.33  & 18:22:35.9  & --14:27:46  &  0.30 & G326.47+00.88 &  15:42:29.3 &  --53:59:03  &  0.95   \\  
G016.93+00.24  & 18:20:46.0  & --14:05:28  &  0.30  & G326.62+00.61 &  15:44:26.9 &  --54:06:02  &  0.95   \\  
G017.19+00.81  & 18:19:12.9  & --13:35:18  &  0.37  & G326.77--00.12 &  15:48:22.3 &  --54:35:29  &  0.85  \\  
G018.11--00.30  & 18:25:00.0  & --13:18:15  &  0.26 & G326.80+00.37 &  15:46:24.8 &  --54:10:57  &  0.67   \\  
G018.15--00.39  & 18:25:24.8  & --13:18:34  &  0.28 & G326.95--00.16 &  15:49:30.7 &  --54:30:42  &  0.44  \\  
G018.26--00.24  & 18:25:05.3  & --13:08:26  &  0.40 & G326.97--00.02 &  15:49:01.0 &  --54:23:15  &  0.77  \\  
G018.48--00.18  & 18:25:16.9  & --12:54:56  &  0.26 & G327.16--00.24 &  15:50:59.1 &  --54:26:28  &  0.46  \\  
G018.63--00.06  & 18:25:09.3  & --12:43:43  &  0.29 & G327.29--00.58 &  15:53:05.3 &  --54:37:13  &  0.46  \\  
G018.84--00.49  & 18:27:05.1  & --12:44:30  &  0.29 & G327.40--00.41 &  15:52:57.9 &  --54:25:05  &  0.68  \\  
G018.87--00.42  & 18:26:54.1  & --12:41:05  &  0.27 & G327.85--00.56 &  15:56:01.5 &  --54:15:02  &  0.53  \\  
G018.99--00.03  & 18:25:42.8  & --12:23:51  &  0.48 & G328.06+00.38 &  15:53:02.6 &  --53:23:27  &  0.55   \\  
G019.29+00.06  & 18:25:56.6  & --12:05:11  &  0.32  & G328.11+00.61 &  15:52:20.6 &  --53:11:00  &  0.46   \\  
G019.37--00.03  & 18:26:26.1  & --12:03:31  &  0.36 & G328.26--00.53 &  15:58:01.3 &  --53:57:44  &  0.56  \\  
G019.62--00.66  & 18:29:12.2  & --12:07:52  &  0.30 & G328.81+00.64 &  15:55:46.8 &  --52:43:01  &  0.80   \\  
G019.89--00.54  & 18:29:16.5  & --11:50:27  &  0.27 & G329.03--00.20 &  16:00:30.6 &  --53:12:37  &  0.93  \\  
G019.91--00.79  & 18:30:14.3  & --11:56:07  &  0.25 & G329.06--00.30 &  16:01:06.9 &  --53:16:05  &  0.67  \\  
G020.10--00.70  & 18:30:14.1  & --11:43:18  &  0.26 & G329.46+00.51 &  15:59:38.0 &  --52:23:28  &  0.60   \\  
G022.06+00.21  & 18:30:40.5  & --09:34:11  &  0.35  & G329.72+00.81 &  15:59:37.6 &  --51:59:53  &  0.60   \\  
G022.57--00.02  & 18:32:25.6  & --09:13:08  &  0.26 & G330.78+00.25 &  16:07:09.3 &  --51:42:48  &  1.01   \\  
G022.85--00.45  & 18:34:30.3  & --09:10:11  &  0.28 & G330.87--00.37 &  16:10:17.6 &  --52:06:33  &  1.10  \\  
G022.96+00.03  & 18:32:59.5  & --08:51:08  &  0.26  & G330.99+00.34 &  16:07:45.7 &  --51:30:39  &  0.55   \\  
G022.98--00.19  & 18:33:49.6  & --08:55:50  &  0.26 & G331.25--00.44 &  16:12:23.9 &  --51:53:57  &  0.49  \\  
G023.09--00.15  & 18:33:53.3  & --08:49:05  &  0.30 & G331.38+00.15 &  16:10:27.1 &  --51:23:04  &  0.55   \\  
G023.25--00.36  & 18:34:54.9  & --08:46:36  &  0.32 & G331.41--00.36 &  16:12:48.9 &  --51:44:25  &  0.64  \\  
G023.29--00.06  & 18:33:56.7  & --08:36:09  &  0.35 & G331.53--00.08 &  16:12:09.7 &  --51:27:08  &  0.50  \\  
G023.35--00.21  & 18:34:34.5  & --08:37:04  &  0.35 & G331.63+00.53 &  16:09:58.6 &  --50:56:03  &  0.48   \\  
G023.38--00.12  & 18:34:19.2  & --08:32:53  &  0.41 & G331.71+00.59 &  16:10:04.2 &  --50:50:23  &  0.73   \\  
G023.44--00.06  & 18:34:13.1  & --08:27:47  &  0.31 & G332.15+00.05 &  16:14:26.3 &  --50:55:29  &  0.50   \\  
G023.45--00.51  & 18:35:52.4  & --08:39:48  &  0.28 & G332.19--00.02 &  16:14:58.1 &  --50:56:49  &  0.43  \\  
G023.47+00.09  & 18:33:43.9  & --08:22:18  &  0.27  & G333.02+00.76 &  16:15:20.8 &  --49:48:32  &  0.50   \\  
G023.57+00.12  & 18:33:48.9  & --08:16:00  &  0.26  & G333.08--00.56 &  16:21:22.0 &  --50:42:51  &  0.59  \\  
G024.02+00.14  & 18:34:34.9  & --07:51:38  &  0.26  & G333.19--00.09 &  16:19:47.5 &  --50:18:26  &  0.56  \\  
G024.07+00.18  & 18:34:32.8  & --07:47:49  &  0.28  & G333.20--00.36 &  16:21:03.0 &  --50:29:19  &  0.43  \\  
G024.18+00.03  & 18:35:17.0  & --07:46:11  &  0.27  & G333.22--00.41 &  16:21:19.3 &  --50:30:37  &  0.46  \\  
G024.37--00.15  & 18:36:16.5  & --07:40:56  &  0.32 & G333.31--00.36 &  16:21:32.2 &  --50:24:41  &  0.51  \\  
G024.43--00.24  & 18:36:42.2  & --07:40:06  &  0.31 & G333.47--00.15 &  16:21:18.9 &  --50:08:51  &  0.44  \\  
G024.50+00.09  & 18:35:39.2  & --07:27:13  &  0.27  & G333.49--00.24 &  16:21:46.8 &  --50:12:19  &  0.57  \\  
G024.61--00.33  & 18:37:21.6  & --07:33:08  &  0.31 & G333.60--00.22 &  16:22:09.9 &  --50:06:32  &  0.44  \\   
G024.64+00.15  & 18:35:41.4  & --07:18:22  &  0.38 &  G333.66+00.37 &  16:19:50.9 &  --49:38:47  &  0.68   \\    
G024.82--00.11  & 18:36:57.6  & --07:16:02  &  0.26 & G333.75--00.33 &  16:23:17.6 &  --50:04:46  &  0.56  \\    
G024.94--00.15  & 18:37:19.4  & --07:10:34  &  0.35 & G333.76+00.35 &  16:20:25.0 &  --49:35:26  &  0.54   \\    
G025.15--00.28  & 18:38:09.6  & --07:02:52  &  0.32 & G334.20--00.20 &  16:24:42.2 &  --49:39:56  &  0.43  \\    
G025.63--00.12  & 18:38:30.1  & --06:32:55  &  0.32 & G334.45--00.24 &  16:25:58.1 &  --49:31:03  &  0.50  \\    
G025.79+00.81  & 18:35:26.6  & --05:58:57  &  0.19 &  G335.06--00.42 &  16:29:22.5 &  --49:11:57  &  0.77  \\    
G028.53+00.21  & 18:42:39.1  & --03:49:20  &  0.30 &  G335.25--00.30 &  16:29:38.2 &  --48:59:04  &  0.68  \\    
G030.48--00.38  & 18:48:20.1  & --02:21:16  &  0.29 & G335.28--00.13 &  16:29:00.9 &  --48:50:43  &  0.69  \\    
G030.62+00.18  & 18:46:34.4  & --01:58:31  &  0.31 &  G335.44--00.23 &  16:30:07.0 &  --48:47:56  &  0.82  \\    
G030.73+00.12  & 18:47:00.0  & --01:54:22  &  0.32 &  G337.15--00.39 &  16:37:46.4 &  --47:38:49  &  0.79  \\    
G030.90+00.00  & 18:47:43.1  & --01:48:24  &  0.30 &  G337.45--00.40 &  16:39:00.2 &  --47:26:07  &  0.45  \\    
G031.05+00.27  & 18:47:02.7  & --01:33:08  &  0.30 &  G337.50--00.19 &  16:38:18.9 &  --47:15:09  &  0.48  \\    
G034.03--00.33  & 18:54:36.9  & +00:49:43  &  0.30 &  G337.77--00.34 &  16:40:00.7 &  --47:08:50  &  0.53  \\    
G034.11+00.06  & 18:53:22.0  & +01:04:37  &  0.30 &   G337.93--00.51 &  16:41:23.3 &  --47:08:51  &  0.64  \\    
G034.34--00.90  & 18:57:12.8  & +00:50:40  &  0.30 &  G339.26--00.41 &  16:46:01.8 &  --46:04:40  &  0.53  \\    
G034.35--00.72  & 18:56:35.7  & +00:56:11  &  0.34 &  G339.58--00.13 &  16:46:00.1 &  --45:38:54  &  0.52  \\    
G034.71--00.63  & 18:56:55.6  & +01:17:44  &  0.41 &  G339.62--00.12 &  16:46:05.1 &  --45:37:07  &  0.41  \\    
G034.77--00.81  & 18:57:40.0  & +01:15:57  &  0.39 &  G340.06--00.24 &  16:48:14.9 &  --45:21:32  &  0.47  \\    
G034.85+00.43  & 18:53:23.6  & +01:54:07  &  0.34 &   G340.26--00.24 &  16:48:59.8 &  --45:12:03  &  0.57  \\    
G034.98+00.30  & 18:54:06.3  & +01:57:43  &  0.31 &   G340.77--00.12 &  16:50:18.0 &  --44:44:05  &  0.44  \\    
G035.19--00.75  & 18:58:13.9  & +01:40:07  &  0.59 &  G340.93--00.23 &  16:51:21.4 &  --44:40:36  &  0.61  \\   
G035.49--00.30  & 18:57:10.4  & +02:08:27  &  0.38 &  G341.12--00.42 &  16:52:51.9 &  --44:39:26  &  0.60  \\    
G036.42--00.15  & 18:58:20.4  & +03:02:04  &  0.27 &  G341.12--00.42 &  16:52:51.6 &  --44:39:17  &  0.61\\      
G037.26+00.09  & 18:59:01.9  & +03:53:40  &  0.36 &   G341.21--00.24 &  16:52:25.3 &  --44:28:11  &  0.47\\     
G037.44+00.14  & 18:59:09.6  & +04:04:18  &  0.28 &   G342.57+00.18 &  16:55:21.5 &  --43:09:13  &  0.56\\       
G037.48+00.07  & 18:59:29.3  & +04:05:03  &  0.29 &   G343.40--00.33 &  17:00:21.5 &  --42:49:13  &  0.52\\      
G037.54+00.20  & 18:59:08.7  & +04:11:44  &  0.29 &   G343.75--00.16 &  17:00:48.1 &  --42:26:19  &  0.56\\     
G037.65+00.12  & 18:59:38.5  & +04:15:22  &  0.29 &   G343.76--00.15 &  17:00:49.1 &  --42:25:24  &  0.56\\      
G038.93--00.36  & 19:03:40.9  & +05:10:19  &  0.48 &  G343.84--00.08 &  17:00:45.1 &  --42:19:04  &  0.66\\      
G044.30+00.03  & 19:12:17.0 & +10:07:02  &  0.31 &    G344.10--00.65 &  17:04:03.9 &  --42:27:36  &  0.81\\     
G046.33--00.24  & 19:17:05.9  & +11:47:23  &  0.32 &  G344.21--00.61 &  17:04:14.8 &  --42:21:09  &  0.44\\      
G048.90--00.27  & 19:22:08.9  & +14:02:57  &  0.41 &  G344.99--00.23 &  17:05:10.4 &  --41:29:38  &  0.87\\      
G049.39--00.31  & 19:23:16.0  & +14:27:17  &  0.34 &  G345.00--00.23 &  17:05:12.5 &  --41:29:35  &  0.50 \\     
G049.48--00.38  & 19:23:40.6  & +14:30:25  &  0.60 &  G345.04--00.21 &  17:05:15.2 &  --41:26:59  &  1.20\\      
G050.06+00.06  & 19:23:13.0  & +15:13:28  &  0.36 &   G345.26--00.04 &  17:05:14.5 &  --41:10:16  &  0.80\\      
G050.39--00.41  & 19:25:34.4  & +15:17:32  &  0.35 &  G345.49+00.31 &  17:04:29.1 &  --40:46:27  &  0.64 \\      
G053.14+00.07  & 19:29:18.0  & +17:56:08  &  0.46 &   G345.50+00.34 &  17:04:24.5 &  --40:44:38 &  1.21   \\     
G053.21--00.09  & 19:30:01.7  & +17:55:28  &  0.34 &  G345.67+00.34 &  17:04:58.5 &  --40:36:44  &  0.82  \\     
G053.24+00.06  & 19:29:31.8  & +18:01:14  &  0.57 &   G348.18+00.47 &  17:12:10.4 &  --38:31:19  &  0.95  \\     
G053.57+00.06  & 19:30:12.6  & +18:18:35  &  0.37 &   G350.02--00.51 &  17:21:39.5 &  --37:35:45  &  0.95  \\    
G053.63+00.03  & 19:30:26.3  & +18:20:40  &  0.33 &   G350.52--00.36 &  17:22:27.3 & - -37:05:32  &  0.79  \\    
G053.81--00.00  & 19:30:54.4  & +18:29:32  &  0.41 &  G350.69--00.48 &  17:23:26.9 &  --37:01:05  &  0.52  \\    
G058.48+00.42  & 19:39:01.4  & +22:46:46  &  0.30 &   G350.94+00.75 &  17:19:06.7 &  --36:07:04  &  0.45   \\    
G059.63--00.18  & 19:43:47.6  & +23:28:45  &  0.30 &   G350.94+00.66 &  17:19:30.2 &  --36:09:52  &  0.83   \\    
G059.79+00.06  & 19:43:13.2  & +23:44:22  &  0.29 &  G350.96+00.55 &  17:19:58.7 &  --36:12:56  &  0.57  \\     
\multicolumn{4}{c|}{{\it 4th quadrant clouds}}   &     G351.16+00.71 &  17:19:53.7 &  --35:57:29  &  0.62  \\     
G300.91+00.88 &  12:34:13.0 &  --61:55:39  &  1.00 &  G351.25+00.66 &  17:20:23.3 &  --35:54:39  &  0.82  \\     
G305.36+00.19 &  13:12:33.8 &  --62:35:12  &  0.58 &  G351.44+00.66 &  17:20:54.7 &  --35:45:11  &  0.92  \\     
G309.13--00.14 &  13:45:14.1 &  --62:21:36  &  0.52 & G351.47--00.45 &  17:25:30.5 &  --36:21:33  &  0.75 \\     
G309.37--00.12 &  13:47:18.1 &  --62:17:35  &  0.22 & G351.52--00.56 &  17:26:07.7 &  --36:22:40  &  0.40 \\     
G309.42--00.62 &  13:48:37.0 &  --62:46:00  &  0.51 & G351.53--00.56 &  17:26:06.6 &  --36:22:05  &  1.30 \\     
G310.22+00.39 &  13:53:21.6 &  --61:36:15  &  0.66 &  G351.59--00.36 &  17:25:26.9 &  --36:12:41  &  0.65 \\     
G311.57+00.31 &  14:04:24.5 &  --61:19:45  &  0.53&   G351.78--00.54 &  17:26:44.9 &  --36:09:17  &  0.58 \\     
G311.60+00.41 &  14:04:25.5 &  --61:13:36  &  0.50&   G351.81+00.65 &  17:21:59.2 &  --35:27:34  &  0.94   \\    
G311.95+00.15 &  14:07:48.4 &  --61:22:56  &  0.49&   G351.96--00.27 &  17:26:08.4 &  --35:51:15  &  0.71  \\    
G312.11+00.27 &  14:08:47.1 &  --61:13:09  &  0.47 &  &&&\\
\noalign{\smallskip}
\hline \hline
\end{longtable}
}}

\section{Observations and data reduction}

\subsection{Millimeter bolometer observations and calibration}

The 1.2 mm dust continuum of the selected extinction peaks was imaged with the
117-element Max-Planck Millimeter BOlometer array (MAMBO-2) installed at the IRAM 30m telescope. The observations were performed
in four sessions between 2006 October and 2007 March. The MAMBO passband has an
equivalent width of approximately 80 GHz centered on an effective frequency of
240~GHz ($\lambda \sim 1.25~$mm). The full width at half maximum (FWHM) beam size at this frequency is $10\rlap{$.$}\,\arcsec5$.
The maps were taken in the dual-beam on-the-fly mapping mode, where the
telescope was scanning row by row in azimuth at a constant speed, while the
secondary beam was wobbling in azimuth with a throw of 92\arcsec. The map sizes
were 6\arcmin$\times$6\arcmin, resulting in $\sim$ 10\arcmin$\times$10\arcmin\ images with a reduced
sensitivity at the edges. Each scan was separated by  8\arcsec\ in elevation. We used a
scanning velocity of 8\arcsec$~\mathrm{s^{-1}}$ with a wobbler rate of 2 Hz. The observation
time per map was $\sim27$ minutes. In total 25 high extinction clouds listed
in Table \ref{ta:cloud} were observed, the respective center positions used
for the observations are given in Table \ref{ta:ext}. 

\begin{table*}
\begin{minipage}[]{\textwidth}
\renewcommand{\footnoterule}{}  
\begin{center}
\caption{Properties of the high extinction clouds\label{ta:cloud}}
\begin{tabular}{l l r@{/}l c c c c r  r@{/}l  l}
\hline\hline
\noalign{\smallskip}
HEC name & & \multicolumn{2}{c}{Radius \footnote{Radius is the square root of
    the area divided by $\pi$. For the conversion to parsecs we use the
    kinematic distance based on the \amm(1,1) line.}} & $<A_V>_\mathrm{ext}$ & $<N_\mathrm{H_2}>_\mathrm{ext}$ & $F^i$ & $M_\mathrm{ext}$ & $M_\mathrm{1.2\,mm}$ & Clump&Cloud \footnote{Peak of the 1.2\,mm emission in the clump and the mean 1.2\,mm emission of the cloud without the clumps.}&Class\\
    & & (\arcsec&pc) & (mag) & ($10^{22}$ cm$^{-2}$) & (Jy) & ($\mathrm{M_\odot}$) & ($\mathrm{M_\odot}$) & (mJy~beam$^{-1}$ & mJy) &\\
\noalign{\smallskip}
\hline
\noalign{\smallskip}
G012.73--00.58... &   & 109&0.26 & 32 & 3.0 & 1.89 & 119 & 109 & 63&52 & diffuse\\
G013.28--00.34... &   & 59&1.15  & 42 & 3.9 & 3.85 & 3,039 & 1,693& 129&66 & diffuse \\
G013.91--00.51... &  &  49&0.65  & 44 & 4.2 & 2.84 & 1,015 & 602  & 203&66 & peaked\\ 
G013.97--00.45... &  &  52&0.6   & 22 & 2.1 & 2.29 & 457   & 308  & 99&56 & diffuse \\ 
G014.39--00.75... & A & 62&0.66  & 34 & 3.2 & 4.88 & 507 & 796    &  64&195 & peaked\\
              & B & 25&0.31  & 16 & 1.5 & 0.69 & 86  & 165    & 103&66 & diffuse\\
G014.63--00.57... &   & 98&1.04  & 35 & 3.3 & 17.47 & 2,084 & 1,822& 907&96 & multiply p.\\   
G016.93+00.24... &   & 52&0.61  & 32 & 3.0 & 2.20 & 632 & 294&118&56 & peaked\\    
G017.19+00.81... &   & 71&0.79  & 34 & 3.2 & 6.78 & 1,258 & 785& 547&63&multiply p.\\
G018.26--00.24... &   & 85&1.93  & 32 & 3.0 & 11.27 & 6,540 & 5.896& 317&92& multiply p.\\
G022.06+00.21... &   & 45&0.81  & 28 & 2.6 & 4.05 & 1,000 & 1,018& 1151&72 & multiply p.\\
G023.38--00.12... &   & 68&1.84  & 29 & 2.8 & 5.16 & 5,498 & 3,470&211&67 & peaked\\
G024.37--00.15... &   & 43&0.81  & 26 & 2.5 & 0.56 & 921 & 561 &168&52& multiply p.\\ 
G024.61--00.33... &   & 40&0.61  & 25 & 2.4 & 1.78 & 496 & 424 &214&69 & multiply p.\\
G024.94--00.15... &   & 49&0.79  & 32 & 3.0 & 2.17 & 1,095 & 650 & 160&62 & peaked\\
G025.79+00.81... &   & 38&0.63  & 24 & 2.3 & 1.05 & 494 & 351 &-&46 & diffuse \\
G030.90+00.00 ...&A   & 36&0.80 & 24 & 2.2 & 1.45 & 826 & 619 &188&61 & peaked\\
              &B   & 39&1.35 & 20 & 1.8 & 1.57 & 1,960 & 2,232 & 157&57 & peaked\\ 
              &C   & 30&0.83 & 21 & 2.0 & 0.83 & 799 &636 & 129&71 & diffuse\\
              &D   & 14&0.17 & 22 & 2.1 & 0.17 & 37 & 31 & 115&68 & diffuse\\ 
G034.03--00.33... &   & 6&- & 21 &1.9 & 0.02 & - & - & -&60 & diffuse\\
G034.34--00.90... &   & -   & -  &-   &-    &-     & -  &-&- &-& diffuse \\
G034.71--00.63... &   & 51&0.74 & 31 & 3.0 & 3.05 & 968 & 642&206&54 & multiply p.\\
G034.77--00.81... &   & 9&0.13 & 32 &3.0 & 0.05 & 30 & 12 & -&- & diffuse\\
G034.85+00.43... &    & 16&0.29 & 37 & 3.5 & 0.17 & 171 & 73 &-&43 & diffuse \\
G035.49--00.30... &A   & 19&0.34 & 11 &1.1 & 0.39 & 73 & 103 &198&58 & peaked \\
              &B   & 63&0.91 & 33 & 3.1 & 4.1 & 1,495 & 1,300 & 118&56 & peaked\\
G037.44+00.14... &A   & 26&0.16 & 33 & 3.1 & 0.58 & 47 & 37 & -&67 & diffuse\\
              &B   & 9&0.11  & 25 & 2.4 & 0.05 & 17 & 9 & -&54 & diffuse\\
G050.06+00.06... &   & 51&1.18 & 29 & 2.8 & 2.92 & 2,257 & 2,227&140&54 & peaked\\
G053.81--00.00... &    & 40&0.37 & 31 & 2.9 & 1.54 & 236 & 192 & 140&56 & peaked\\
\noalign{\smallskip}
\hline
\noalign{\smallskip}
\end{tabular}
\end{center}
{\bf Notes.} The first two columns give the HEC name and the subgrouping of clouds based on the kinematic distance, the following columns represent (in order of appearance): radius of the cloud, average visual extinction derived from the extinction maps, average hydrogen column density derived from the extinction maps, integrated 1.2\,mm flux, cloud mass derived from the extinction maps, cloud mass derived from the 1.2\,mm, clump peak flux and median cloud flux, class of HEC.
\end{minipage}
\end{table*}

The project was observed in the bolometer pool as a
backup project resulting in strongly varying weather conditions 
from one session to another. The zenith atmospheric opacity, $\tau$, was
monitored by sky dips, performed every $\sim$1.5 hour, and varied between 0.1
and 0.5. The calibration was performed every $\sim$3 hours mainly on H {\sc ii} regions
G34.3+0.2 and G10.2--0.4, and was found to be accurate within $\sim$10\%. Pointing checks
were made roughly every half hour to hour, before and usually also
after each map. Bolometer observations are dominated by the sky noise, the
variation of the brightness of the sky, which usually exceeds the intensity of astronomical
sources. The sky noise was reduced by subtracting the correlated noise
between the bolometers in the array. The average r.m.s. noise signal in the individual maps was better than $\sim$$15\,\mathrm{mJy~beam}^{-1}$ after reduction of the sky noise.
The data were reduced using the MOPSI software package developed by R.~Zylka.

\subsection{Ammonia  observations}

For each cloud, we selected by eye the mm emission peaks for follow-up
with pointed ammonia observations. All mm emission
peaks above two times the mean emission of the cloud, that is the r.m.s. emission in the cloud where we omit the bright mm peaks, were observed. When no clear emission peak was present, such as in a very diffuse
cloud, the center of the diffuse emission was targeted. Even in cases where
the mm emission was below $3\sigma$, where $\sigma$ is the noise in the bolometer map ($\sigma=0.015$\,Jy~beam$^{-1}$), we choose a few positions to observe the ammonia lines to search for a cloud so cold and diffuse that is was missed by the bolometer. For clouds that had
several continuum emission peaks separated by more than 20\arcsec\ from each other, more than one position was observed, denoted
by MM1, MM2 etc. 
Several high exinction clouds had very weak or no mm emission - in this case the name extension, which was MM1, MM2 for the mm sources, was changed to 1 or 2, e.g., G034.34--00.90~1.

The ammonia observations were performed with the MPIfR 100m Effelsberg Telescope. We observed 54 positions, given in Table \ref{ta:flux} and \ref{ta:noclump}, between 2007 April and 2008 December. The Effelsberg beam is $40\arcsec$ FWHM at the ammonia inversion line frequencies of $\sim$23.7\,GHz.
The 2007 observations used the 8192 channel AK90 auto correlator backend. The correlator was configured into eight spectral windows with 20\,MHz
bandwidth and 1024 channels each, where every window could be set to a
different frequency. This provided the opportunity to simultaneously observe
the ($J,K$) = (1,1), (2,2) and (3,3) inversion lines of ammonia in both polarizations. The spectral
resolution with this setup was 0.25\,km\,s$^{-1}$. 
The 2008 observations used the Fast Fourier Transform Spectrometer (FFTS) with
a 500\,MHz bandwidth and 8192 channels. This bandwidth is sufficient to simultaneously observe the three
ammonia inversion lines with a spectral resolution of 0.7\,km s$^{-1}$.
All the observations were performed in frequency switching mode with
a throw of 7.5 MHz.

\begin{table*}
\begin{minipage}[t]{\textwidth}
\renewcommand{\footnoterule}{}
\caption{\label{ta:flux} Positions (J2000) and fluxes of the clumps from the 1.2 mm continuum data}                            
\begin{tabular}{lllrrrrrrrc}
\hline\hline
\noalign{\smallskip}
HEC name &  & \multicolumn{1}{c}{R.A.}     & \multicolumn{1}{c}{Declination} & \multicolumn{1}{c}{$F^p$}  &\multicolumn{1}{c}{$F^i$} &\multicolumn{1}{c}{$F^i_{\mathrm{0.25 pc}}$}& \multicolumn{1}{c}{Major Axis}& \multicolumn{1}{c}{Minor Axis} & \multicolumn{1}{c}{P.A.}& \multicolumn{1}{c}{$\mathrm{H_2O}$\footnote{Water maser; `+'means a detection, `--' a non-detection and `..' not observed.}}\\
      &   & \multicolumn{1}{c}{(h:m:s)}  & \multicolumn{1}{c}{(\degr:\,\arcmin:\,\arcsec)} &(mJy beam$^{-1}$)&(mJy)&(mJy)&(arcsec)& (arcsec)& (deg)&\\
\noalign{\smallskip}
\hline
\noalign{\smallskip}
G012.73--00.58... &MM1   & 18:15:41.3   & --18:12:44   &  42(8)  & 165 & 316 & 28.4 & 15.2 & 26.9  &..\\
G013.28--00.34... &MM1    & 18:15:39.9   & --17:34:37  &  59(12) & 212 & 141 & 22.7 & 17.5 & -49.6 &--\\
G013.91--00.51... &MM1    & 18:17:34.8   & --17:06:52  & 142(15) & 685  & 535 & 27.3 & 17.1 & -22.2 &--\\%
G014.39--00.75A.. &MM1   & 18:19:19.0   & --16:43:49  & 142(17) & 503  & 456  & 27.1 & 14.4 & 57.1  &--\\%
G014.39--00.75B..              &MM3   & 18:19:33.3   & --16:45:01   &  81(9)  & 259  & 377  & 21.5 & 16.4 &  6.7 &..\\%
G014.63--00.57... &MM1   & 18:19:15.2   & --16:29:59   & 689(8)  &3,111 &2.780   &  28.5 &  17.5  & 10.3      & +\\
              &MM2   & 18:19:14.3   & --16:30:41   & 332(9)  & 584  & 650  & 16.2 & 12.0 & -57.3 &..\\%
              &MM3   & 18:19:02.9   & --16:30:29   & 112(11) & 269  & 276  & 21.5 & 12.3 & -42.0 &.. \\%
              &MM4   & 18:19:20.5   & --16:31:42   &  86(9)  & 215  & 265  & 22.1 & 12.5 & -49.3 &..\\%
G016.93+00.24... &MM1    & 18:20:50.8  & --14:06:01  &  76(11) & 357  & 296  & 27.2 & 19.1 & -80.4 &..\\%
G017.19+00.81... &MM1   & 18:19:08.9   & --13:36:29   & 120(10) & 386  & 409  & 21.6 & 16.4 & 65.3 &.. \\%
              &MM2   & 18:19:12.9   & --13:33:46 & 475(11) &1,304 & 1,430 & 17.8 & 16.9 & 87.8 & +\\%
              &MM3   & 18:19:12.1   & --13:33:32  & 157(11) & 725  & 1,080 & 31.3 & 16.2 & -72.9&..\\%
              &MM4   & 18:19:15.2   & --13:39:29  &  87(10) & 426  & 407  & 28.2 & 19.1 & -42.0 &..\\%
G018.26--00.24... &MM1   & 18:25:11.8   & --13:08:04  & 257(16) & 666  &  438 & 19.4 & 14.7 & 26.5 &--\\%
              &MM2   & 18:25:06.4   & --13:08:51  & 221(17) & 887  &  429         & 31.2 & 14.2 & 30.2 &--\\%
              &MM3   & 18:25:05.6   & --13:08:20  & 152(16) & 231  &  201 & 14.8 & 11.3 & -42.8  &..\\%
              &MM4   & 18:25:04.5   & --13:08:27  & 133(14) & 211  &  168 & 16.7 & 10.5 & 47.7 &..\\%
              &MM5   & 18:25:01.8   & --13:09:06  & 141(19) & 559  &  291 & 24.1 & 18.1 & 64.6 &..\\%
G022.06+00.21... &MM1   & 18:30:34.7   &  --9:34:46  & 995(14) &2,019  & 1,820 & 15.6 & 14.3 & 70.6 &+\\%
              &MM2   & 18:30:38.5   &  --9:34:29  & 166(15) & 259  & 208 & 14.6 & 11.8 & 64.0 &..\\%
G023.38--00.12... &MM1   & 18:34:23.5   &  --8:32:20  & 171(15) & 521  & 246 & 22.3 & 15.1 & 29.1 &..\\%
G024.37--00.15... &MM1   & 18:36:27.8   &  --7:40:24  & 133(12) & 529  & 350  & 26.8 & 14.4 & 33.0  &--\\%
                 &MM2   & 18:36:18.3   &  --7:41:00  &  94(11) & 254  & 202  & 17.3 & 15.1 & 42.7 &+\\%
G024.61--00.33... &MM1   & 18:37:23.1   &  --7:31:39  & 147(8)  & 655  & 496  & 25.3 & 19.4 & 84.2 &-- \\%
                 &MM2   & 18:37:21.3   &  --7:33:07  &  75(8)  & 276  & 187  & 26.1 & 15.6 & 20.6 &--\\%
G024.94--00.15... &MM1   & 18:37:19.7   &  --7:11:41   & 133(15) & 350  & 318 & 19.4 & 13.2 &-32.8 &+ \\%
              &MM2   & 18:37:12.2   &  --7:11:23  & 114(15) & 278  & 218  & 19.2 & 12.4 & 23.2 &--\\%
G030.90+00.00A..&MM1   & 18:47:28.9   &  --1:48:07   & 141(10) & 728  & 307 & 32.1 & 17.7 & -2.5 &.. \\%
G030.90+00.00B..&MM2   & 18:47:41.9   &  --1:52:13  & 116(19) & 494  & 130 & 26.9 & 17.4 & 43.7 &-- \\%
G030.90+00.00C..&MM3   & 18:47:48.2   &  --1:51:30  &  98(20) & 369  & 164 & 24.3 & 17.1 &-10.5 &-- \\%
G030.90+00.00D..&MM4   & 18:47:51.5   &  --1:49:24  &  106(10) & 260   & 279 & 20.9 & 13.0 & 32.4  &..\\
G034.71--00.63... &MM1   & 18:56:48.3   &   1:18:49 & 148(9)  &1,174 & 678 & 36.0 & 24.3 &  2.2  &--\\%
              &MM2   & 18:56:58.2   &   1:18:44 &  79(11) & 368  & 250 & 30.2 & 17.1 &-81.7  &..\\%
              &MM3   & 18:57:06.5   &   1:16:52 &  60(11) & 203  & 179 & 23.9 & 15.6 & 78.9  &--\\%
G035.49--00.30A.. &MM1   & 18:57:05.2   &   2:06:29 & 180(11) & 374  & 339 & 16.8 & 13.6 &-38.8  &+\\%
G035.49--00.30B.. &MM2   & 18:57:08.4   &   2:09:01 &  76(10) & 204  & 139 & 19.1 & 15.5 & 28.0 &--\\%
              &MM3   & 18:57:08.1   &   2:10:47 &  73(14) & 437  & 279 & 33.3 & 19.9 & -1.6  &+ \\%
              &MM4   & 18:57:09.0   &  2:08:23  &  60(10) & 178   & 118     & 24.1 & 13.5 & -1.4&..\\
              &MM5   & 18:57:06.7   &  2:08:27  & 62(10) & 211 & 154 & 26.4 & 14.3 & -28.9&..\\ 
              &MM6   & 18:57:11.5   &  2:07:27  & 55(11) & 274 & 184 & 29.0 & 19.1 & -6.1&..\\
G050.06+00.06... &MM1   & 19:23:12.4   &  15:13:35 &  89(8)  & 332  & 167 & 26.9 & 15.4 & 43.1 &--\\%
              &MM2   & 19:23:09.2   &  15:12:42  &  84(9)  & 246  & 157 & 30.5 & 14.8 & 83.0 &..\\ %
G053.81--00.00... &MM1    & 19:30:55.7  &  18:29:55 & 106(11) & 226  & 291 & 16.5 & 14.3 &-50.8 &--  \\%
\noalign{\smallskip}
\hline
\noalign{\smallskip}
\end{tabular}
\\
{\bf Notes.} The first two columns give the HEC name and the millimeter clump number, the following columns represent (in order of appearance): right ascension, declination, 1.2\,mm peak flux, integrated 1.2\,mm flux, integrated 1.2\,mm flux within 0.25\,pc diameter, clump major axis, clump minor axis, position angle and water maser detection.
\end{minipage}
\end{table*} 
\begin{table*}
\begin{minipage}[t]{\textwidth}
\renewcommand{\footnoterule}{}
\begin{center}
\caption{\label{ta:noclump} Positions (J2000) and determined properties toward the positions without clumps }                           
\begin{tabular}{l l r r r r r r r }     
\hline\hline
HEC name &  & \multicolumn{1}{c}{R.A.}     & \multicolumn{1}{c}{Declination} & \multicolumn{1}{c}{$d_\mathrm{kin}$}  &\multicolumn{1}{c}{$F^i_{\mathrm{0.25 pc}}$}&\multicolumn{1}{c}{$M_{0.25pc}$} &\multicolumn{1}{c}{$T_{\mathrm{rot}}$} &\multicolumn{1}{c}{${N_{\mathrm{NH_3}}}$} \\
      &   & \multicolumn{1}{c}{(h:m:s)}  & \multicolumn{1}{c}{(\degr:\,\arcmin:\,\arcsec)} &(kpc)&(mJy)& \multicolumn{1}{c}{($\mathrm{M_\odot}$)}  &\multicolumn{1}{c}{(K)} & \multicolumn{1}{c}{($10^{15}$\,cm$^{-2}$)}\\
\noalign{\smallskip}
\hline
\noalign{\smallskip}
G012.73--00.58... &MM2   & 18:15:32.7   & --18:10:15  & 1.1&178  &8    & 11.4(0.7)  & ..    \\
G013.97--00.45... &MM1    & 18:17:16.5   & --17:01:16 & 2.4&189  &25   & 16.6(0.9)   & ..    \\
G014.39--00.75A.. &MM2   & 18:19:17.4   & --16:44:04  & 2.1&382  &32   & 20.0(2.3)  & ..     \\
G023.38--00.12... &MM2   & 18:34:20.4   &  --8:33:16  & 5.6&..   & 44  &17.8(2.0).    &..    \\
G025.79+00.81... &MM1   & 18:35:20.5   &  --5:56:36  & 3.4 &196 & 77  &  13.1(0.7) & .. \\%
              &MM2   & 18:35:26.3   &  --5:59:21  & 3.4 & 63 &  31  & ..   &..\\
G034.03--00.33... &MM1   & 18:54:25.1   &   0:49:56  & ..  &..  & .. & .. &..  \\
              &2     & 18:54:39.2   &   0:51:37  & ..  &..  & .. & .. & .. \\
G034.34--00.90... &1      & 18:57:16.5   &   0:50:48 &..   &..  &..  &..\\
G034.77--00.81... &MM1    & 18:57:40.7  &   1:16:09 & 2.9   & 49 & 11  & ..   & ..\\
G034.85+00.43... &MM1    & 18:53:23.2  &   1:53:16 &3.6  & 73 & 31   &13.2(1.5)    & ..\\
G037.44+00.14A.. &MM1   & 18:59:14.0   &   4:07:37 &1.3  & 163 &  8  & ..   &..  \\
G037.44+00.14B..              &MM2   & 18:59:10.2   &   4:04:32 &1.7  & 130 &  ..  &  ..  &..\\
\noalign{\smallskip}
\hline
\end{tabular}
\end{center}
{\bf Notes.} The first two columns give the HEC name and the millimeter clump
number, the following columns represent (in order of appearance): right
ascension, declination, kinematic distance, integrated 1.2\,mm flux within
0.25\,pc diameter, mass within 0.25\,pc diameter, \amm\ rotational temperature, and \amm\ column density.\\
\end{minipage}
\end{table*}

\subsubsection{Data calibration}

During the observations, pointings on a nearby compact continuum source  were performed every hour for determining pointing corrections. 
For the flux calibration, we observed a well-known flux calibrator, NGC\,7027 or 3C\,286, in every run.
The post-observational calibration to obtain the main beam brightness
temperature, $T_\mathrm{MB}$, consisted of the opacity correction, elevation
correction, and flux calibration: 
\begin{equation}
T_{\mathrm{MB}} = \frac{A \times T_{\mathrm{ant}} e^{-\tau/\sin\theta}}{G(\theta)}\,,
\label{eq:tmb}
\end{equation}
where $A$ is a scaling factor, $T_{\mathrm{ant}}$ is the antenna temperature,
$\tau$ the zenith opacity, and G the function of the gain with
elevation $\theta$. 

The opacity, $\tau$, was calculated by fitting a linear function to the system temperature against airmass ($\sin^{-1}
\theta$) and taking the slope of the fit.
We found a $\tau$ of 0.031 for good weather. When it was not
possible to retrieve the $\tau$ by this method the
average $\tau$ of 0.054 at 23\,GHz was assumed (the averaged water vapor
radiometer value for 2007). 

For all parabolic dish telescopes the gain decreases at very low and very high elevations. We corrected for this by
dividing by $G(\theta)$ taken from the Effelsberg website\footnote{http://www.mpifr-bonn.mpg.de/div/effelsberg/calibration/1.3cmsf.html} (see Eq.~\ref{eq:tmb}), which is given by
\begin{equation}
G(\theta) = a_0 +a_1\theta + a_2\theta^2\,,
\end{equation}
where $a_0=0.88196$, $a_1=6.6278\times10^{-3}$, $a_2=-9.2334\times10^{-5}$, and $\theta$ the elevation. 
After the opacity and gain corrections, the scaling factor $A$ was found by comparing the measured uncalibrated intensities of absolute flux density calibrator sources with the literature
values calculated from the formulae given by \citet{baars:1977} for 3C\,286, and \citet{ott:1994} for NGC\,7027.

The calibrated spectra were baseline subtracted, and the ammonia lines were
fitted by a Gaussian. Only the
\amm(1,1) line, for which the hyperfine structure was clearly detectable given
the signal to noise ratios, was fitted by special routine `method nh3(1,1)' of
the GILDAS/CLASS software. This method calculates the optical depth from the
hyperfine structure and returns optical depth corrected line widths. The
observed ammonia parameters, such as the velocity in the local standard of
rest (LSR), $V_{\mathrm{LSR}}$, main beam temperatures, $T_{\mathrm{MB}}$, line widths, $\Delta v$, and the main group optical depth, $\tau_{\mathrm{main}}$ are listed in Table \ref{ta:nh3}. 

\onltab{5}{
\begin{table*}
\begin{center}
\caption{ Observed ammonia parameters with uncertainties (in parenthesis) from
  hyperfine and Gaussian fits with CLASS}             
\label{ta:nh3}      
\begin{tabular}{l l r r r r r r r r}     
\hline\hline 
\noalign{\smallskip}    
HEC name && \multicolumn{4}{c}{\amm(1,1)}& \multicolumn{2}{c}{\amm(2,2)}& \multicolumn{2}{c}{\amm(3,3)} \\ 
       & & $V_{LSR}$&$T_{\mathrm{MB}}$ &  $\Delta v$ & $\tau_{\mathrm{main}}$ & $T_{\mathrm{MB}}$ & $\Delta v$& $T_{\mathrm{MB}}$ & $\Delta v$\\
       & &(km\,s$^{-1}$)& (K)        & (km\,s$^{-1}$)   &               & (K)        & (km\,s$^{-1}$)& (K)        & (km\,s$^{-1}$)\\
\noalign{\smallskip}
\hline
\noalign{\smallskip}
G012.73--00.58.. &MM1  & 6.48(0.01) & 1.3(0.3) & 0.7(0.1) & 4.1(0.4) & 0.2(0.1) & 0.8(0.1) & .. & ..\\  
             &MM2  & 6.19(0.01) & 1.8(0.3) & 1.0(0.1) & 1.3(0.1) & 0.3(0.1) & 1.6(0.2) & ..&..\\   
G013.28--00.34..  &MM1 & 41.30(0.06) & 3.2(0.7) & 1.6(0.2) & 1.8(0.5) & 1.3(0.3) & 1.6(0.2) & ..&..\\
G013.91--00.51.. &MM1  & 22.94(0.02) & 2.5(0.3) & 1.3(0.1) & 1.8(0.3) & 0.8(0.2) & 1.3(0.2) & ..&..\\
G013.97--00.45..  &MM1 & 19.76(0.02) & 1.4(0.2) & 2.3(0.1) & 0.7(0.1) & 0.5(0.1) & 3.0(0.1) & 0.3(0.1) & 4.5(0.3)\\   
G014.39--00.75A.. &MM1  & 17.84(0.04) & 1.0(0.1) & 1.2(0.1) & 0.8(0.4) & 0.4(0.1) & 2.9(0.4) & 0.1(0.1) & 3.0(0.9)\\  
             &MM2  & 17.49(0.03) & 1.1(0.1) & 1.2(0.1) & 0.5(0.3) & 0.6(0.1) & 2.2(0.3) & 0.2(0.1) & 2.9(0.5)\\   
G014.39--00.75B..             &MM3  & 21.29(0.03) & 1.5(0.2) & 0.9(0.1) & 2.3(0.5) & 0.4(0.1) & 1.3(0.4) & .. & ..\\
G014.63--00.57.. &MM1  & 18.75(0.01) & 4.0(0.4) & 1.8(0.1) & 2.2(0.1) & 2.4(0.1) & 2.4(0.1) & 1.2(0.1) & 2.7(0.1)\\  
             &MM2  & 18.45(0.02) & 3.0(0.3) & 1.3(0.1) & 1.8(0.2) & 1.3(0.1) & 1.8(0.1) & 0.4(0.1) & 2.4(0.4)\\   
             &MM3  & 17.64(0.04) & 1.0(0.2) & 1.4(0.1) & 2.1(0.4) & 0.5(0.1) & 1.9(0.2) & 0.2(0.1) & 1.5(0.3)\\   
             &MM4  & 19.13(0.05) & 0.6(0.2) & 0.8(0.1) & 2.0(0.9) & 0.4(0.1) & 1.0(0.3) & .. & ..\\
G016.93+00.24.. &MM1  & 23.80(0.01) & 1.5(0.1) & 0.9(0.1) & 1.7(0.2) & 0.5(0.1) & 1.7(0.3) & .. & ..\\
G017.19+00.81.. &MM1 & 25.04(0.01) & 3.2(0.1) & 1.2(0.1) & 1.5(0.1) & 1.5(0.1) & 1.5(0.1) & 0.2(0.1) & 2.5(0.7)\\ 
             &MM2 & 22.75(0.01) & 3.1(0.1) & 1.3(0.1) & 1.3(0.1) & 1.7(0.2) & 1.6(0.1) & 0.5(0.1) & 2.3(0.3)\\ 
             &MM3 & 22.73(0.01) & 3.1(0.2) & 1.3(0.1) & 1.6(0.1) & 2.0(0.2) & 1.5(0.1) & 0.5(0.1) & 2.4(0.2)\\
             &MM4 & 21.64(0.06) & 0.5(0.1) & 2.8(0.1) & 0.9(0.3) & 0.3(0.1) & 3.6(0.3) & 0.2(0.1) & 2.6(0.4)\\
G018.26--00.24.. &MM1 & 68.07(0.01) & 2.6(0.1) & 1.5(0.1) & 2.4(0.2) & 1.6(0.1) & 2.3(0.1) & 0.6(0.1) & 2.8(0.3)\\ 
             &MM2 & 67.75(0.02) & 2.6(0.1) & 2.0(0.1) & 2.8(0.1) & 1.7(0.2) & 2.6(0.2) & 0.3(0.1) & 2.8(0.3)\\
             &MM3 & 68.32(0.01) & 3.6(0.7) & 2.2(0.1) & 2.6(0.1) & 1.8(0.1) & 2.7(0.1) & 0.6(0.1) & 3.3(0.1)\\
             &MM4 & 68.31(0.02) & 3.4(0.4) & 2.1(0.1) & 2.2(0.1) & 1.8(0.1) & 2.8(0.1) & 0.8(0.1) & 2.9(0.2)\\
             &MM5 & 66.28(0.03) & 1.9(0.2) & 2.0(0.1) & 2.6(0.2) & 1.1(0.1) & 2.0(0.2) & 0.2(0.1) & 5.0(1.1)\\
G022.06+00.21.. &MM1 & 51.17(0.02) & 1.8(0.1) & 1.7(0.1) & 1.9(0.2) & 1.5(0.1) & 2.2(0.2) & 0.5(0.1) & 3.6(0.3)\\
             &MM2 & 51.45(0.02) & 1.1(0.1) & 1.2(0.1) & 1.9(0.3) & 0.5(0.1) & 2.9(0.5) & 0.1(0.1) & 5.3(1.0)\\
G023.38--00.12.. &MM1 & 98.42(0.02) & 1.7(0.1) & 1.6(0.1) & 2.3(0.2) & 1.1(0.1) & 1.9(0.1) & 0.2(0.1) & 3.4(0.3)\\
             &MM2 & 98.86(0.02) & 1.0(0.1) & 1.1(0.1) & 1.4(0.3) & 0.5(0.1) & 1.1(0.1) & ..&..\\ 
G024.37--00.15.. &MM1 & 58.83(0.03) & 0.9(0.1) & 2.1(0.1) & 2.8(0.3) & 0.6(0.1) & 2.6(0.2) & 0.2(0.1) & 4.2(0.4)\\  
             &MM2 & 55.96(0.03) & 1.3(0.1) & 1.5(0.1) & 2.2(0.3) & 0.6(0.1) & 2.0(0.3) & 0.1(0.1) & 3.2(1.0)\\   
G024.61--00.33.. &MM1 & 42.80(0.02) & 1.3(0.1) & 1.4(0.1) & 1.2(0.2) & 0.6(0.1) & 1.5(0.2) & 0.1(0.1) & 4.9(1.1)\\ 
             &MM2 & 43.58(0.01) & 1.9(0.1) & 0.8(0.1) & 1.7(0.2) & 0.8(0.1) & 1.3(0.1) & 0.2(0.1) & 3.6(0.8)\\  
G024.94--00.15.. &MM1 & 47.10(0.01) & 2.8(0.1) & 1.5(0.1) & 2.3(0.1) & 1.3(0.2) & 1.9(0.2) & 0.4(0.1) & 2.4(0.3)\\  
             &MM2 & 48.03(0.02) & 1.8(0.1) & 1.4(0.1) & 2.2(0.2) & 0.8(0.1) & 2.1(0.2) & 0.2(0.1) & 3.7(0.7)\\  
G025.79+00.81.. &MM1 & 49.70(0.01) & 2.0(0.1) & 1.1(0.1) & 2.1(0.1) & 0.6(0.1) & 1.7(0.2) & .. & ..\\ 
             &MM2 & 49.81(0.02) & 0.9(0.2) & 0.9(0.1) & 1.3(0.4) & ..  & .. & .. & ..\\  
G030.90+00.00A.. &MM1 & 74.82(0.03) & 1.6(0.1) & 1.5(0.1) & 1.7(0.3) & 0.9(0.1) & 2.0(0.1) & 0.3(0.1) & 2.1(0.4)\\  
G030.90+00.00B..             &MM2 &109.55(0.06) & 1.4(0.1) & 2.5(0.1) & 1.9(0.3) &  ..  & .. & .. & ..\\
G030.90+00.00C..             &MM3 & 93.78(0.03) & 1.2(0.1) & 1.1(0.1) & 2.7(0.4) & 0.7(0.2) & 2.3(0.4) & 0.1(0.1) & 1.7(0.9)\\  
G030.90+00.00D..             &MM4 & 37.61(0.05) & 0.8(0.1) & 1.0(0.1) & 1.9(0.8) & .. & .. & .. & .. \\  
G034.71--00.63.. &MM1 & 44.61(0.02) & 1.7(0.1) & 1.7(0.1) & 1.1(0.1) & 0.8(0.1) & 2.2(0.2) & 0.3(0.1) & 4.4(0.3)\\  
             &MM2 & 45.43(0.02) & 1.9(0.4) & 1.3(0.1) & 3.0(0.3) & 0.6(0.1) & 1.5(0.1) & 0.2(0.1) & 6.4(0.8)\\    
             &MM3 & 46.16(0.02) & 1.1(0.1) & 1.4(0.1) & 0.6(0.2) & 0.4(0.1) & 1.3(0.2) & ..&..\\    
G034.77--00.81.. &MM1 & 43.36(0.08) & 0.6(0.1) & 1.6(0.2) & 0.5(0.7) & .. & .. & .. & .. \\ 
G034.85+00.43.. &MM1 & 55.52(0.03) & 1.1(0.3) & 0.8(0.1) & 1.8(0.5) & 0.3(0.1) & 0.8(0.1) & ..&..\\ 
G035.49--00.30A.. &MM1 & 55.23(0.06) & 0.7(0.1) & 1.9(0.1) & 0.7(0.4) & 0.3(0.1) & 2.2(0.4) & 0.2(0.1) & 4.1(0.6)\\    
G035.49--00.30B.. &MM2 & 45.31(0.01) & 2.7(0.3) & 1.0(0.1) & 3.5(0.3) & 0.8(0.1) & 1.3(0.1) & 0.1(0.1) & 2.4(1.2)\\    
             &MM3 & 45.71(0.02) & 2.1(0.1) & 1.6(0.1) & 2.5(0.2) & 0.8(0.1) & 1.6(0.1) & 0.1(0.1) & 1.9(0.7)\\    
G037.44+00.14A.. &MM1 & 18.37(0.02) & 1.3(0.2) & 0.8(0.1) & 0.6(0.3) & 0.4(2.0) & 0.5(2.2) & ..&..\\  
G037.44+00.14B..             &MM2 & 40.17(0.11) & 0.4(0.1) & 1.3(0.3) &0.5(1.1) & .. & .. & ..&..\\
G050.06+00.06.. &MM1 & 54.11(0.03) & 0.9(0.1) & 1.3(0.1) & 1.5(0.3) & 0.3(0.1) & 1.3(0.3) &  ..&..\\ 
             &MM2 & 54.53(0.03) & 0.8(0.1) & 1.3(0.1) & 1.2(0.3) & 0.2(0.1) & 3.2(0.7) &  ..&..\\ 
G053.81--00.00.. &MM1 & 24.13(0.02) & 1.3(0.1) & 1.4(0.1) & 1.3(0.2) & 0.3(0.1) & 3.1(0.9) &  ..&..\\  
\noalign{\smallskip}             
\hline  
\end{tabular}
\end{center}
\end{table*} 
} 
\subsection{Water maser observations}

Several positions with a peak in the 1.2\,mm emission above twice the
mean cloud emission, indicating that they are possibly harboring evolved clumps, were searched for water maser emission using the 100m
Effelsberg telescope on 5 and 24 of February 2008. In total 24 positions in common with \amm\ were observed (marked in Table \ref{ta:flux}).
We performed on-off
observations using the FFTS backend with a
bandwidth of 20\,MHz centered on 22.235\,GHz. This setup afforded a high
spectral resolution of 0.04\,km~s$^{-1}$,
while allowing the water maser to have line widths of up to 100\,km~s$^{-1}$. The bandwidth was 270\,km~s$^{-1}$ and was centered on the \amm\ $V_\mathrm{LSR}$ of the observed clump.
We applied the same data reduction as described above in Sect. 3.2.1. for the
ammonia observations. Table \ref{ta:water} lists the $V_\mathrm{LSR}$ and peak intensity of the water maser detections. In case of multiple maser components we give the $V_\mathrm{LSR}$ range.

\begin{table}
\centering
\caption{\label{ta:water} Detected water masers}
\begin{tabular}{llcc}
\hline
\hline
\noalign{\smallskip}
HEC name & & Peak Intensity & $V_\mathrm{LSR}$\\
         & & (Jy~beam$^{-1}$) & (km~s$^{-1}$)\\
\noalign{\smallskip}
\hline
\noalign{\smallskip}
G014.63--00.57 & MM1 & 9.3 &22\\
G017.19+00.81 & MM2 & 22.3&$-4-37$\\
G022.06+00.21 & MM1 & 11.7 &$42-51$\\
G024.37--00.15 & MM2 & 6.0 &65\\
G024.94--00.15 & MM1 & 6.6 &$52-72$\\
G035.49--00.30A & MM1 & 5.6 & $40-77$\\
G035.49--00.30B & MM2 & 2.6 & 40\\
\noalign{\smallskip}
\hline
\end{tabular}
\end{table}

\section{Results}

\subsection{Kinematic distances}

Accurate distance determination within the Galaxy is generally difficult. Nevertheless, the kinematic distance is commonly used as a distance measure. It is based on a model of Galactic rotation, the ``rotation curve'' characterized by the distance between the Sun and the Galactic Center, $R_0$, and the rotation velocity at the Suns orbit, $\Theta_0$.
Toward the inner part of the Galactic plane, kinematic distances are ambiguous: for a given Galactic longitude and LSR velocity, it cannot a priori be determined if the object is at the ``near'' or ``far'' kinematic distance. 
All extinction clouds should be at the near kinematic
distance since at large distances the number of background stars decreases and the percentage of foreground stars increases making it difficult to measure any color excess.
We calculated the kinematic distances for all clumps with \amm\ detections,
using a program of Todd Hunter, which applies the Galactic rotation model of
\citet{fich:1989} assuming a flat rotation curve, $\Theta_0=220\,\mathrm{km
  s^{-1}}$, and $R_0=8.5$\,kpc. The resulting kinematic distances are given in
Table  \ref{ta:noclump} and \ref{ta:mass}.
 
\subsection{Galactic scale extinction}

\begin{figure}[!hptb]
\centering
\includegraphics[height=\columnwidth,angle=-90]{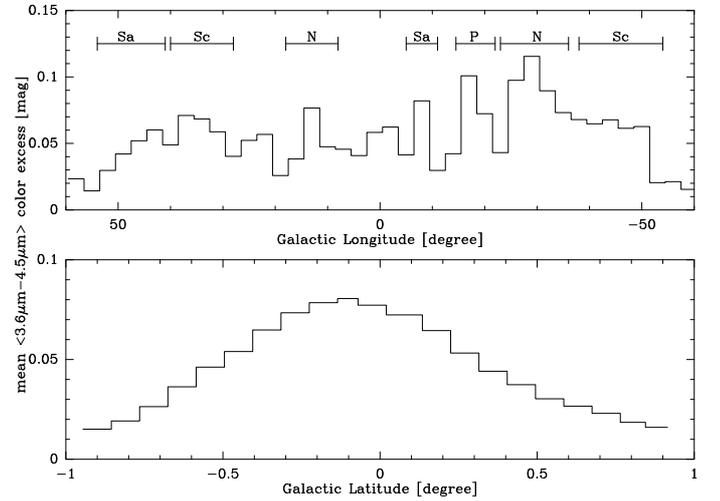}
\caption{Mean color excess along Galactic longitude (upper diagram, bins of 3\degr) and
  Galactic latitude (lower diagram, bins of 0.1\degr). In the longitude
  histogram, encounters with a spiral arm are marked with letters representing
  N for Norma-Cygnus, Sc for Scutum-Crux, Sa for Sagittarius-Carina, and P for Perseus.}
\label{fig:ll}
\end{figure}

The low resolution extinction maps show large extinction structures
across the whole inner Galactic disk from 60\degr\
to $-60$\degr\ longitude (Fig.~\ref{fig:galext}). In longitude, the
3\degr\ averaged distribution (Fig.~\ref{fig:ll}, top panel) shows signs of Galactic structure, meaning that peaks in the distribution can be related to known spiral arms. The column density appears higher in the fourth quadrant, $0\degr>l>-60\degr$, than in the first, $60\degr>l>0\degr$. Additionally, there seems to be a quasi symmetrical distribution around the Galactic Center around longitudes of $\sim$15\degr, --10\degr and $\sim$35\degr, --30\degr. 
In latitude (Fig.~\ref{fig:ll}, lower panel), we found a peak towards
$b\sim-0\rlap{$.$}\,\degr1$, which is also seen for the compact submillimeter
sources found in the ATLASGAL survey \citep{schuller:2009}. The shift of the peak out of the midplane indicates that the Sun is located above the midplane. Similar results have been found by the studies using young open clusters and OB stars \citep[see e.g.][]{joshi:2007}.  
The FWHM of our distribution is $\sim$1\degr, while in ATLASGAL this is more narrow, $\sim$$0\rlap{$.$}\,\degr6$. If one assumes that the Galactic disk has a constant scale height, then objects closer to the Sun should have a wider latitude distribution then ones at larger distances. It therefore appears that the high extinction clouds are on average closer to the Sun than the ATLASGAL sources.  

\begin{figure}
\centering
\includegraphics[angle=-90,width=5cm]{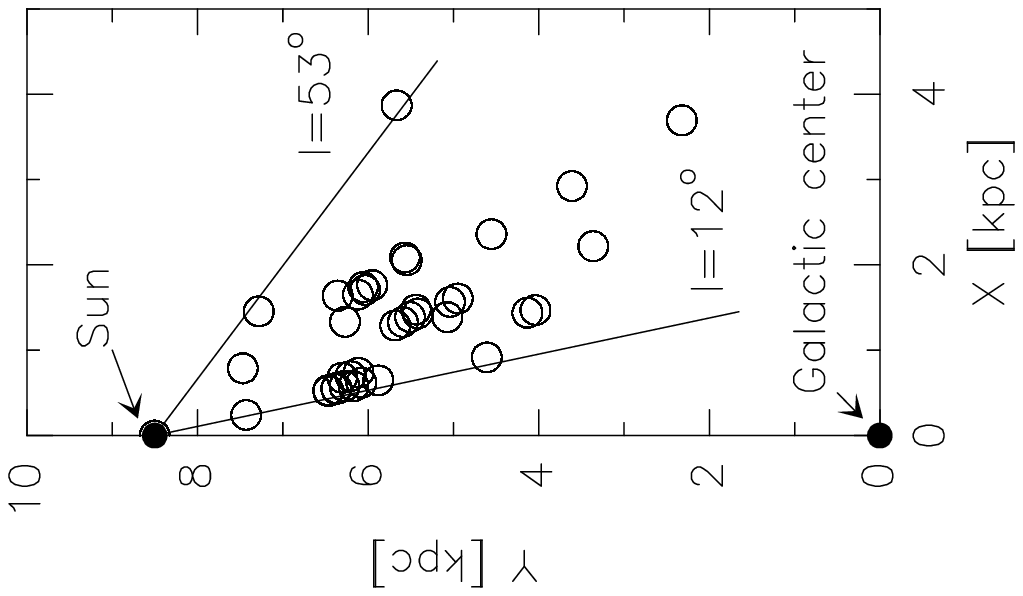}
\caption{\label{fig:galplot} A face-on view of the Galactic plane. The
  Galactic center is located at (0,0). The Galacitic longitudes of 12\degr\
  and 53\degr\ are indicated by solid lines. The locations of the sample of
  high extinction clouds, which were studied in this paper, are marked by open circles.}
\end{figure}

With the kinematic distances we can place our sample of clouds within a face-on view of the Galactic plane (Fig.~\ref{fig:galplot}). They inhabit similar regions as IRDCs \citep[see the IRDCs distribution by][]{jackson:2008}. The extinction method misses nearby ($d<1\,\mathrm{kpc}$) and most of the far away ($d>4\,\mathrm{kpc}$) clouds. The insensitivity to the latter stems from the increasing number of foreground stars at larger distances (see Sect.~\ref{sec:hecmap}). Between distances of one and four kilo parsec the high extinction clouds agree with the IRDCs regions. 

\subsection{Extinction clouds}
\begin{figure}[!htpb]
\includegraphics[height=\columnwidth, angle=-90]{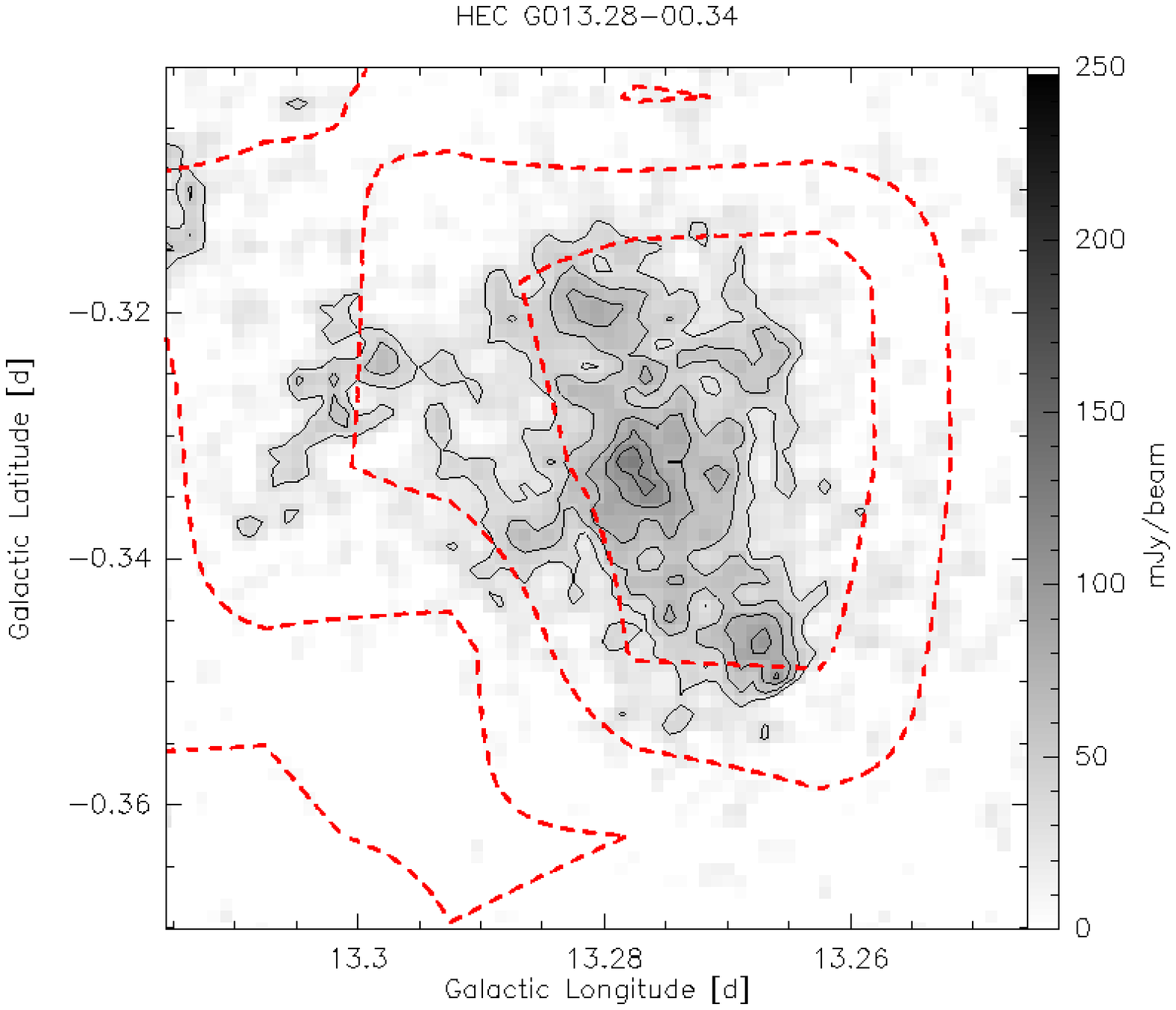}
\includegraphics[height=\columnwidth, angle=-90]{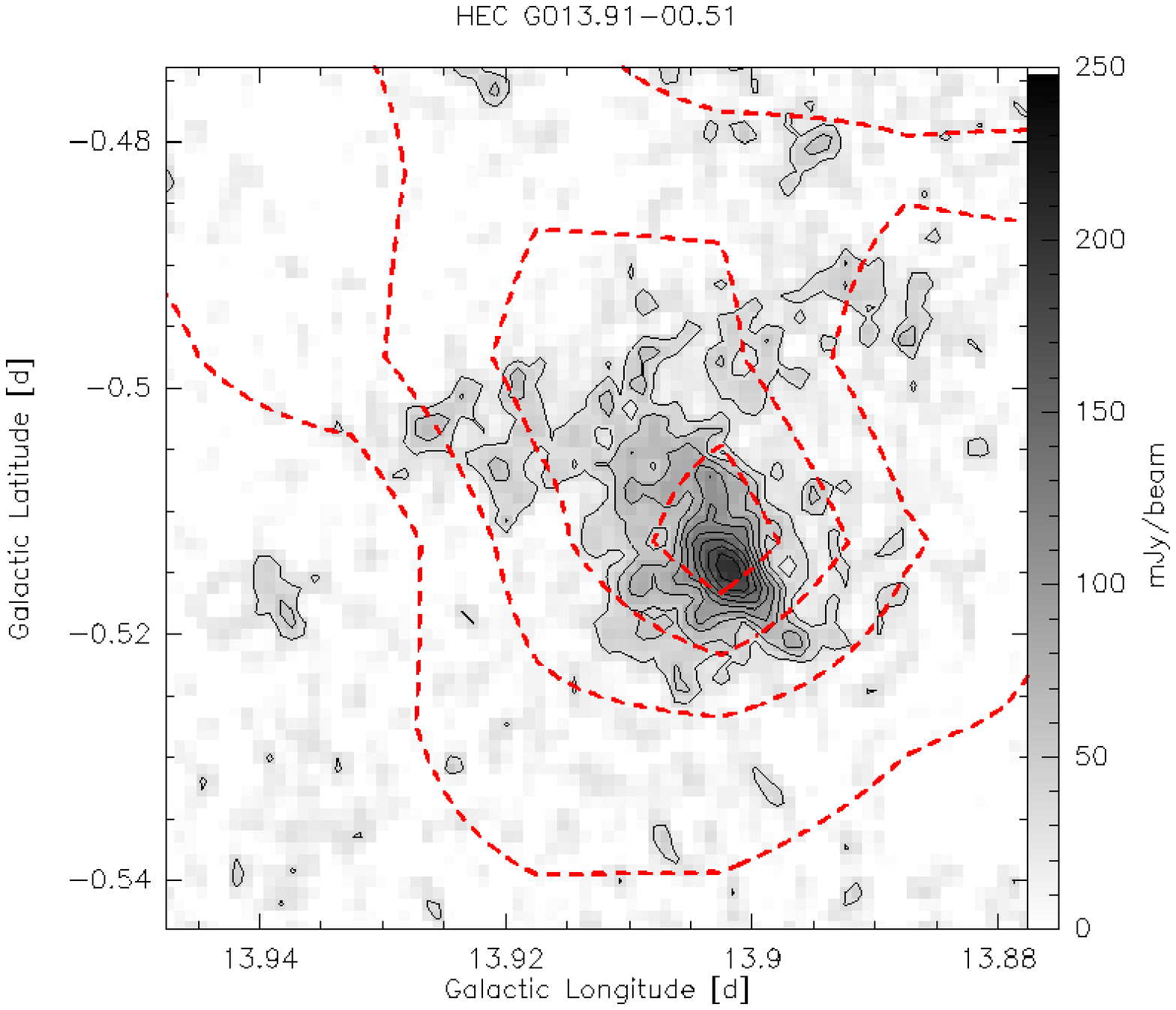}
\includegraphics[height=\columnwidth, angle=-90]{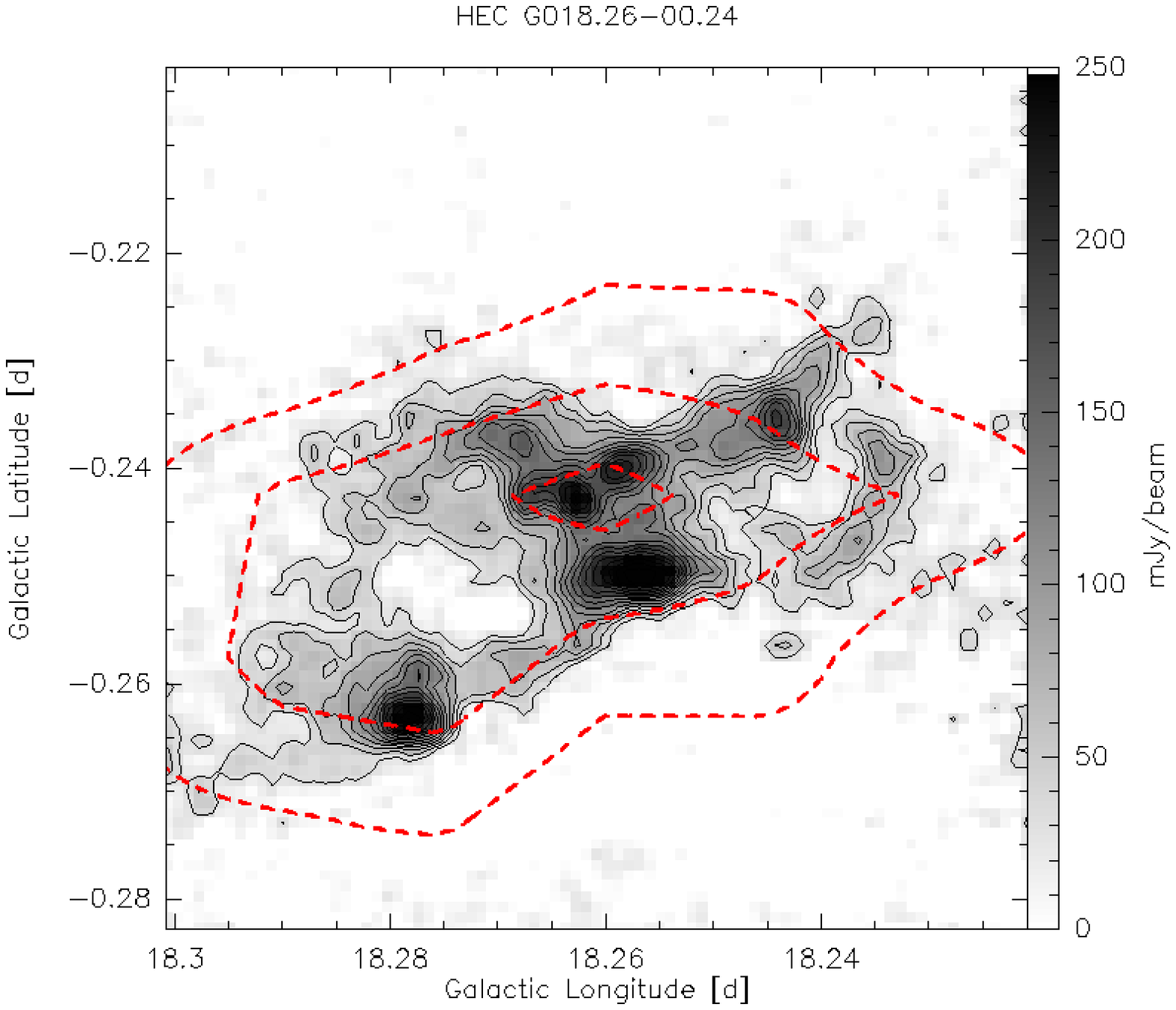}
\caption{Examples of extinction maps (red dashed contour, starting at
  $A_\mathrm{V}$=21\,mag and increasing by 10 magnitudes, beam 54\arcsec) on top of the
  1.2\,mm emission observed with MAMBO-2 (beam $10\rlap{$.$}\,\arcsec5$) in greyscale and solid black contours. }
\label{fig:ext-mambo}
\end{figure}

\begin{figure*}[!htpb]
\centering
\includegraphics[width=7cm, angle=-90]{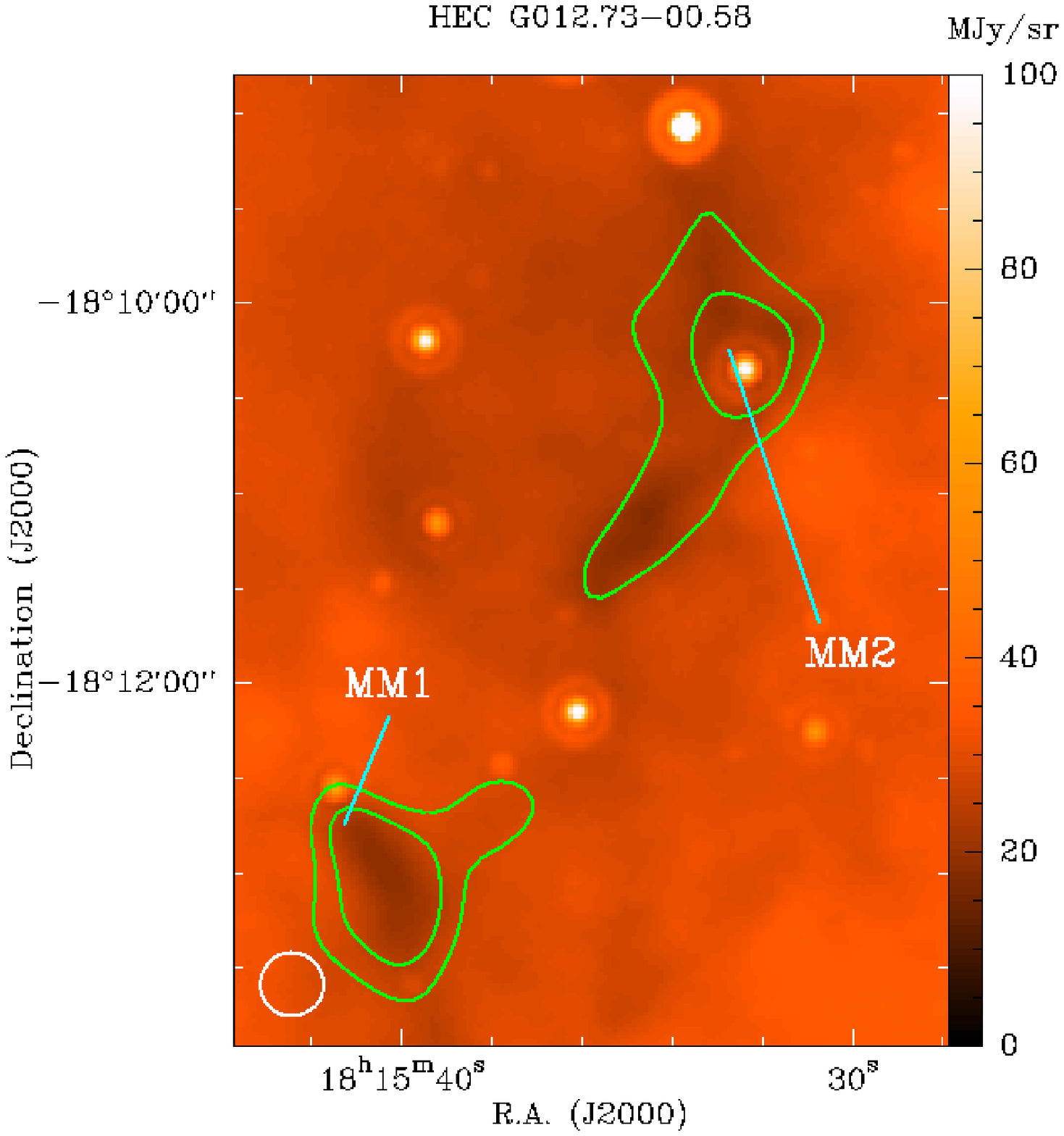}
\includegraphics[width=7cm, angle=-90]{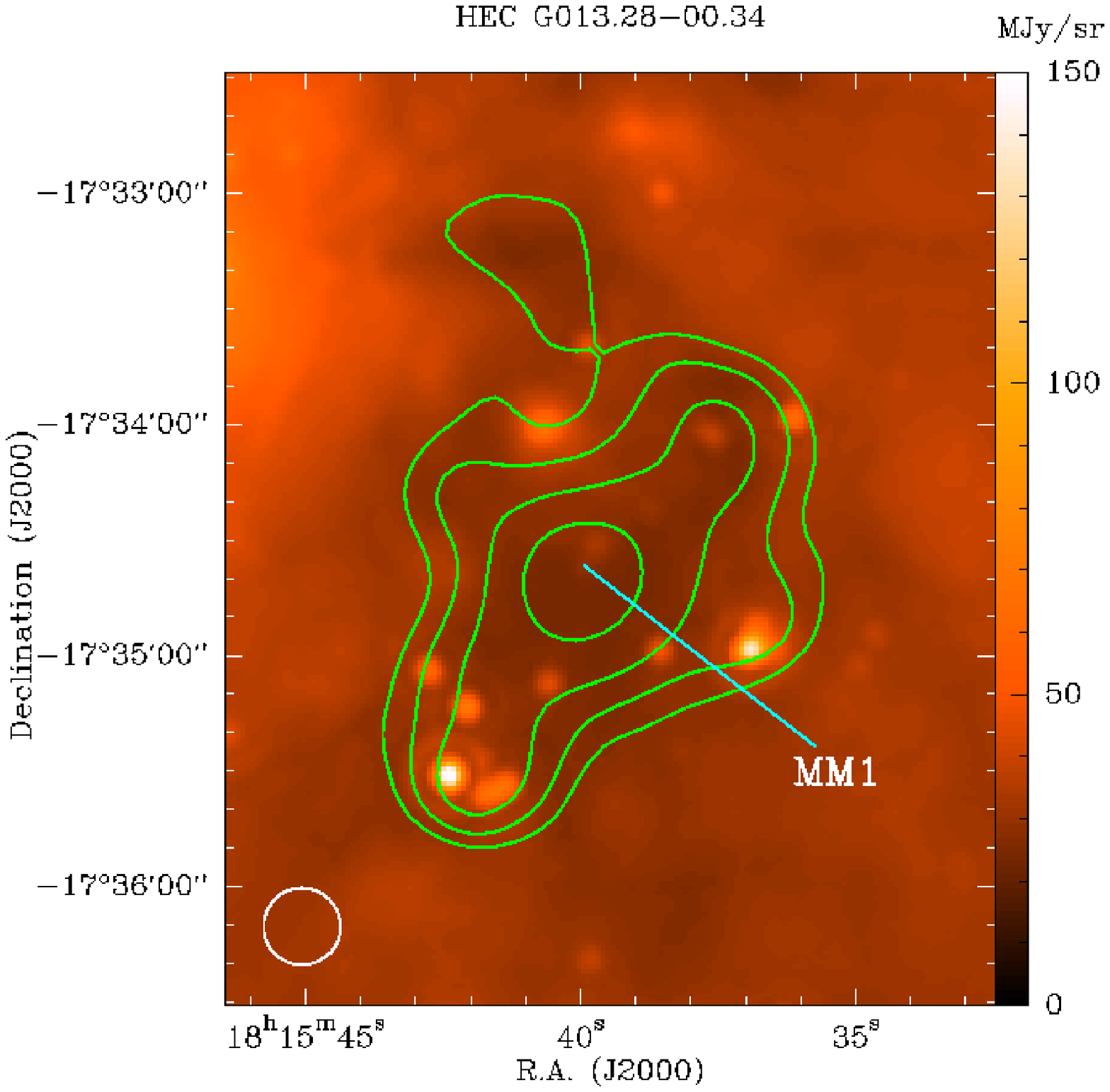}
\includegraphics[width=7cm, angle=-90]{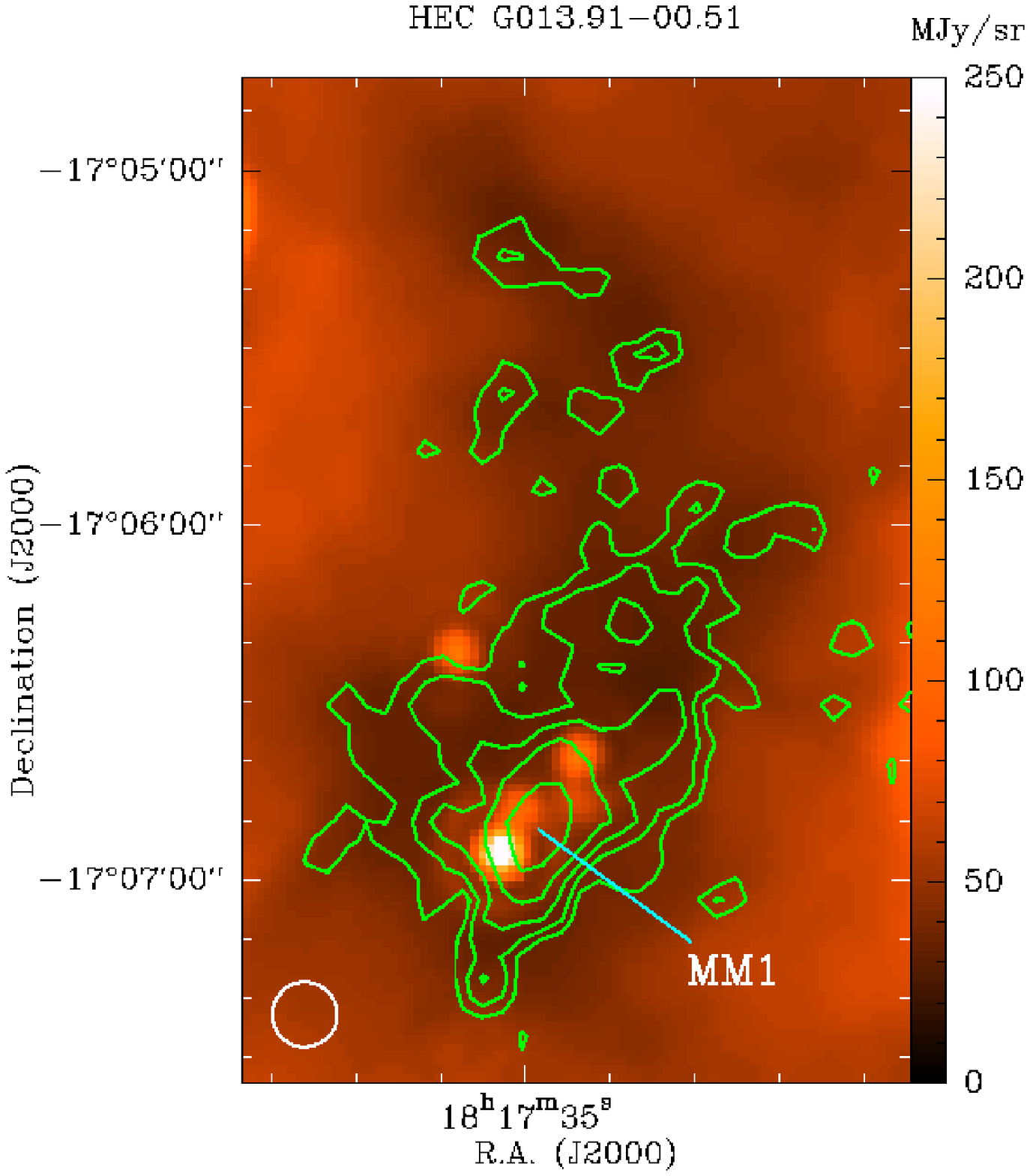}
\includegraphics[width=7cm, angle=-90]{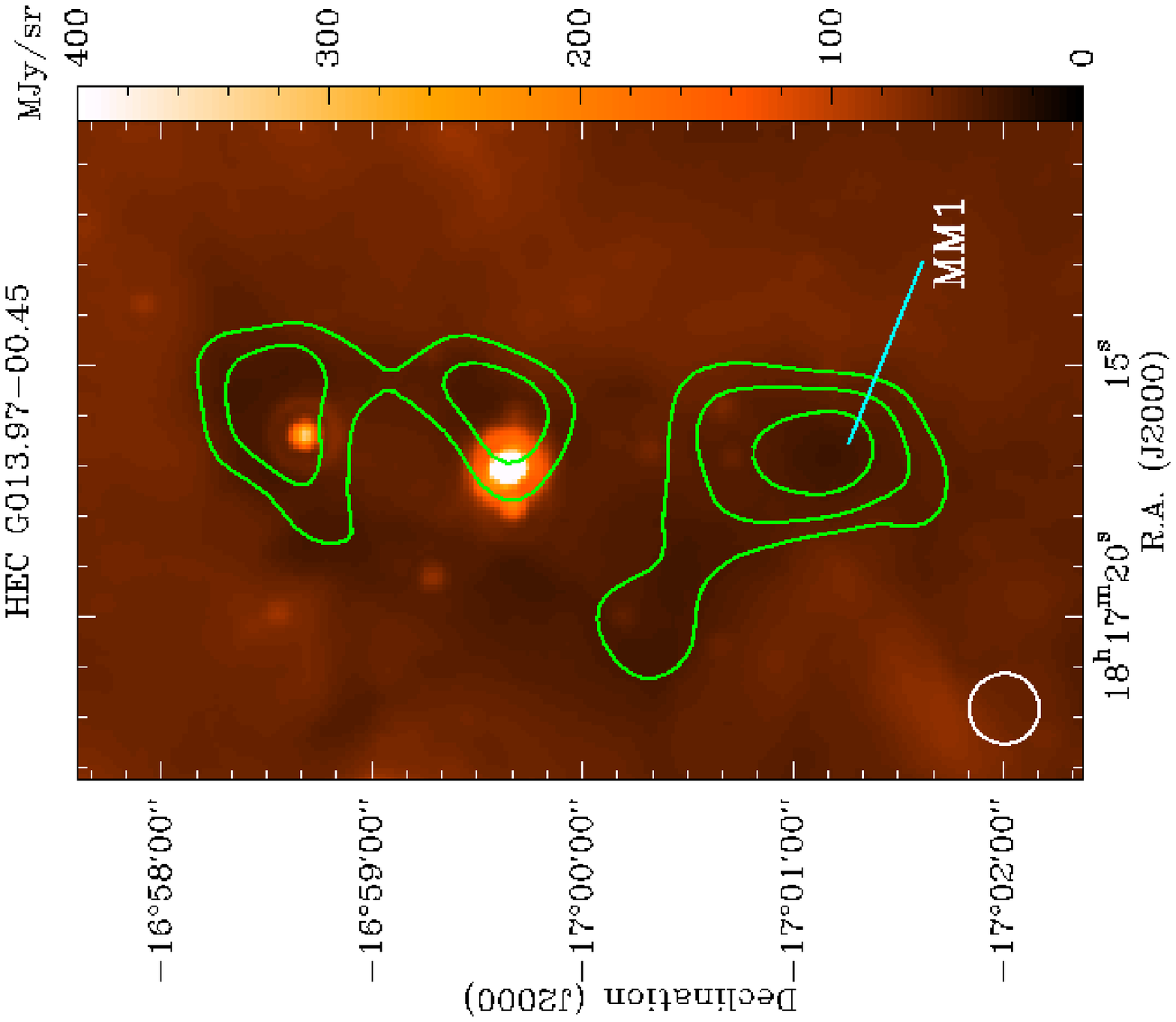}
\includegraphics[width=7cm, angle=-90]{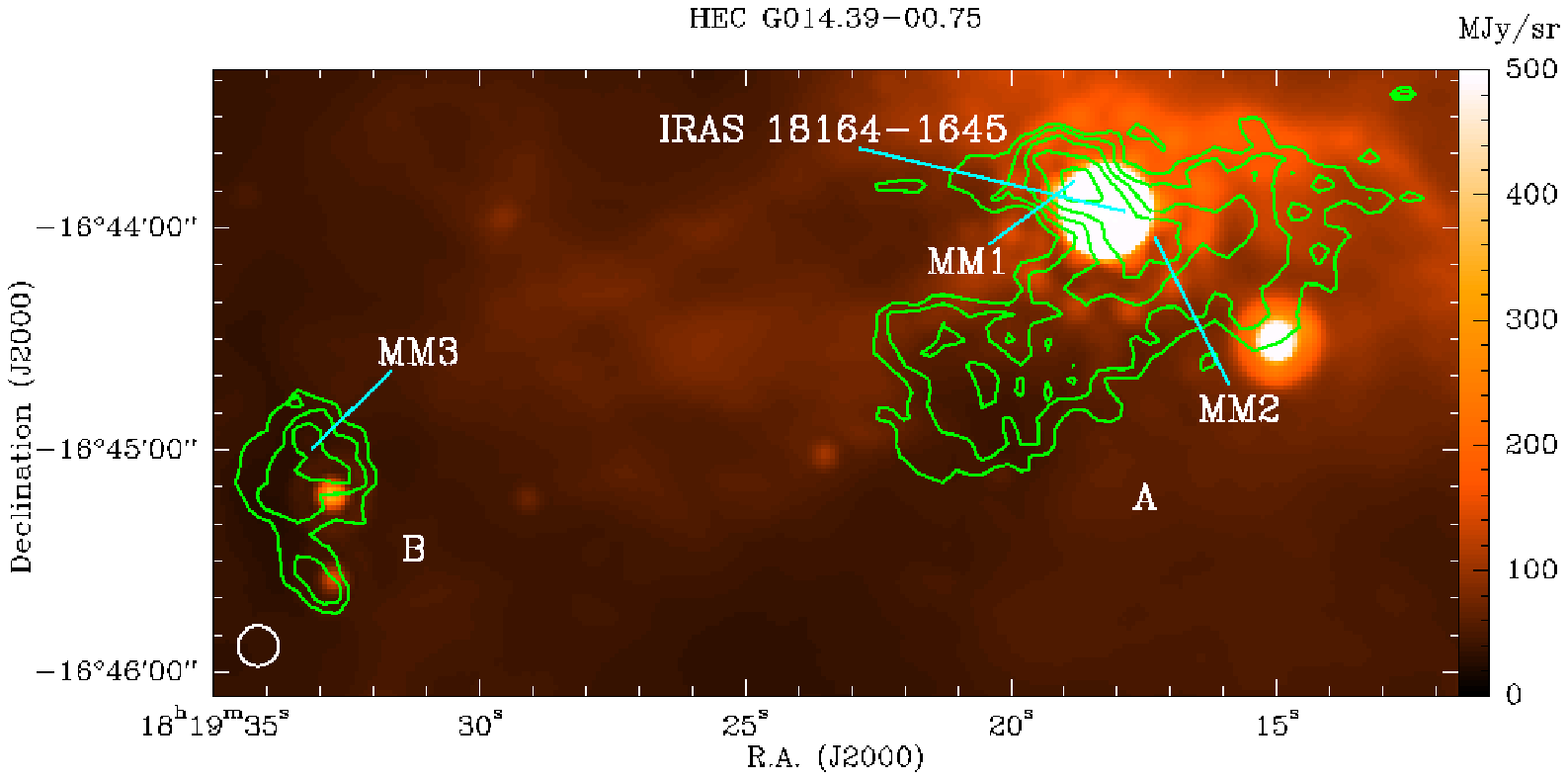}
\caption{\label{fig:mips} Spitzer/MIPS 24 $\mathrm{\mu m}$ images overlaid with
  the 1.2 mm continuum emission contours starting at $2\sqrt{2}\sigma$
  ($\sigma=0.015$\,Jy~beam$^{-1}$) and
  increasing by factors of $\sqrt{2}$. The 1.2\,mm clumps are labeled with their
  designation (as listed in Table \ref{ta:flux}, \ref{ta:mass}
  and \ref{ta:nh3}). In the bottom left corner we indicate the original MAMBO
  beam at 1.2\,mm ($10\rlap{$.$}\,\arcsec5$), or the smoothed beam (20\arcsec) when a smoothed map is shown. Additionally, 6.7\,GHz methanol masers (cyan
  triangles), water masers (blue squares), H{\sc ii} regions (green
  pentagons with black contours), and IRAS sources are marked in the images. The 1.2\,mm continuum emission maps are available in fits format on the CDS.}
\end{figure*}
\addtocounter{figure}{-1}
\begin{figure*}[!htpb]
\centering
\includegraphics[width=7cm, angle=-90]{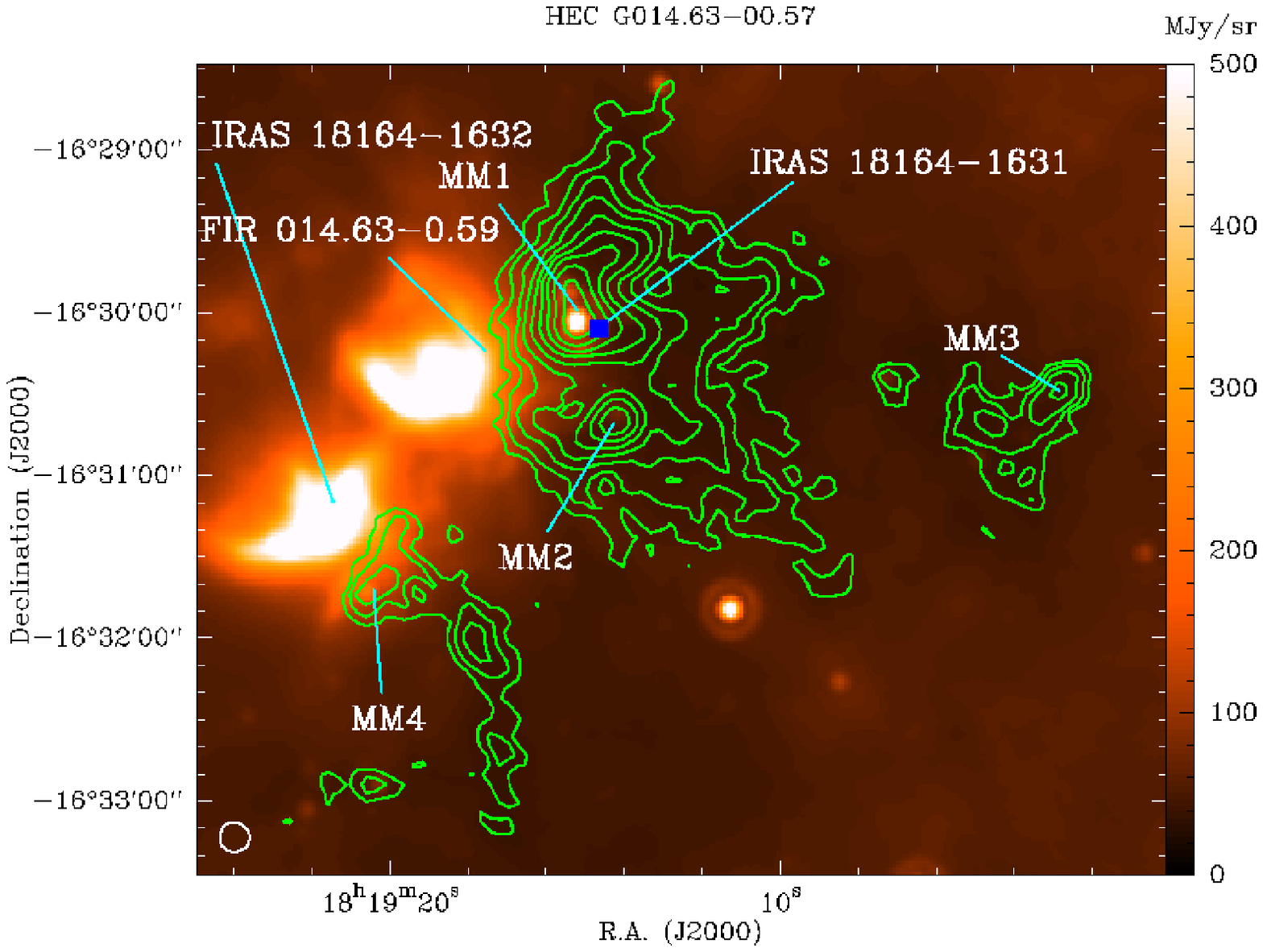}
\includegraphics[width=7cm, angle=-90]{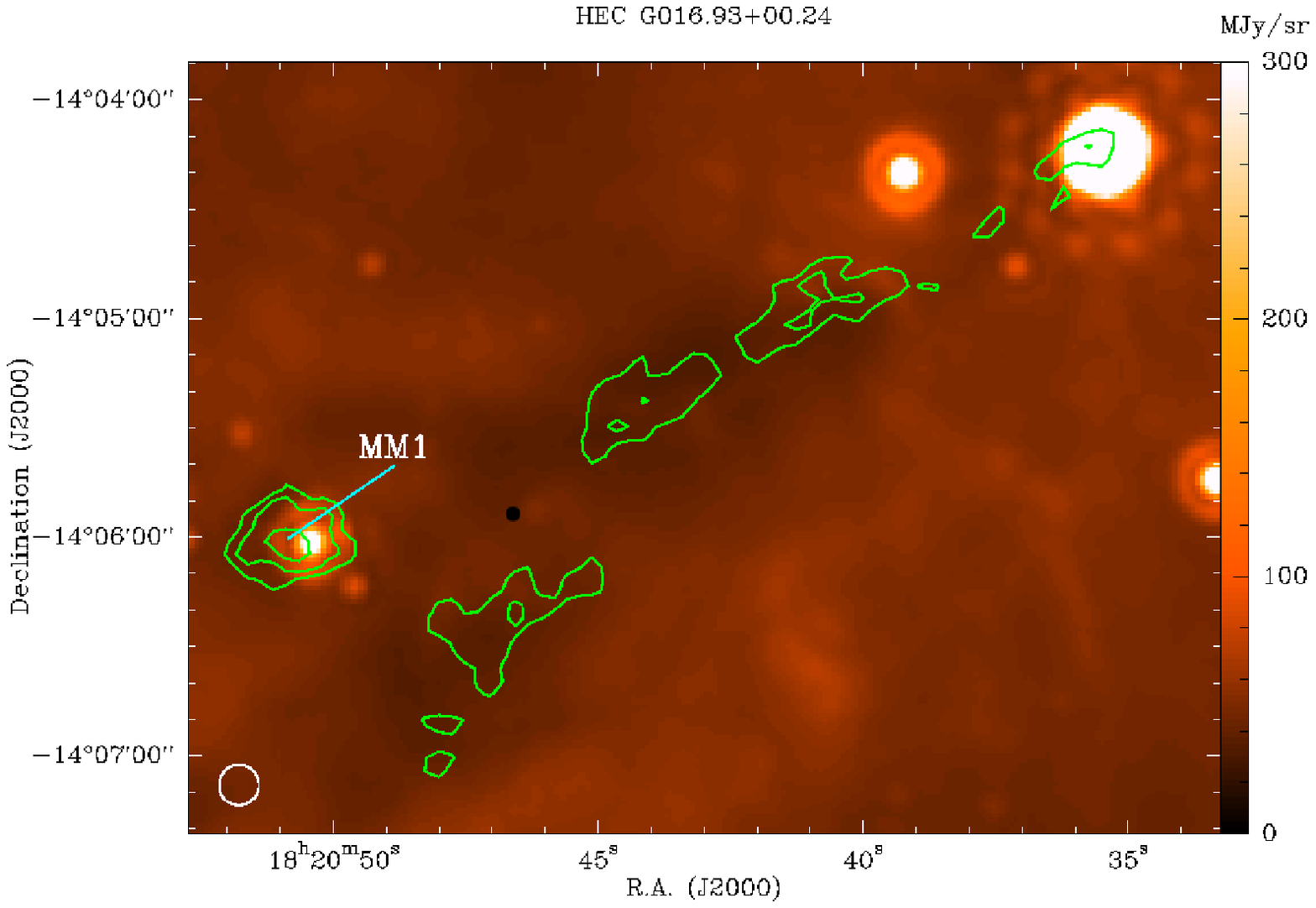}
\includegraphics[width=7cm, angle=-90]{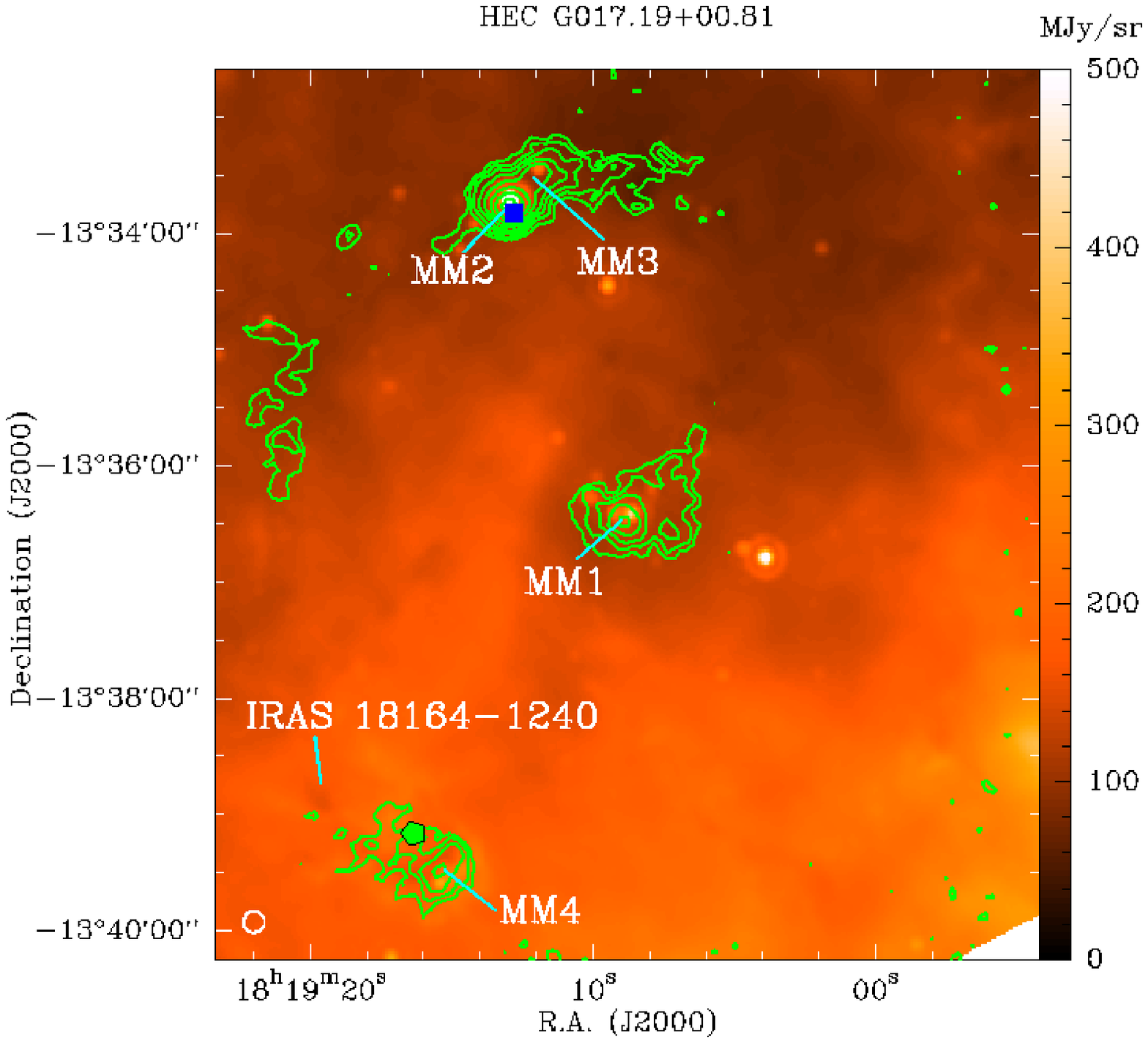}
\includegraphics[width=7cm, angle=-90]{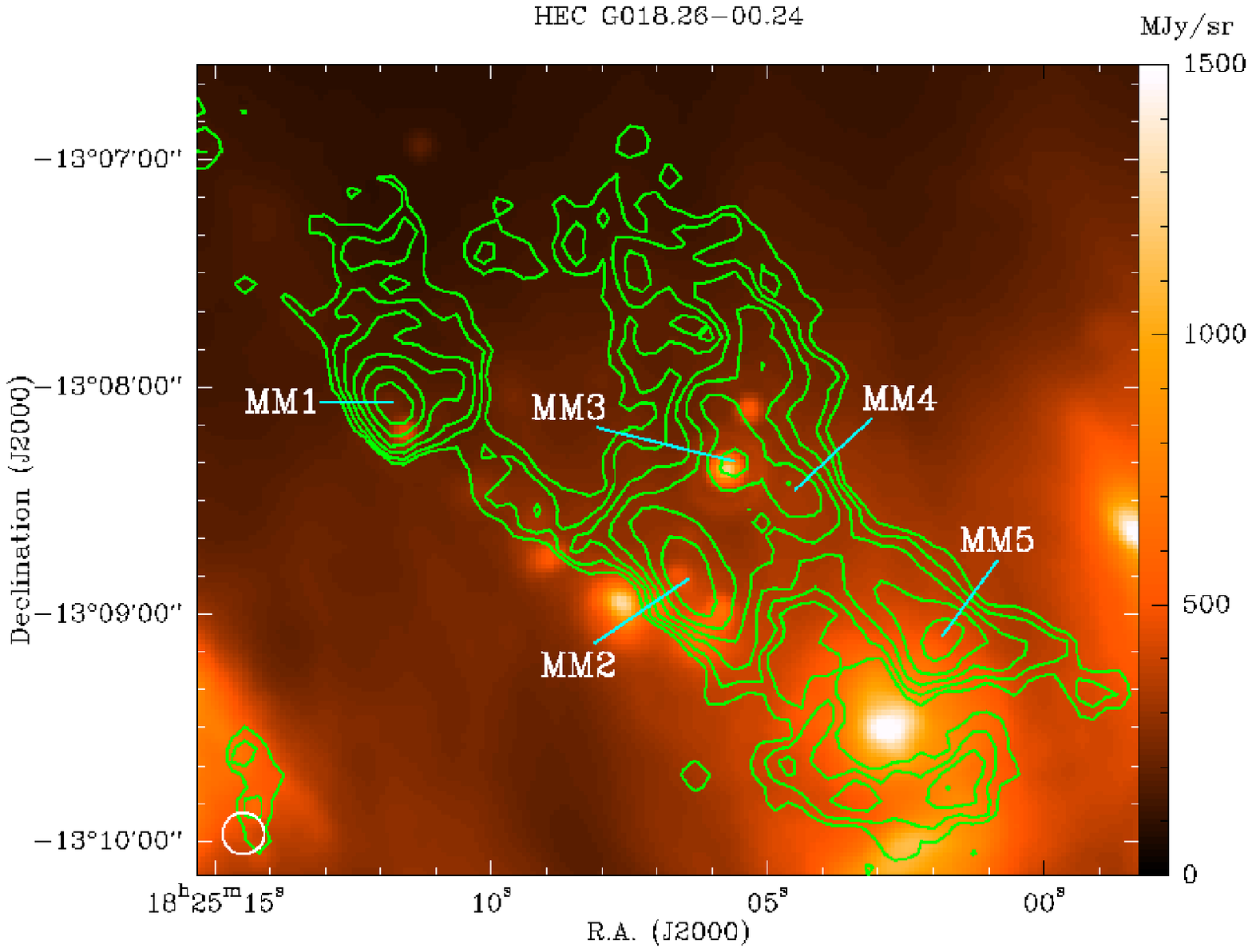}
\caption{--{\em Continued.}}
\end{figure*}
\addtocounter{figure}{-1}
\begin{figure*}[!htpb]
\centering
\includegraphics[width=7cm, angle=-90]{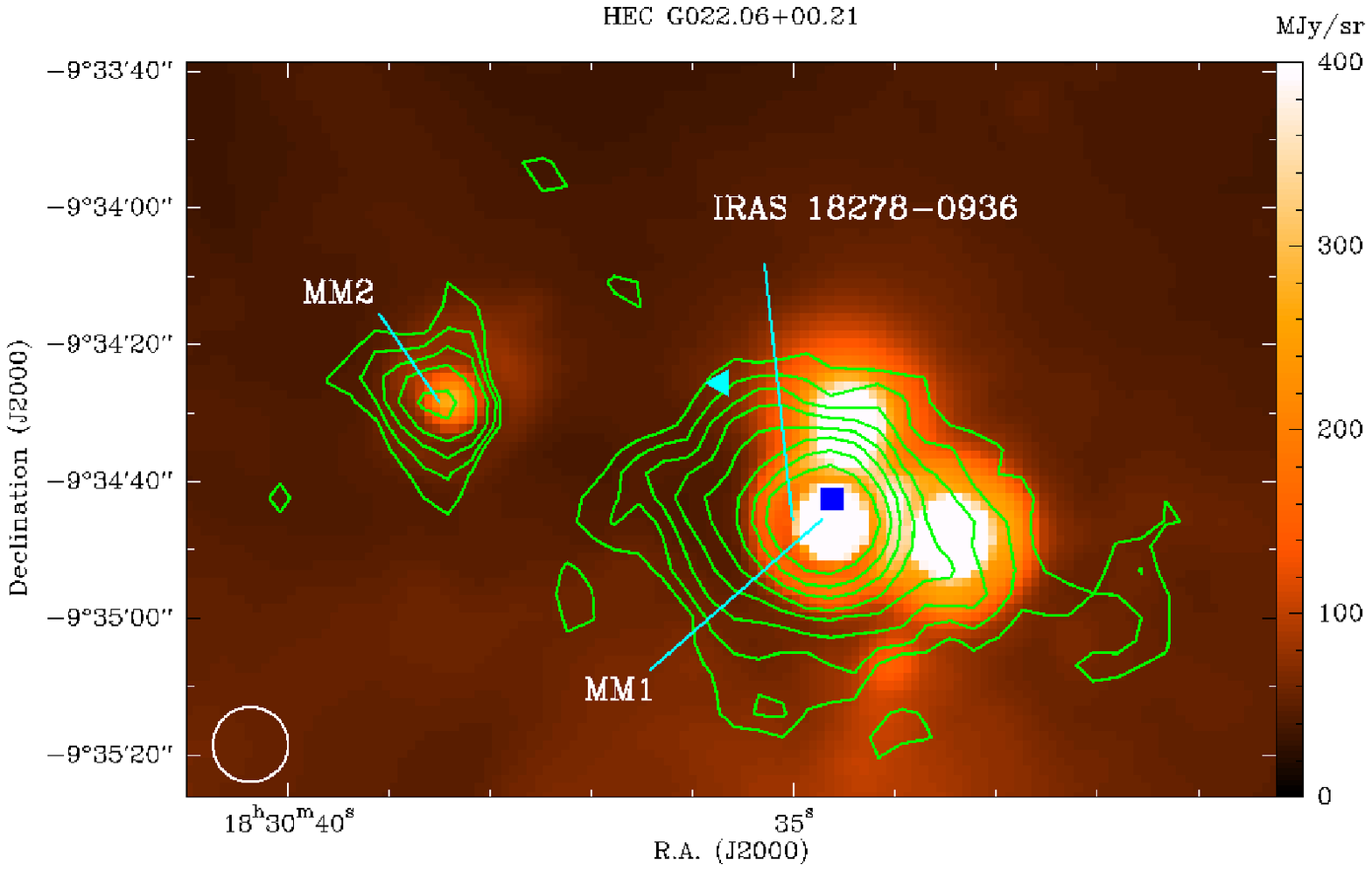}
\includegraphics[width=7cm, angle=-90]{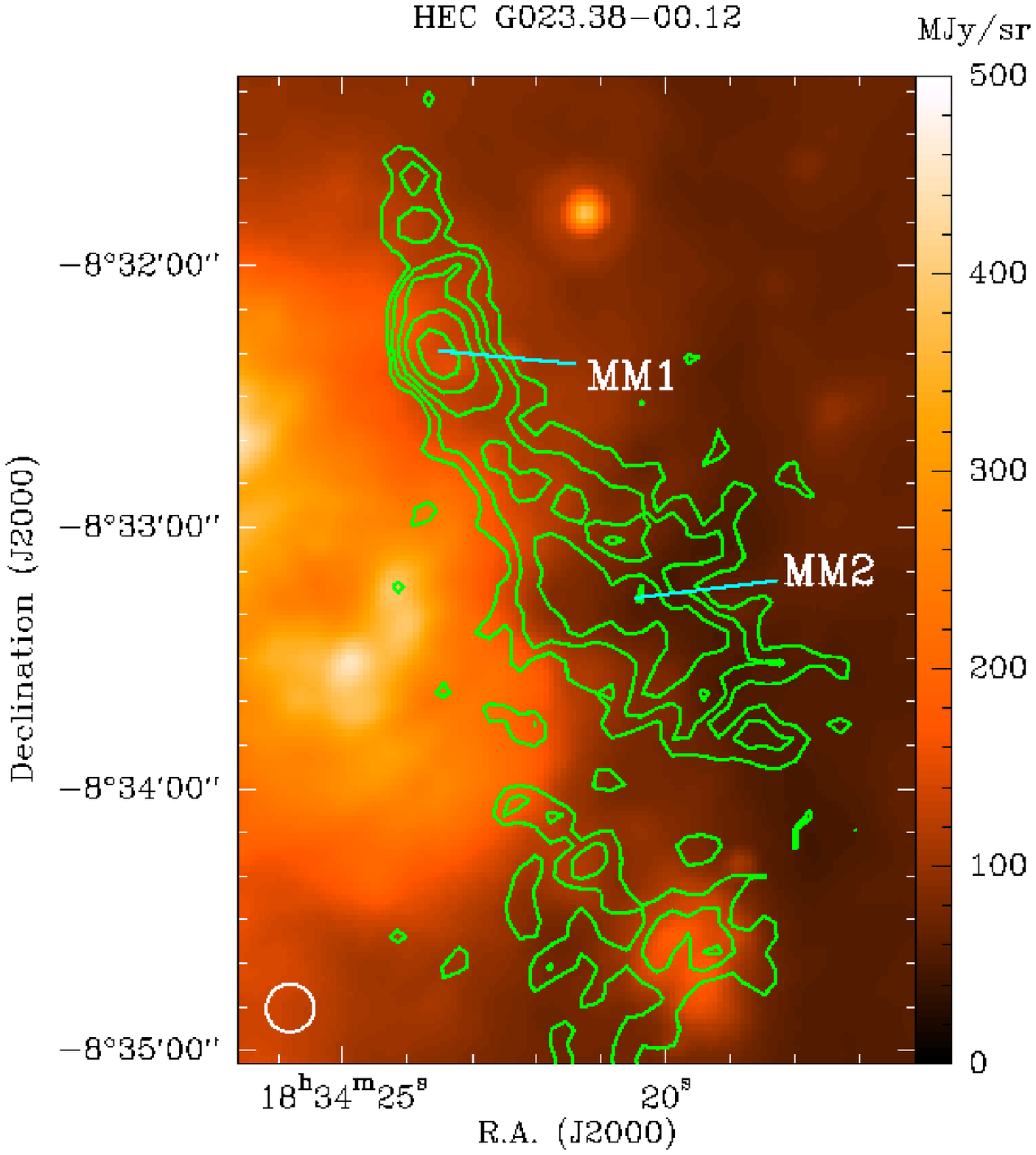}
\includegraphics[width=7cm, angle=-90]{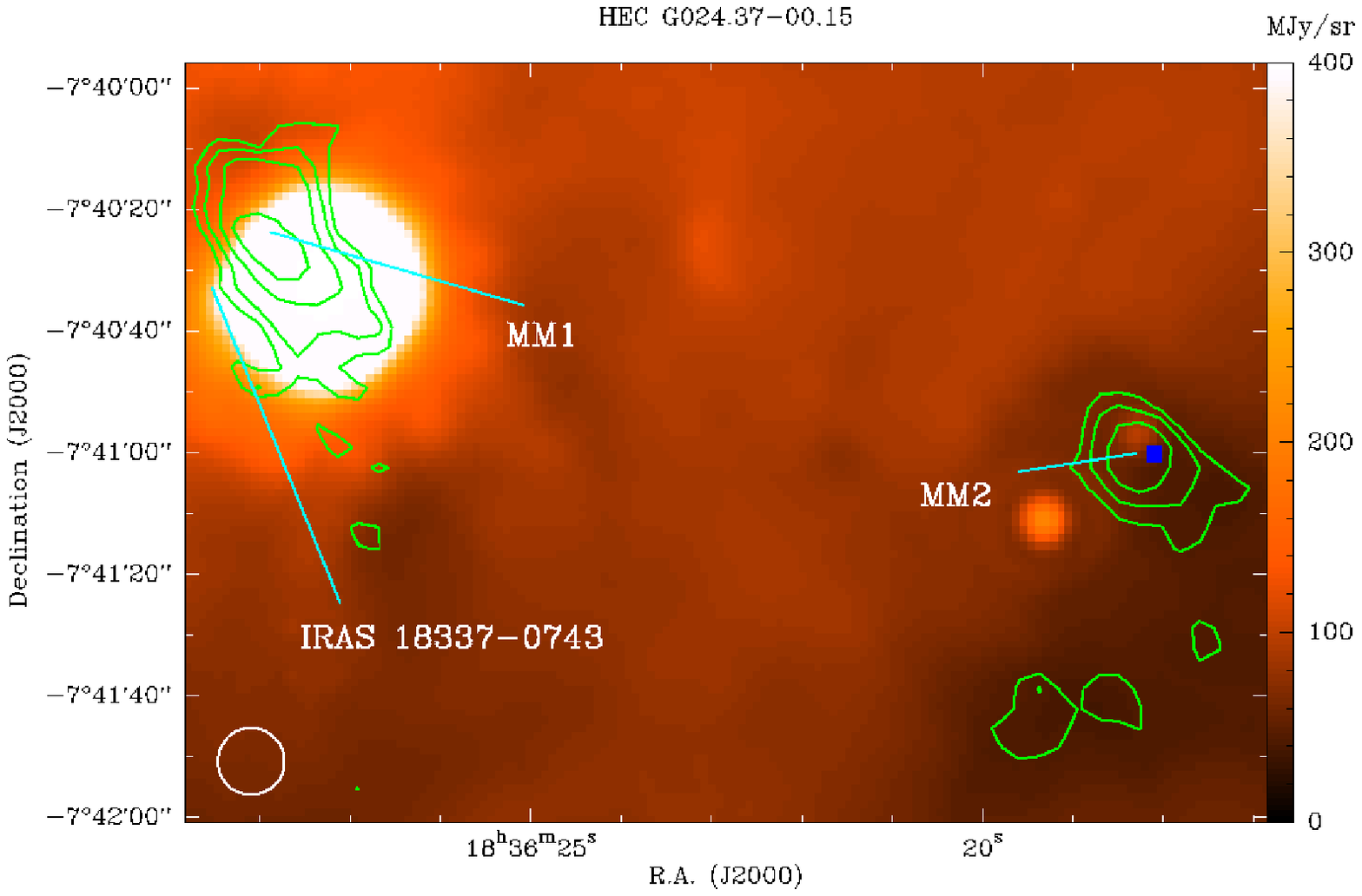}
\includegraphics[width=7cm, angle=-90]{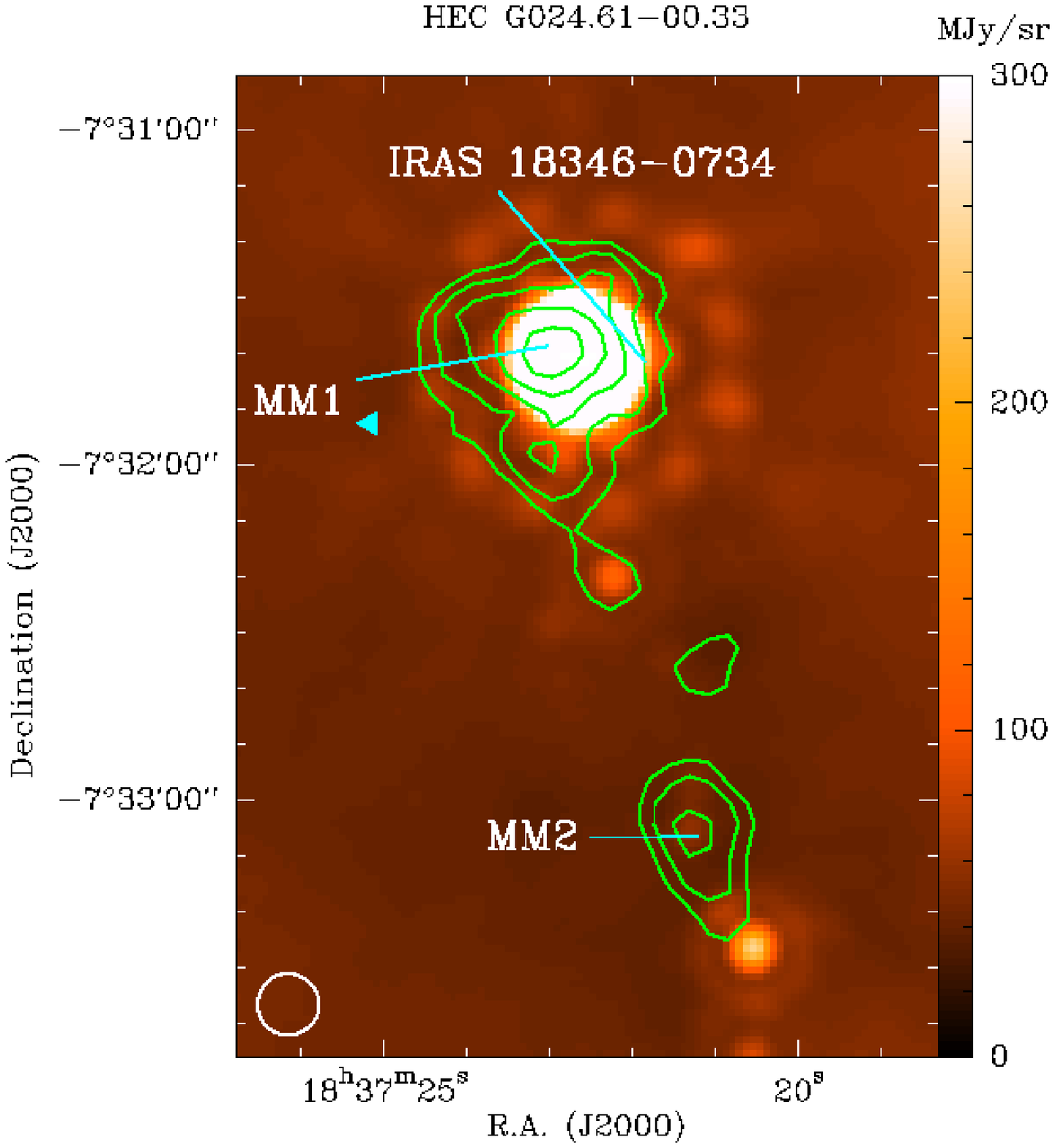}
\includegraphics[width=7cm, angle=-90]{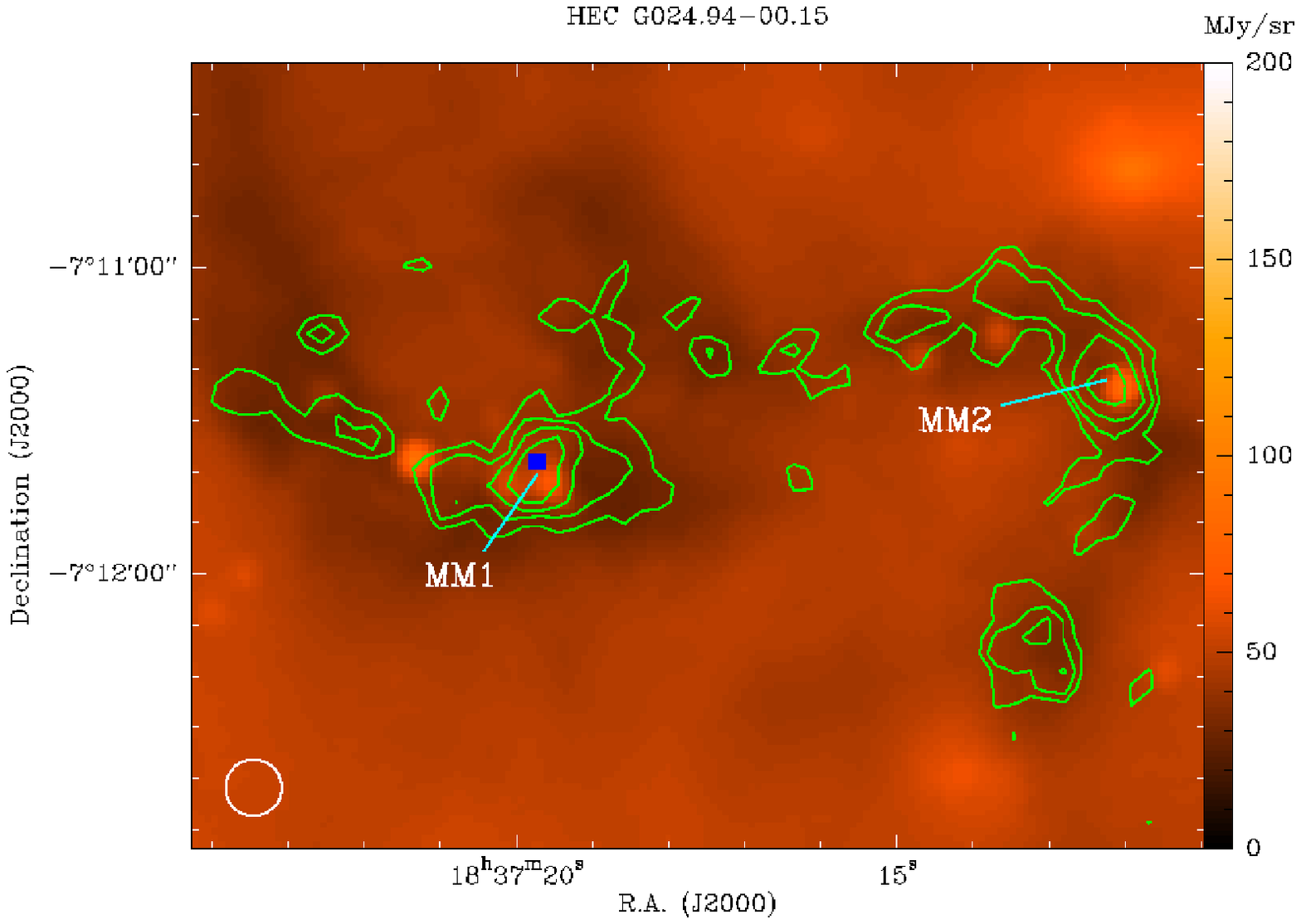}
\caption{--{\em Continued.}}
\end{figure*}
\addtocounter{figure}{-1}
\begin{figure*}[!htpb]
\centering
\includegraphics[width=7cm, angle=-90]{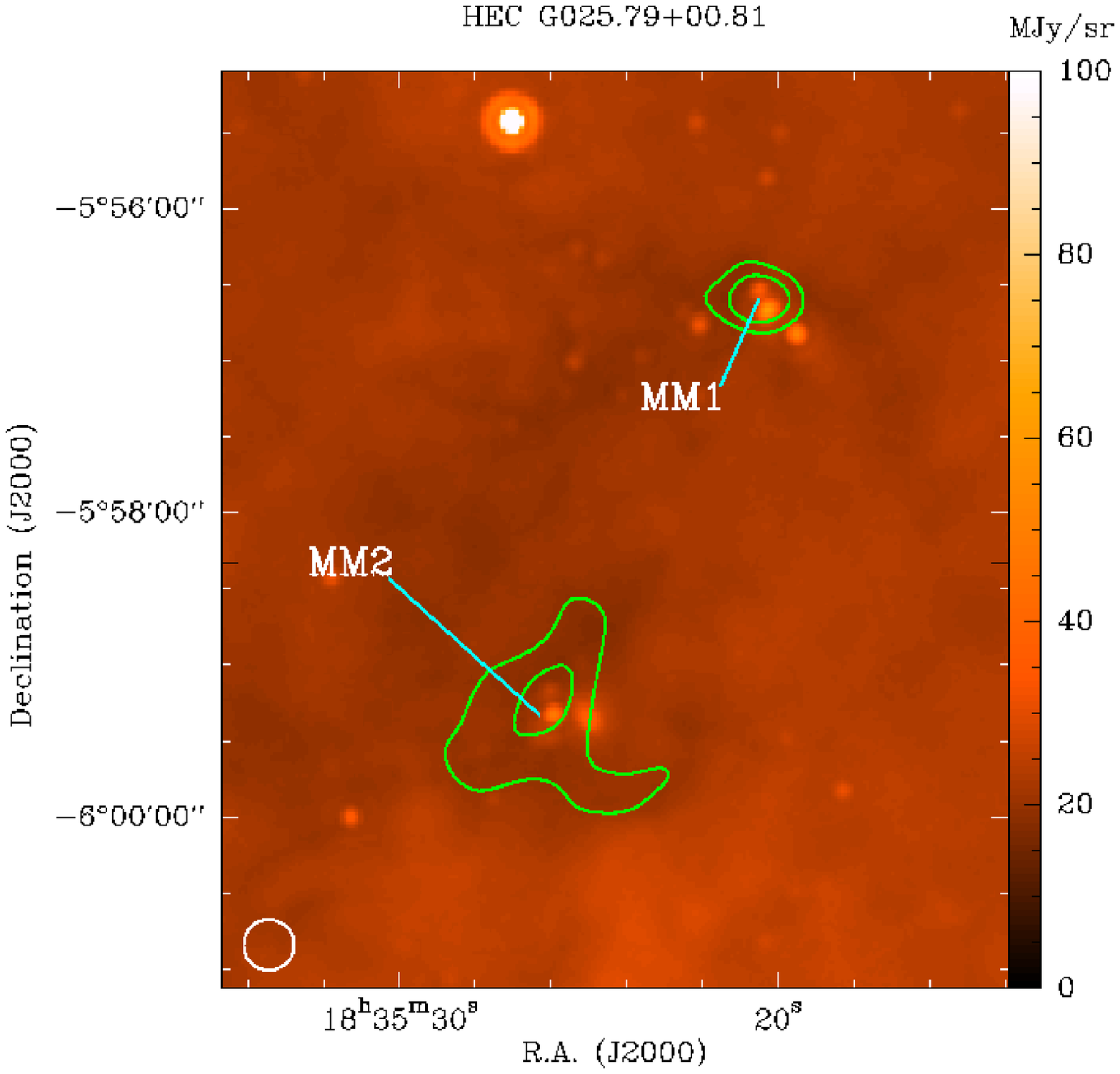}
\includegraphics[width=7cm, angle=-90]{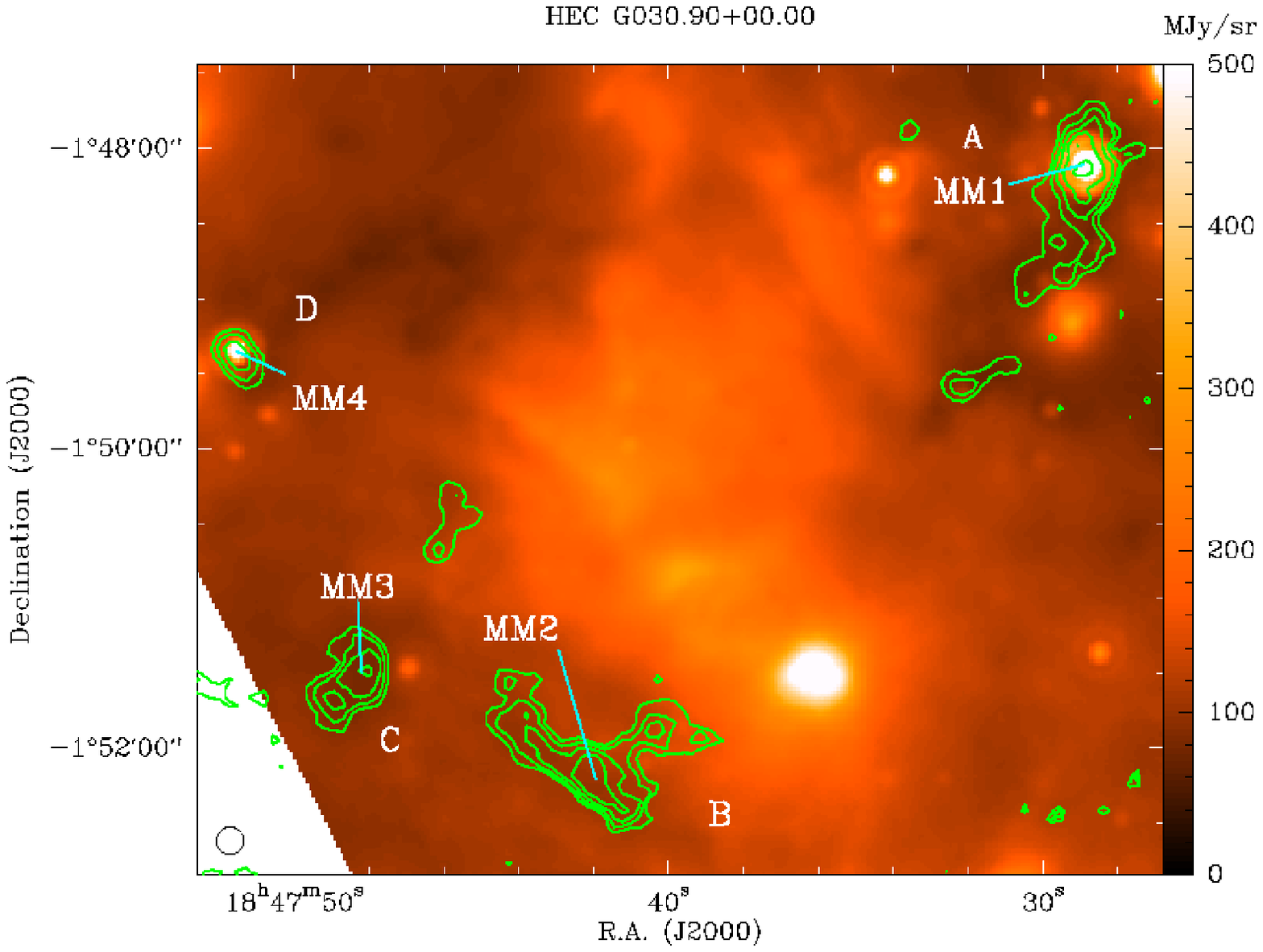}
\includegraphics[width=7cm, angle=-90]{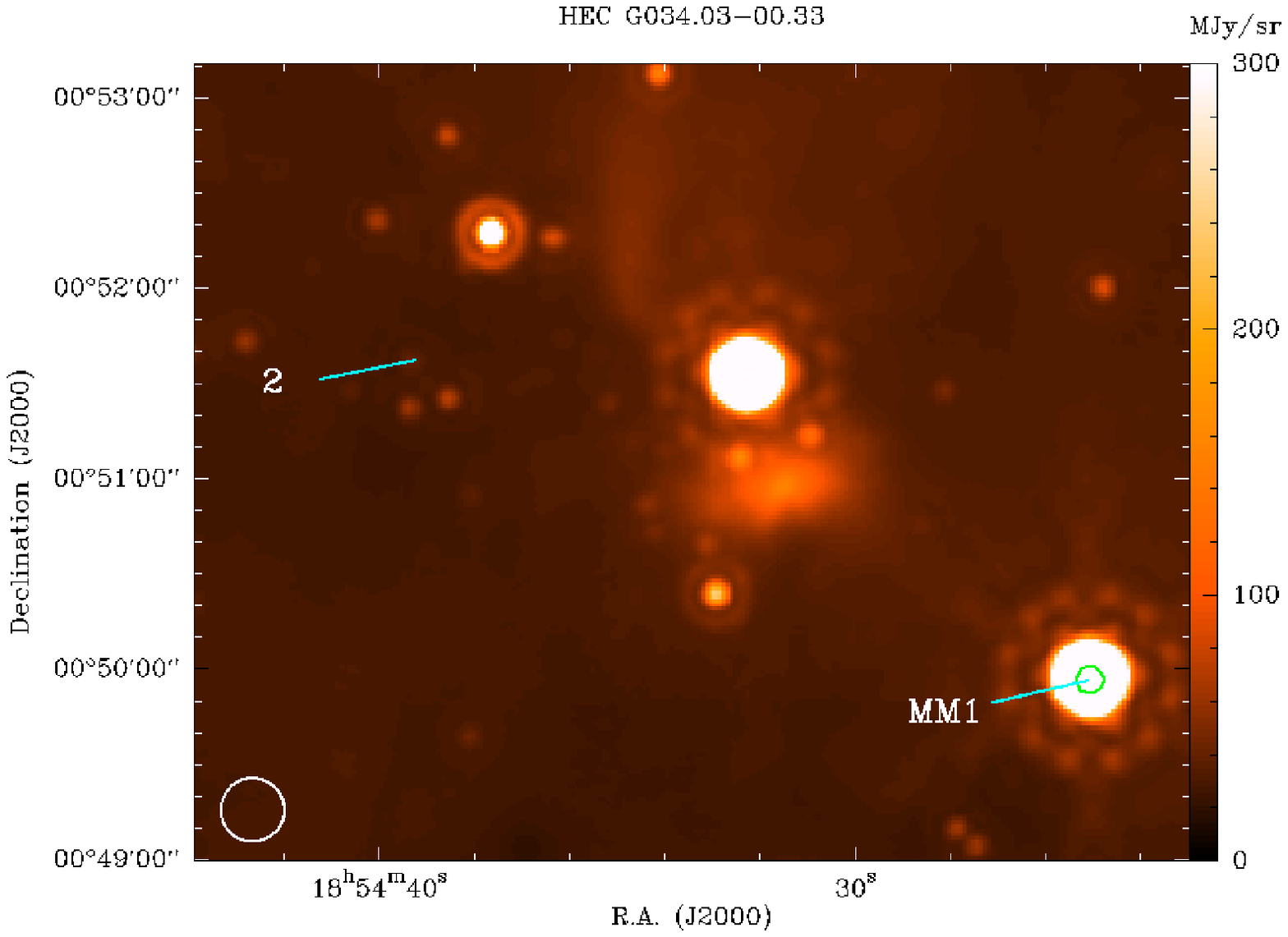}
\includegraphics[width=7cm, angle=-90]{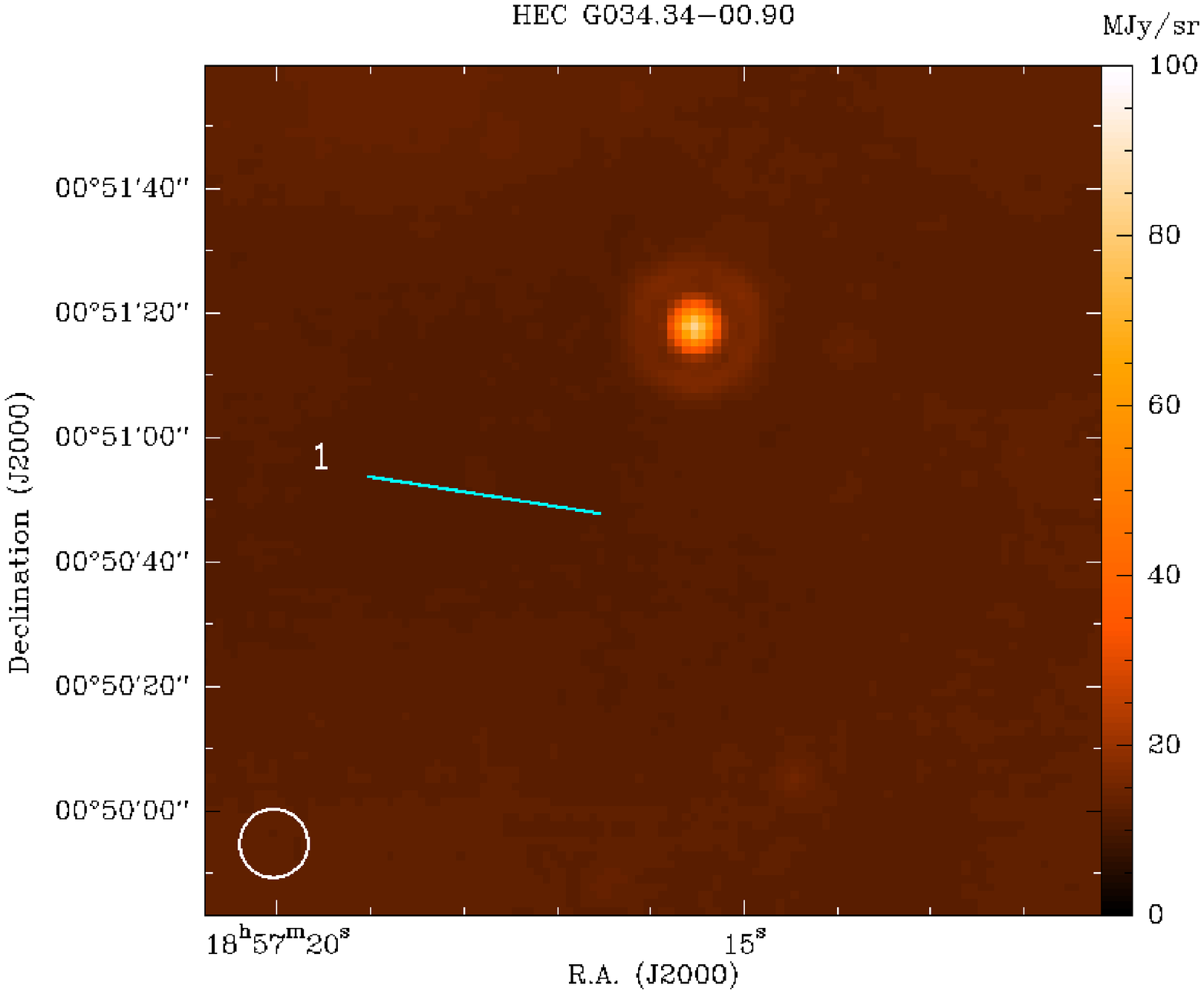}
\includegraphics[width=7cm, angle=-90]{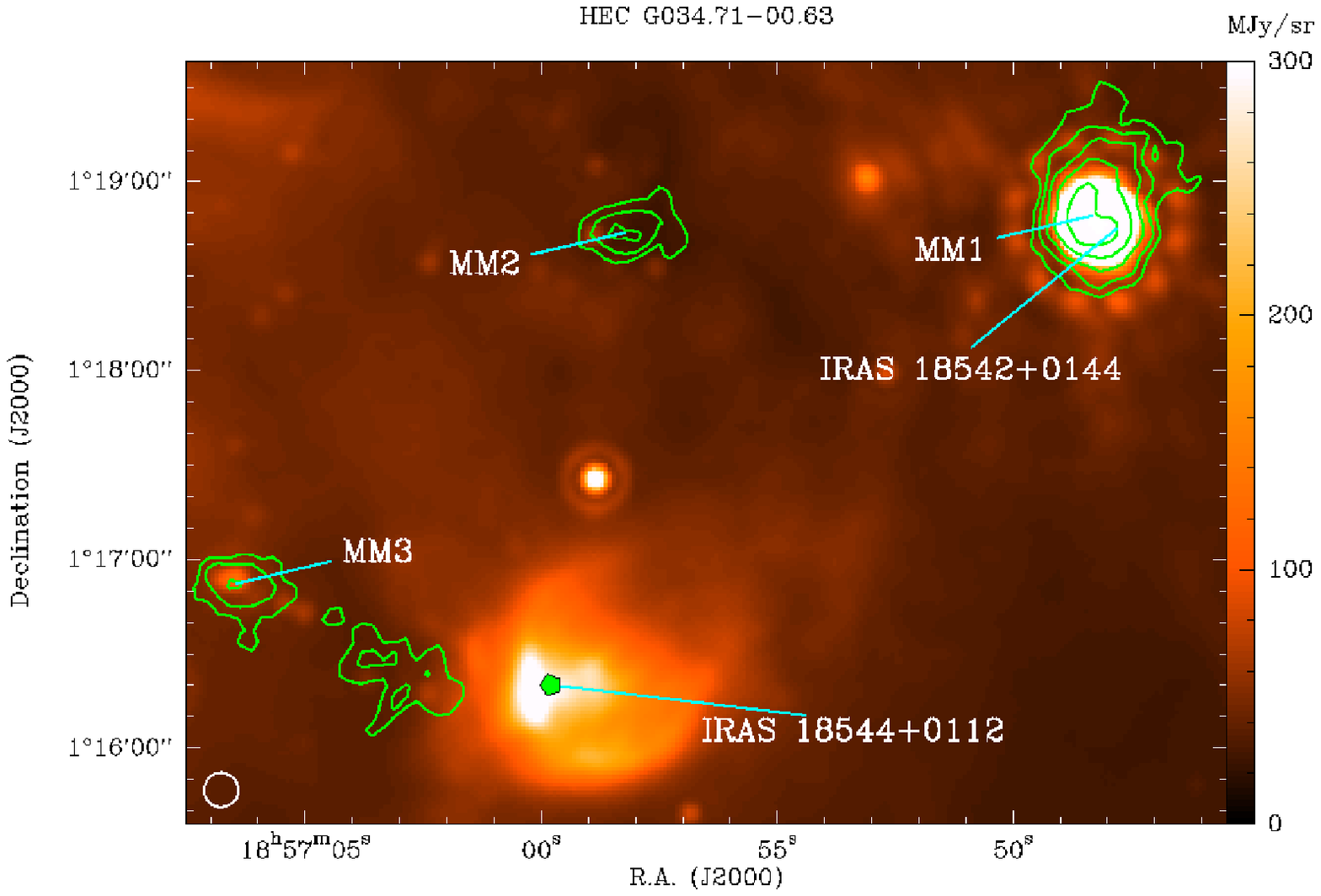}
\caption{--{\em Continued.}}
\end{figure*}
\addtocounter{figure}{-1}
\begin{figure*}[!htpb]
\centering
\includegraphics[width=7cm, angle=-90]{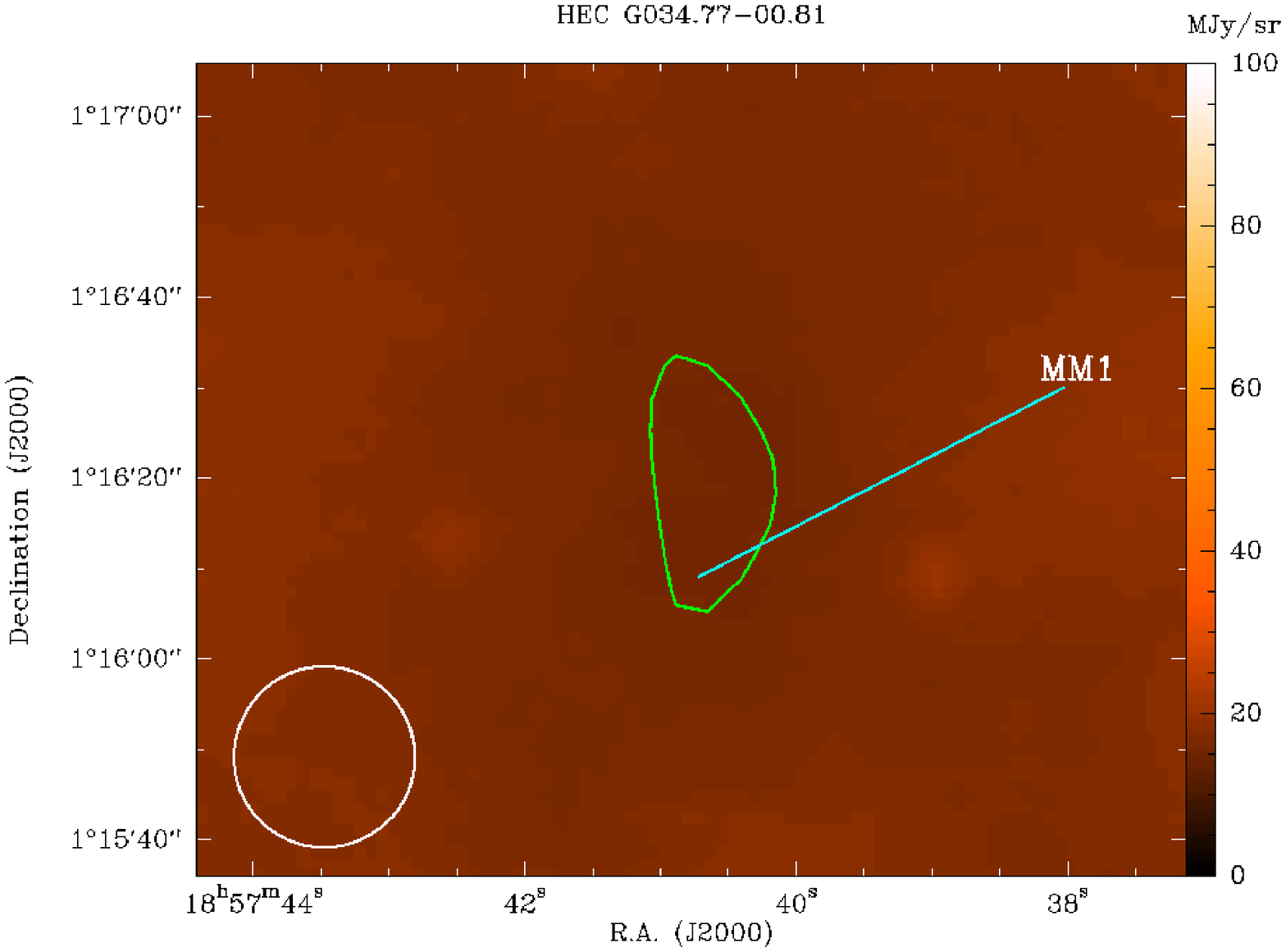}
\includegraphics[width=7cm, angle=-90]{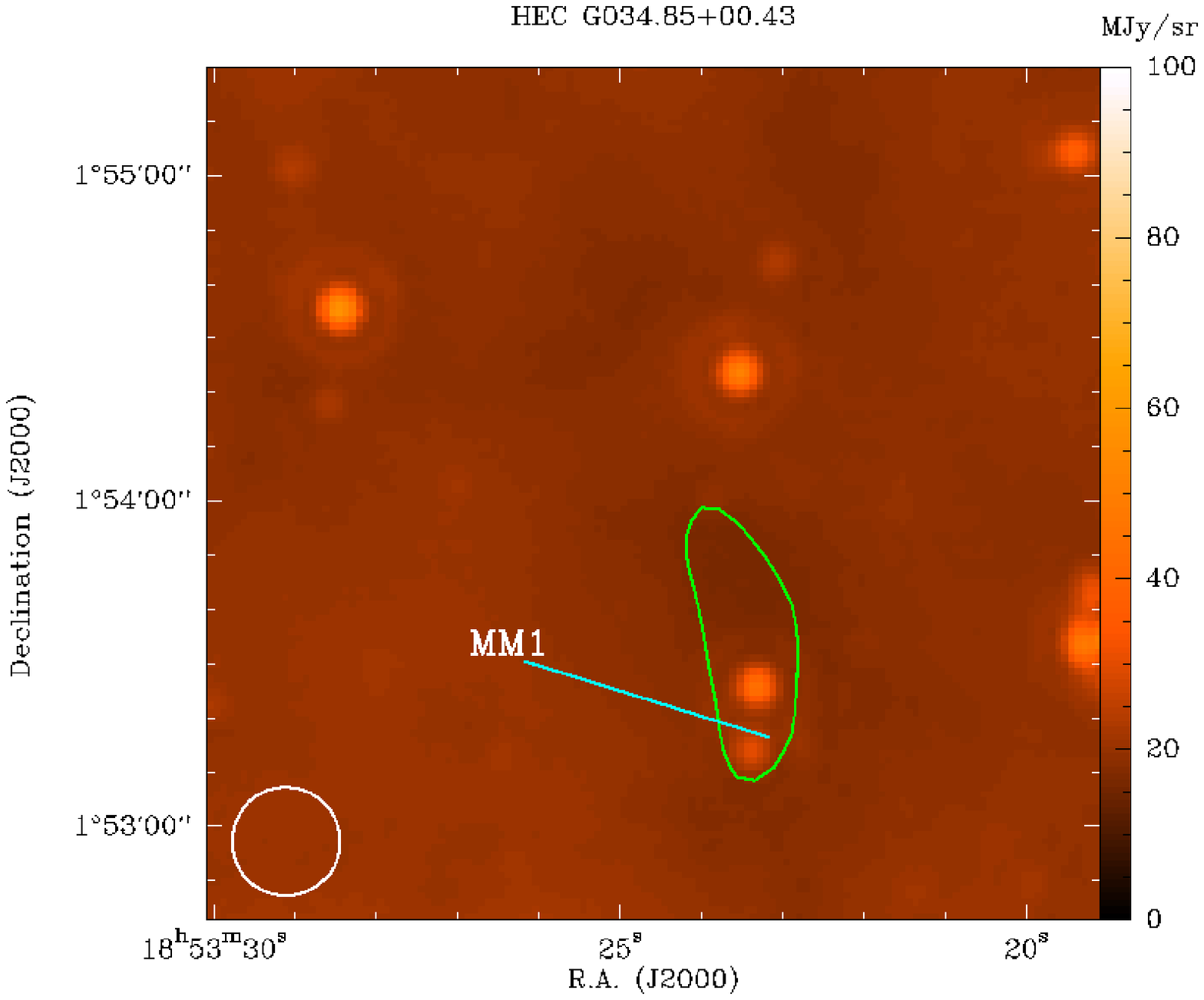}
\includegraphics[width=7cm, angle=-90]{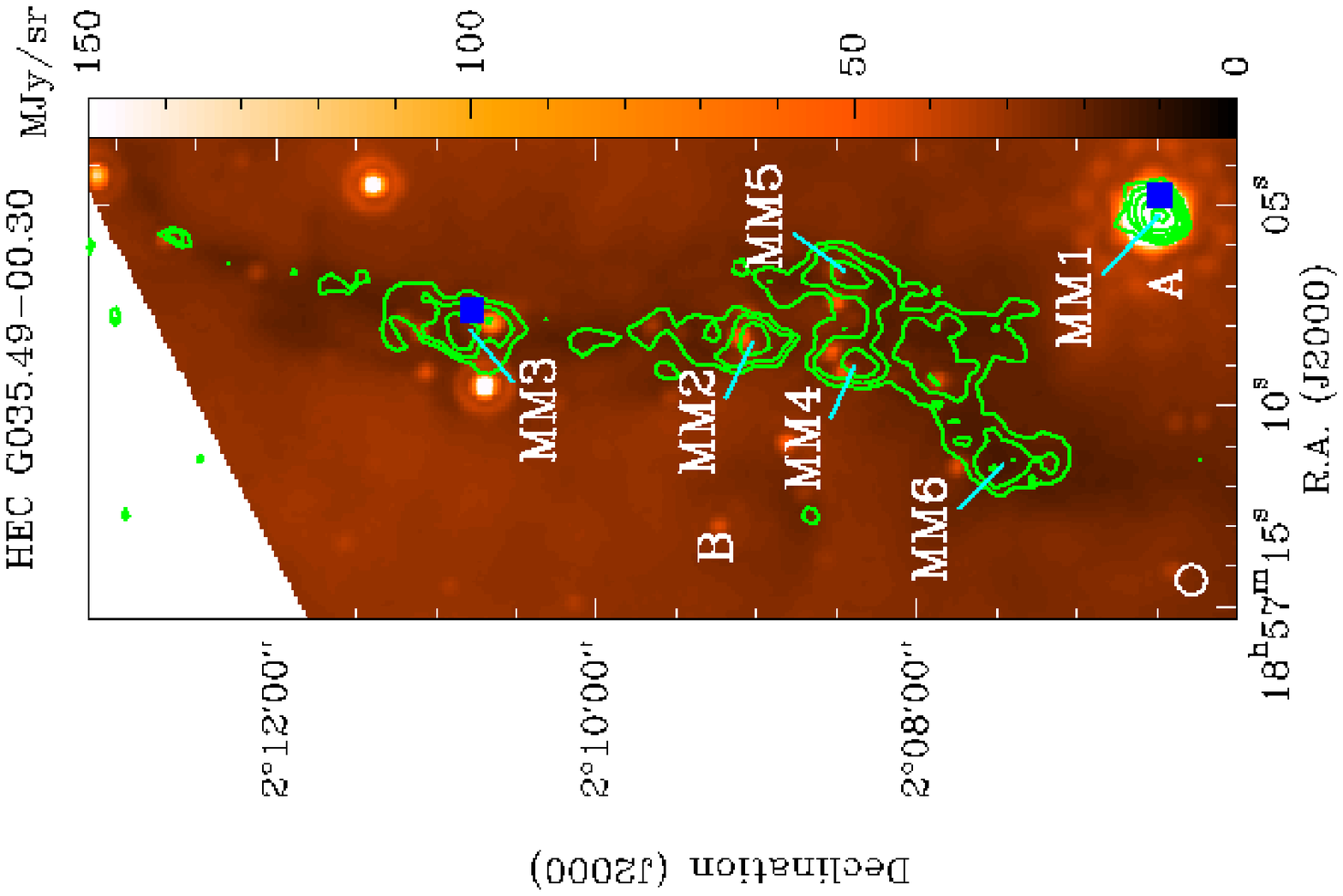}
\includegraphics[width=7cm, angle=-90]{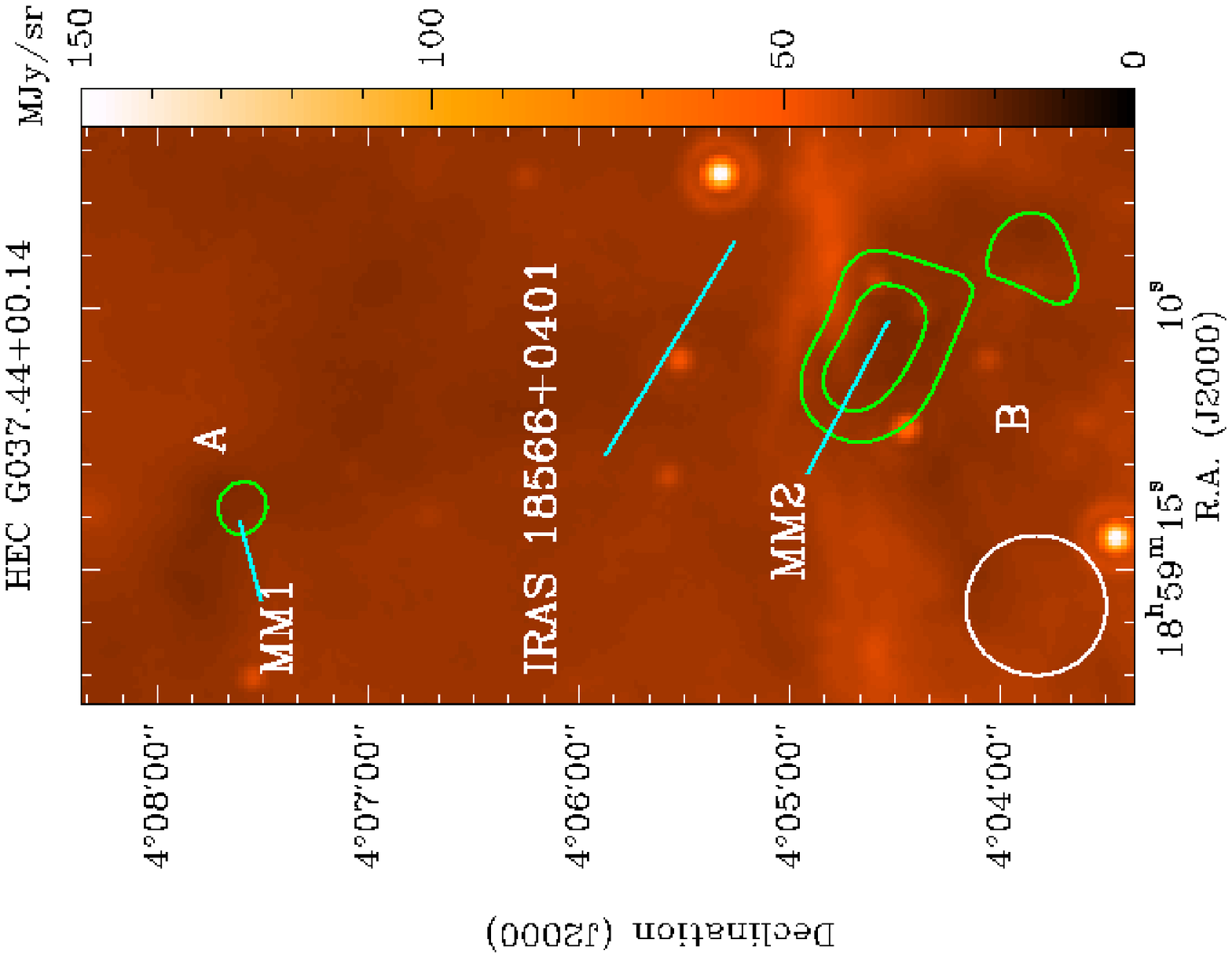}
\includegraphics[width=7cm, angle=-90]{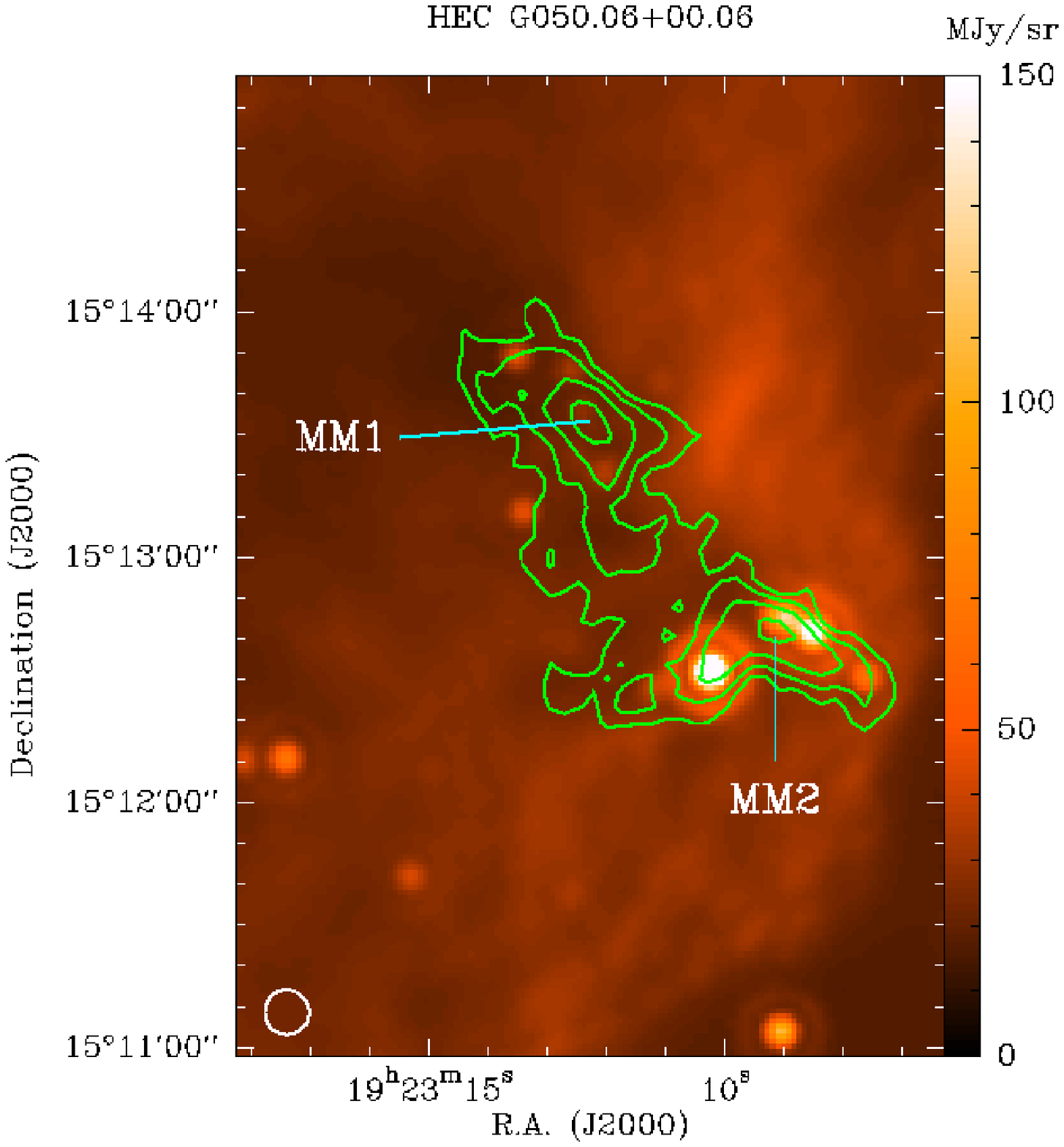}
\includegraphics[width=7cm, angle=-90]{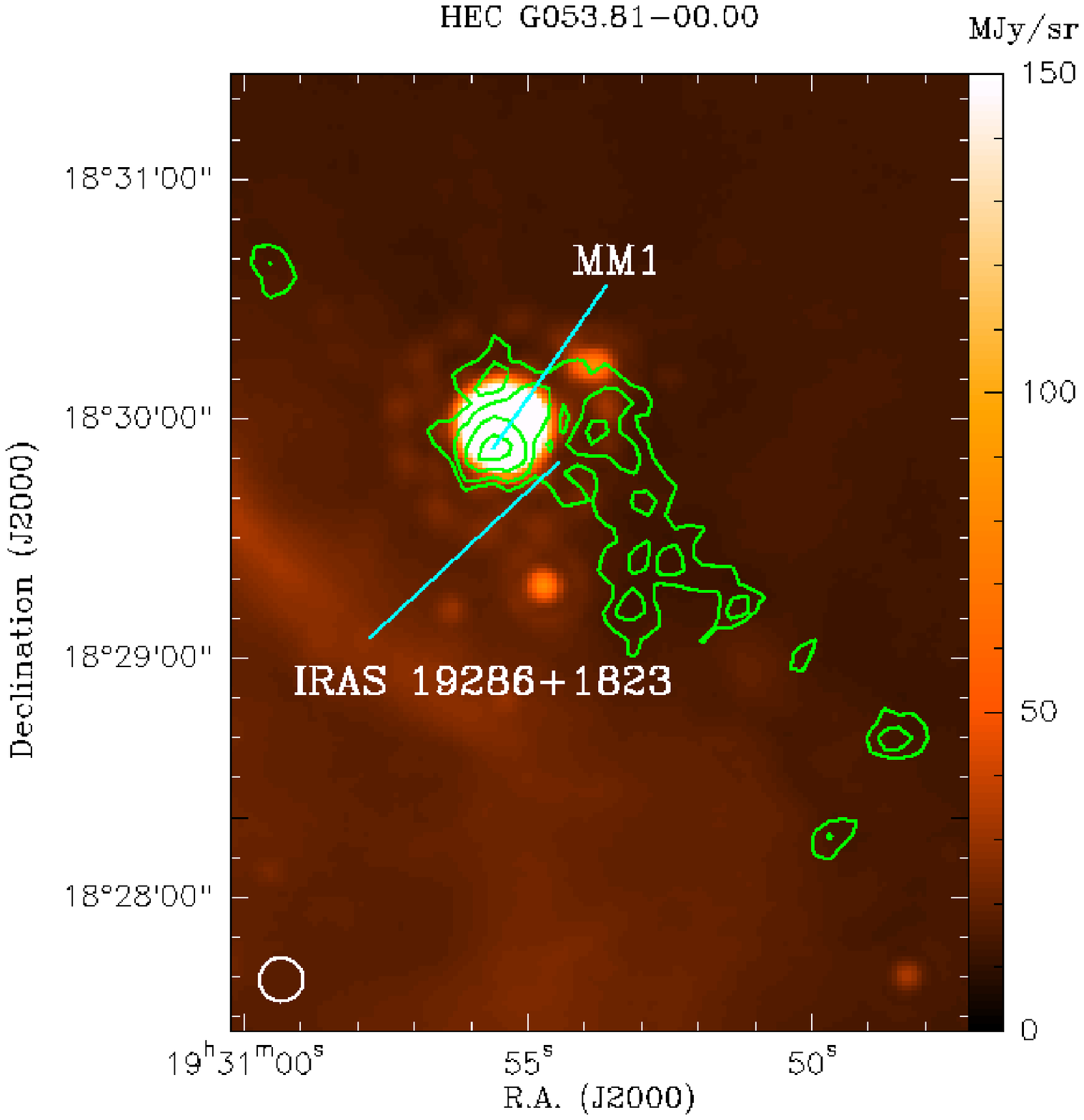}
\caption{--{\em Continued.}}
\end{figure*}
 
Millimeter continuum emission was detected in 88\% of the bolometer
  maps. It usually followed well the 54\arcsec\ resolution extinction maps (Fig.~\ref{fig:ext-mambo}). 
Two examples of weak millimeter emission or non-detections are G034.03--00.33
and G034.34--00.90 (Fig. \ref{fig:mips}). Figure \ref{fig:mips} shows the
24\,$\mu$m maps from the Multiband Imaging Photometer for Spitzer (MIPS)
overlaid with the mm emission. While the high extinction clouds match well with 24\,$\mu$m dark features, the 1.2\,mm peaks often
contain weak compact 24\,$\mu$m emission.  
The bolometer maps were sensitive to structures from $10\rlap{$.$}\,\arcsec5$ (FWHM) to $\sim$
90$\arcsec$; large-scale structures are not faithfully represented due to the sky noise subtraction and chopping. We smoothed the maps, by convolving with a 20\arcsec\ Gaussian beam, to study the
extended cloud structure and increase the signal-to-noise toward weak and diffuse sources.

We find single clouds and clusters of clouds, for example G014.63--00.57 in Fig.~\ref{fig:mips}. Based on the kinematic distance the clouds were identified as belonging together in one large cloud, or being physically separated clouds if the difference in distance was larger than 10\%. The latter clouds were marked as A, B etc. and were treated as distinct clouds throughout the paper. For the total integrated flux of the cloud (Table \ref{ta:cloud}), we defined a cloud edge at a threshold of $3\sigma$ (of the smoothed map) and integrated all the emission within this region using the Gildas Software
package GREG. The corresponding radius of the cloud, $\sqrt{A/\pi}$ with $A$ the surface of the cloud, is given in arcseconds and parsecs (using the kinematic distance) in Table \ref{ta:cloud}. The average cloud radius was 0.7\,pc.
Certain clouds have spherical shapes, for example G013.28-00.34, however, most are filamentary, as can be seen in G024.94-00.15 or G016.93+00.24. 
Cloud sizes are very diverse depending on the geometry; there are elongated
structures up to several parsecs in length. 
The number of clumps per cloud varies
from five (in G018.26-00.24, Fig.~\ref{fig:mips}) to one or none in very
diffuse clouds (G034.85+00.43, Fig.~\ref{fig:mips}). 
We used the SIMBAD
database to investigate if our clouds contain signs of star formation, such as
maser emission, H{\sc ii} regions, and IRAS sources.

\subsection{Clumps in high extinction clouds}

To study the clumps we used the unsmoothed maps, as the extended structure surrounding the clump complicates the
definition of source edge and confuses Gaussian fitting routines. We removed the extended emission from the map by applying a median filter with a box size of 63\arcsec\ (6 $\times$ beamsize) using the MIRIAD task
`immedian'. After median removal, the source-find algorithm of MIRIAD, `sfind', was run on
these images delivering 2D Gaussian fits, the peak flux, and the integrated
fluxes of the clumps. 
The median removal should strip the low density mass reservoir which surrounds the denser part of the clump.

\begin{figure}[!htpb]
\centering
\includegraphics[height=0.8\columnwidth,angle=-90]{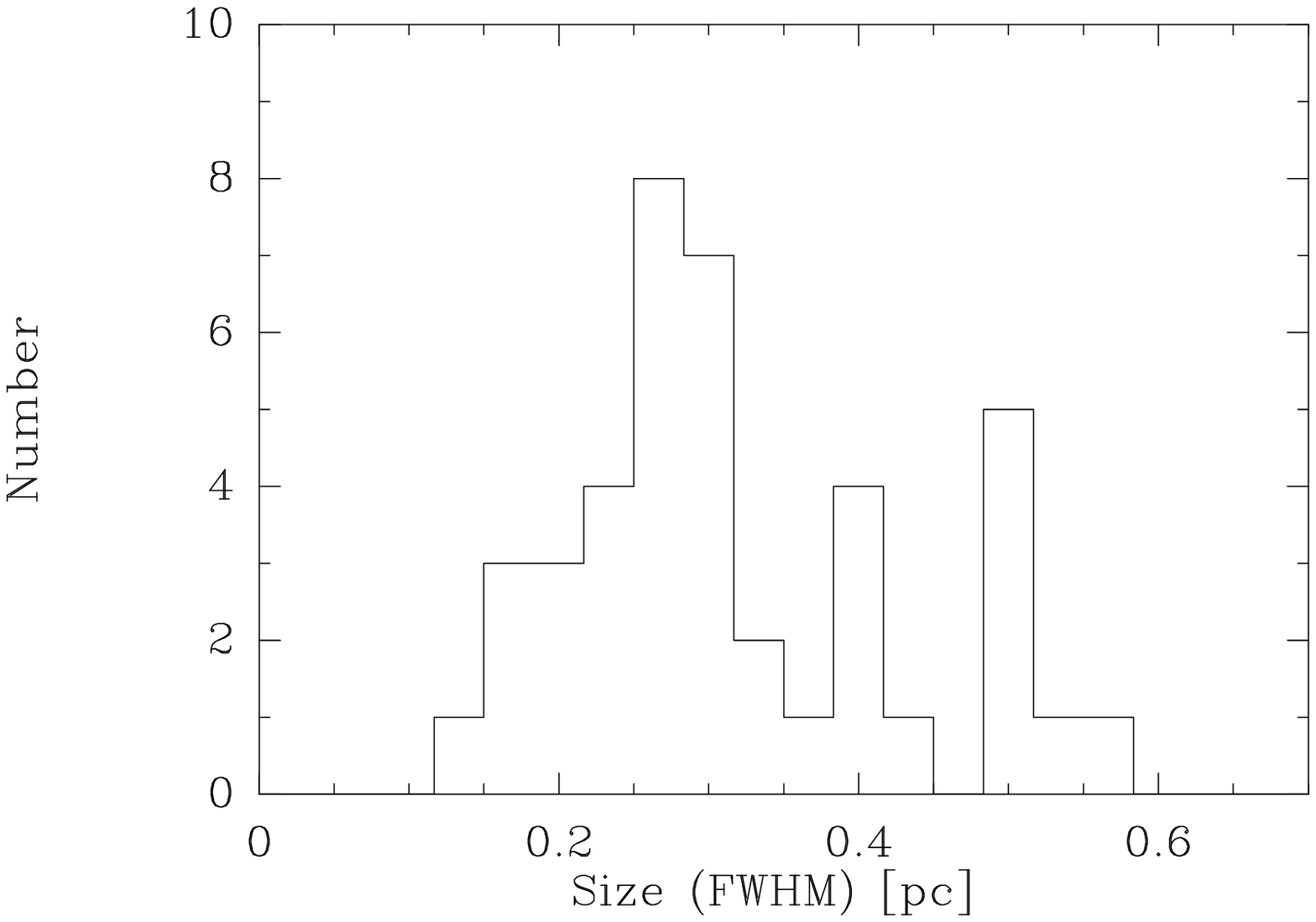}
\caption{\label{fig:fwhm-h}Number distribution of 1.2\,mm clumps with size
  determined from Gaussian fits and converted to physical units using the near kinematic distance.}
\end{figure}

\begin{figure*}[!htbp]
\centering
\includegraphics[height=0.8\textwidth, angle=-90]{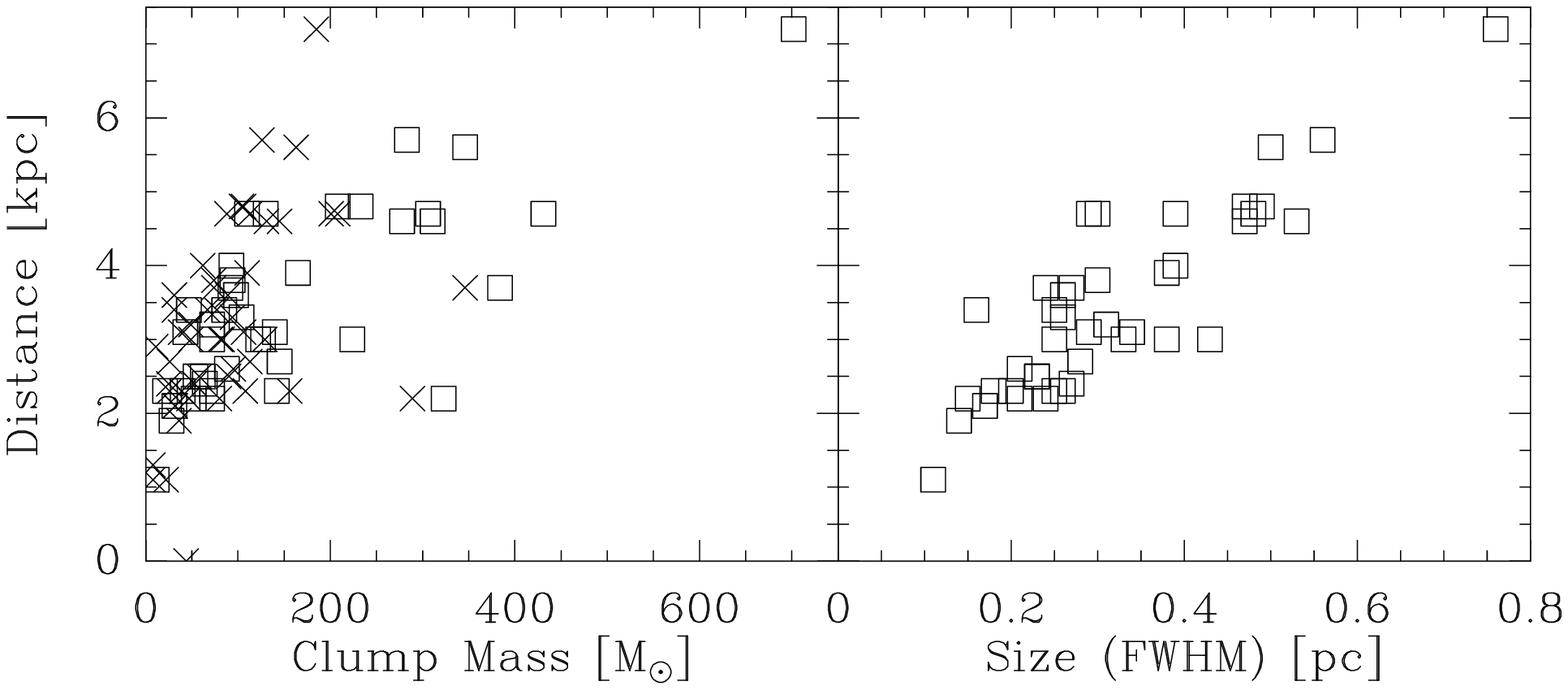}
\caption{\label{fig:distance-m-r}Distance as a function of clump mass (left) and size (right). In the left panel, the clump mass based on Gaussian fitting is marked by squares, and the clump mass within 0.25\,pc around the center of a clump is indicated by crosses.  
There is a linear correlation with distance for the mass and the size. The most massive clumps are at large distances, whereas almost no low-mass clumps are found at distances larger than 4 kpc, which can be expected for a sensitivity limited selection of clouds.}
\end{figure*}

First, we selected the peaks in the mm emission by eye, and then we took the
parameters for the FWHM size, the integrated flux and peak flux of the clump
from `sfind'. The observed parameters of the clumps are presented in Table
\ref{ta:flux}. The clumps have sizes (geometrical mean of the major and minor
axis of the clump) of $10-30$\arcsec\ or $0.11-0.72$\,pc, peaking around 0.25\,pc (Fig.~\ref{fig:fwhm-h}). Most of the clumps are resolved with the $10\rlap{$.$}\,\arcsec5$ beam of
the 30m telescope.  The (physical) clump size follows a linear trend with distance (Fig.~\ref{fig:distance-m-r}, right panel). At far distances we observe only larger, hence brighter clumps, which is expected for our sensitivity limited selection. The angular sizes, however, are more equally distributed (see Table \ref{ta:flux}). For several clouds with very weak mm emission the selected position was not the mm emission peak, but the emission center. In such clouds `sfind' did not find any clumps; these positions without clumps are listed in Table \ref{ta:noclump}.

\begin{figure}[!thbp]
\centering
\includegraphics[height=0.8\columnwidth, angle=-90]{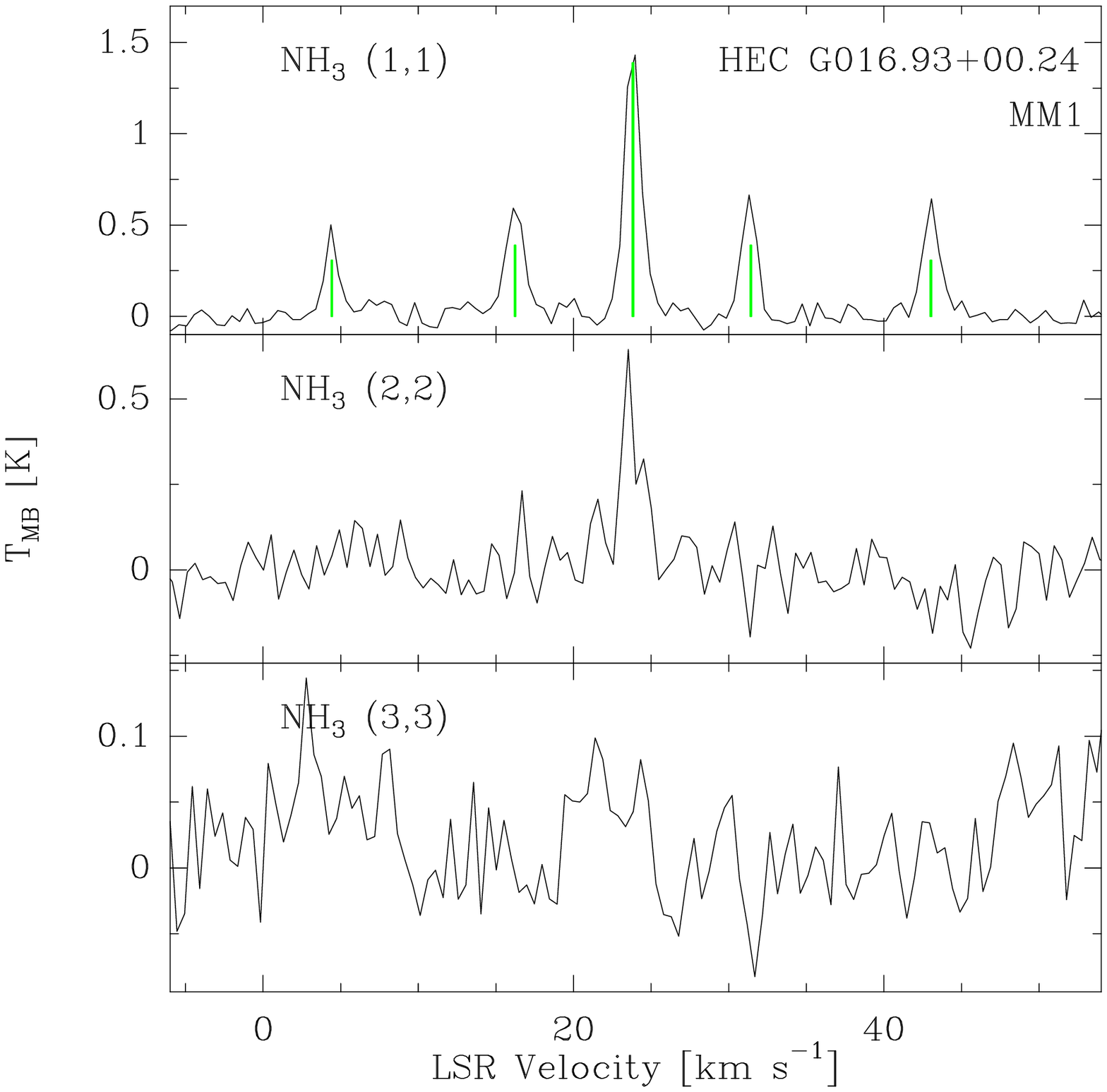}
\includegraphics[height=0.8\columnwidth, angle=-90]{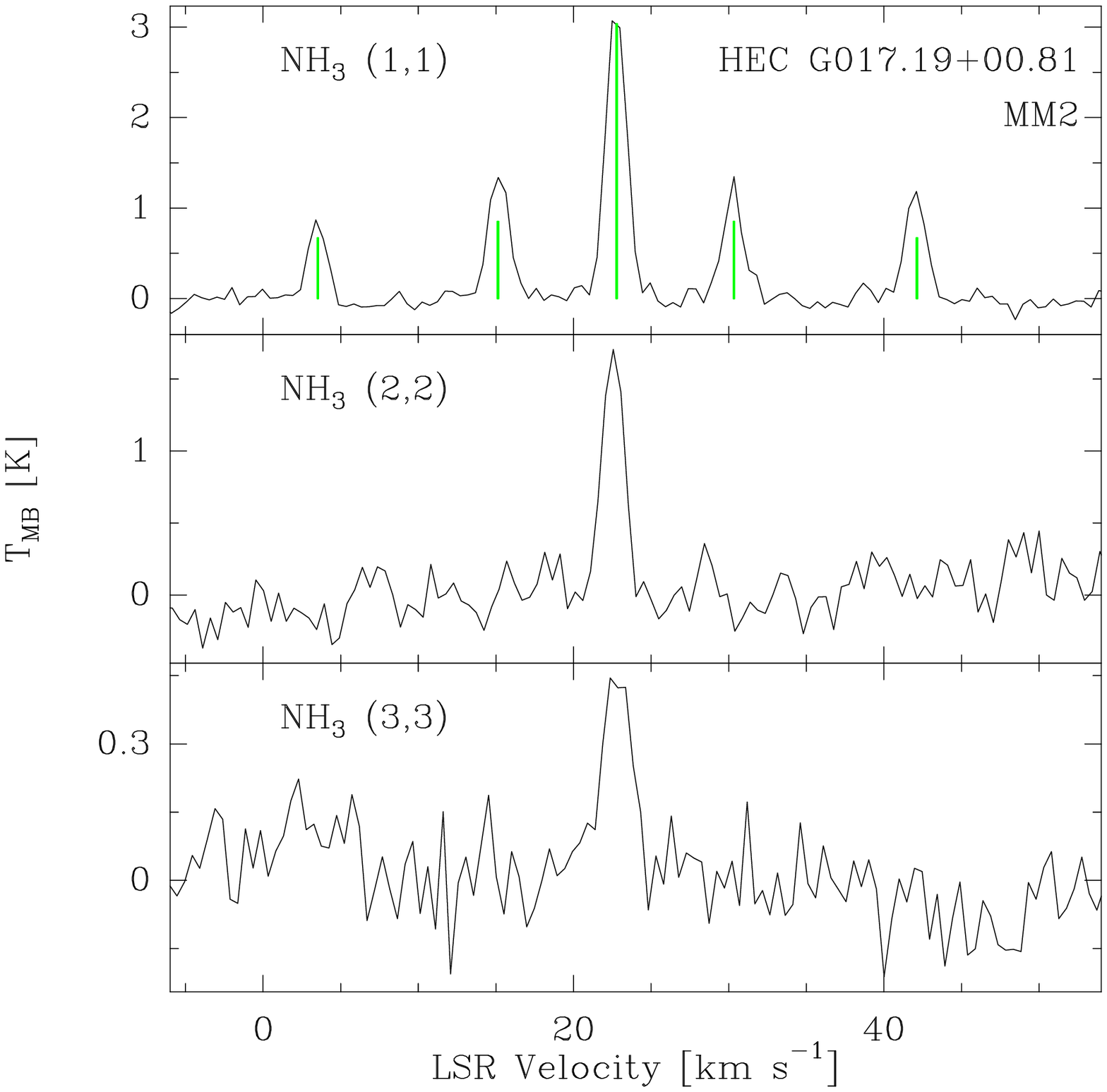}
\caption{\label{fig:amm} The \amm(1,1), \amm(2,2) and \amm(3,3) spectra taken
  with the Effelsberg 100m telescope toward two HEC clumps. The green lines indicate the theoretical intensity of the hyperfine lines given the intenstity of the main line.}
\end{figure}

\begin{figure}[!htbp]
\centering
\includegraphics[height=\columnwidth, angle=-90]{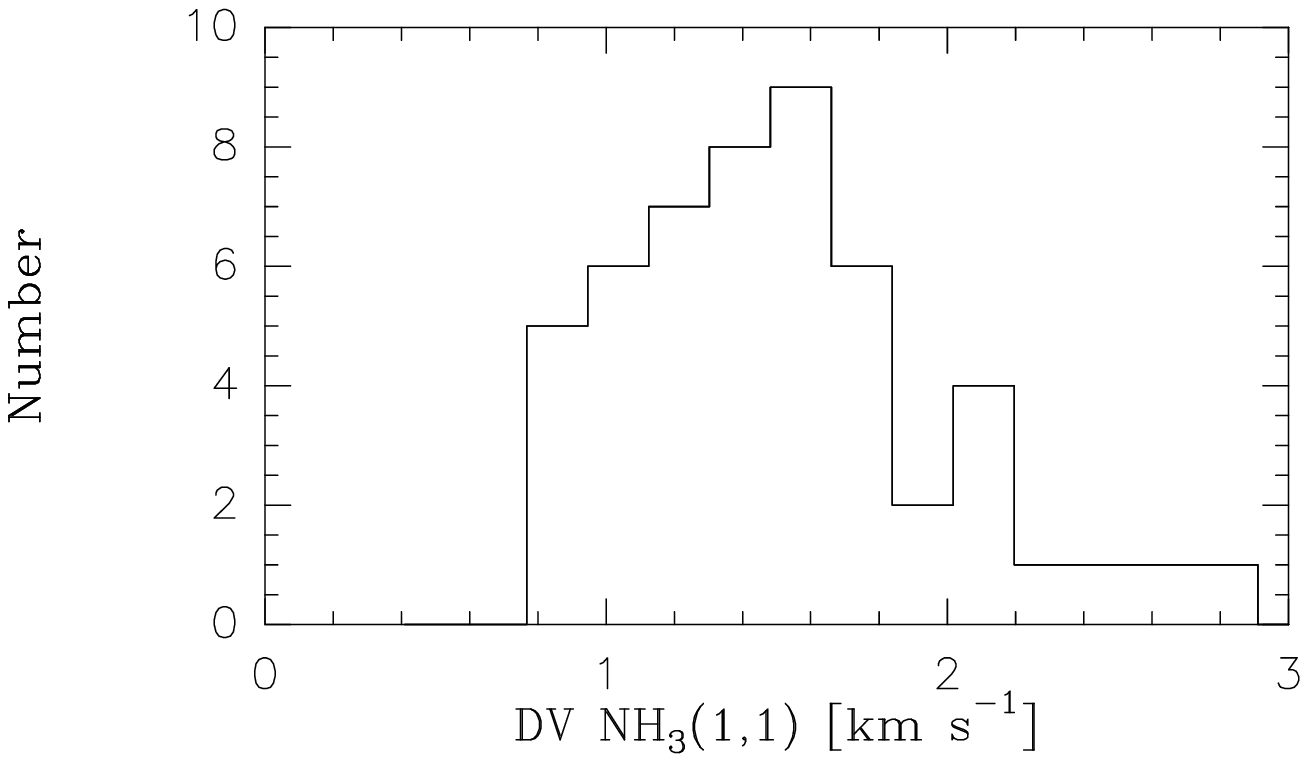}
\caption{\label{fig:dv1-h}Number distribution of the \amm(1,1) line widths of the 1.2 mm clumps.}
\end{figure}

Ammonia emission was detected toward 94\% of the observed positions. Figure \ref{fig:amm}
shows the spectra of the \amm(1,1), \amm(2,2) and \amm(3,3) transitions toward two compact clumps representing typical values for an early stage where the clump is cold (14\,K, G016.93+00.24 MM1) and a more evolved stage, where the temperature has increased (19\,K, G017.19+00.81 MM2). In the early stage the \amm(3,3) line is not detected, while for the warmer clump it is clearly present. 
The results of the ammonia observations are summarized in Table \ref{ta:nh3}. The main beam brightness temperatures of the \amm(1,1) lines are, after baseline removal, between 0.5 and 4.0\,K. 
The baseline r.m.s. is $\sim$0.2\,K, but several spectra are more noisy.  
The line widths from the main component of the \amm(1,1) range between 0.7 and 2.8\,km\,s$^{-1}$, yielding an average of 1.4\,km\,s$^{-1}$(Fig. \ref{fig:dv1-h}). These line widths are
far above a thermal linewidth, which would be around 0.23\,km\,s$^{-1}$ for
temperatures of $\sim$18\,K. For most of the sources, the \amm(1,1) and
\amm(2,2) lines are both detected, while the \amm(3,3) line is often very
weak or not present. The \amm(2,2) and \amm(3,3)
lines are on average both wider than the main \amm(1,1) line. This implies that
these lines do not trace exactly the same volume of gas, meaning that the beam filling
factor is not identical.

\section{Analysis}

\begin{table*}
\begin{center}
\caption{Properties of the clumps derived from \amm\ and the 1.2 mm continuum.\label{ta:mass}} 
\begin{tabular}{l l r r r r r r r r r r }
\hline\hline
 \noalign{\smallskip}      
HEC name &  &\multicolumn{1}{c}{$d_{\mathrm{kin}}$} &\multicolumn{1}{c}{$A_V$} & \multicolumn{1}{c}{$FWHM$} & \multicolumn{1}{c}{${N_{\mathrm{H_2}}}$}&\multicolumn{1}{c}{${n_{\mathrm{H_2}}}$}&\multicolumn{1}{c}{$M_{\mathrm{1.2\,mm}}$}& \multicolumn{1}{c}{$M_{0.25pc}$} &\multicolumn{1}{c}{$M_{\mathrm{vir}}$}& \multicolumn{1}{c}{$T_{\mathrm{rot}}$} &\multicolumn{1}{c}{${N_{\mathrm{NH_3}}}$}\\ 
        &&  \multicolumn{1}{c}{(kpc)} & \multicolumn{1}{c}{(mag)} & \multicolumn{1}{c}{(pc)} &\multicolumn{1}{c}{($10^{22}$\,cm$^{-2}$)} &\multicolumn{1}{c}{($10^5$\,cm$^{-3}$)} &\multicolumn{1}{c}{($\mathrm{M_\odot}$)} & \multicolumn{1}{c}{($\mathrm{M_\odot}$)}& \multicolumn{1}{c}{($\mathrm{M_\odot}$)}  &\multicolumn{1}{c}{(K)} & \multicolumn{1}{c}{($10^{15}$\,cm$^{-2}$)}\\
 \noalign{\smallskip} 
\hline 
 \noalign{\smallskip} 
G012.73-00.58.. &MM1  & 1.1& 47 & 0.11 & 4.5 & 2.8     &  12 & 22 & 6 & 9.3(0.7) & 3.7(0.8)    \\  
G013.28-00.34.. &MM1   & 4.0& 31 & 0.39 & 2.9 & 0.5    & 93 & 62 & 104 & 14.9(2.0) & 3.2(1.1)  \\
G013.91-00.51.. &MM1  & 2.7& 82 & 0.28 & 7.7 & 2.2     & 145 & 113 & 50 & 14.1(1.3) & 2.5(0.5) \\
G014.39-00.75A.. &MM1 & 2.2& 61 & 0.21 & 5.7 & 2.0      & 52 & 47 & 31 & 17.4(2.0) & 1.1(0.6)     \\
G014.39-00.75B..&MM3 & 2.5& 61 & 0.23 & 5.7 & 1.8      & 62 & 90 & 19 & 11.8(1.5) & 2.4(0.8)     \\
G014.63-00.57.. &MM1 & 2.2& 280 & 0.24 & 26.3 & 7.7    & 323 & 289 & 82 & 18.1(1.3) & 4.7(0.5)  \\
             &MM2 & 2.2& 164 & 0.15 & 15.4 & 7.3    & 72 & 80 & 26 & 15.7(1.1) & 2.7(0.4)    \\
             &MM3 & 2.1& 55 & 0.17 & 5.2 & 2.2      & 31 & 32 & 34 & 15.8(1.9) &  3.4(0.8)  \\ 
             &MM4 & 2.3& 32 & 0.18 & 3.0 & 1.2      & 21 & 26 & 12 & 19.1(5.9) & 2.0(1.1)   \\ 
G016.93+00.24.. &MM1  & 2.4& 44 & 0.27 & 4.2 & 1.1     & 64 & 53 & 23 & 14.0(1.3) &  1.7(0.3)  \\
G017.19+00.81.. &MM1 & 2.5& 52 & 0.23 & 4.9 & 1.5      & 54 & 58 & 35 & 17.2(0.9) & 2.1(0.2)   \\   
             &MM2 & 2.3& 186 & 0.20 & 17.4 & 6.3    & 142 & 156 & 35 & 18.7(1.1) & 2.1(0.2) \\
             &MM3 & 2.3& 56 & 0.25 & 5.3 & 1.5      & 72 & 108 & 45 & 20.0(1.5) & 2.6(0.3)  \\ 
             &MM4 & 2.3& 31 & 0.26 & 2.9 & 0.8      & 40 & 39 & 211 & 20.1(2.1) & 3.2(1.1)  \\
G018.26-00.24.. &MM1 & 4.7& 104 & 0.39 & 9.8 & 1.8     & 306 & 201 & 91 & 18.2(1.1) & 4.3(0.4) \\
             &MM2 & 4.7& 95 & 0.48 & 8.9 & 1.3      & 431 & 208 & 201 & 17.4(1.2) & 6.5(0.6)\\
             &MM3 & 4.7& 76 & 0.29 & 7.1 & 1.7      & 130 & 113 & 150 & 15.7(1.5) & 6.4(0.9)\\
             &MM4 & 4.7& 61 & 0.30 & 5.7 & 1.3      & 110 & 88 & 140 & 16.6(1.1) &  5.3(0.6)\\
             &MM5 & 4.6& 63 & 0.47 & 6.0 & 0.9      & 278 & 145 & 197 & 16.8(1.7) & 6.0(1.0)\\
G022.06+00.21.. &MM1 & 3.7& 270 & 0.27 & 25.4 & 6.8    & 384 & 346 & 81 & 24.7(3.3) & 4.9(0.9) \\ 
             &MM2 & 3.7& 84 & 0.24 & 7.9 & 2.4      & 92 & 74 & 36 & 15.5(1.6) &   2.6(0.5) \\
G023.38-00.12.. &MM1 & 5.6& 70 & 0.50 & 6.5 & 0.9      & 346 & 163 & 134 & 18.1(1.2) & 4.4(0.5)  \\
G024.37-00.15.. &MM1 & 3.9& 52 & 0.38 & 4.9 & 1.1      & 165 & 110 & 174 & 18.6(1.9) & 7.1(1.2)  \\
             &MM2 & 3.8& 47 & 0.30 & 4.4 & 1.2      & 94 & 74 & 70 & 15.7(1.5) & 3.8(0.7)    \\
G024.61-00.3.3. &MM1 & 3.1& 63 & 0.34 & 5.9 & 1.2      & 140 & 106 & 69 & 17.5(1.9) & 2.0(0.4)  \\
             &MM2 & 3.2& 38 & 0.31 & 3.5 & 0.8      & 72 & 49 & 21 & 15.5(1.2) &   1.5(0.2)  \\
G024.94-00.15.. &MM1 & 3.3& 69 & 0.26 & 6.5 & 2.0      & 104 & 94 & 61 & 15.2(1.0) & 3.8(0.4)    \\
             &MM2 & 3.4& 59 & 0.25 & 5.5 & 1.8      & 85 & 67 & 52 & 15.2(1.1) &  3.5(0.5)    \\
G030.90+00.00A.. &MM1 & 4.6& 55 & 0.53 & 5.2 & 0.7      & 311 & 131 & 125 & 18.6(1.7) & 3.1(0.6)\\ 
G030.90+00.00B.. &MM2 & 7.2& 61 & 0.76 & 5.8 & 0.5      & 702 & 736 & 497 & .. & ..\\
G030.90+00.00C.. &MM3 & 5.7& 46 & 0.56 & 4.3 & 0.5      & 283 & 126 & 71 & 16.4(2.4) & 3.4(0.9)\\
G030.90+00.00D.. &MM4 & 2.6& 101 & 0.21 & 9.5 & 3.2     & 49 & 95 & 22 & .. & ..\\
G034.71-00.63.. &MM1 & 3.0& 61 & 0.43 & 5.8 & 1.0      & 224 & 129 & 130 & 17.8(1.1) & 2.2(0.3)\\ 
             &MM2 & 3.0& 55 & 0.33 & 5.2 & 1.1      & 122 & 83 & 59 & 12.4(0.9) & 4.4(0.7)  \\
             &MM3 & 3.1& 26 & 0.29 & 2.5 & 0.6      & 43 & 38 & 59 & 17.1(1.9) & 1.0(0.4)   \\
G035.49-00.30A.. &MM1 & 3.6& 71 & 0.26 & 6.6 & 1.8      & 98 & 89 & 100 & 18.6(2.9) & 1.7(1.0) \\
G035.49-00.30B.. &MM2 & 3.0& 57 & 0.25 & 5.3 & 1.5      & 72 & 49 & 26 & 11.9(0.7) & 3.9(0.5)   \\
             &MM3 & 3.0& 44 & 0.38 & 4.2 & 0.8      & 127 & 81 & 102 & 13.6(0.7) & 4.3(0.4)\\
             &MM4 & 3.0& 39 & 0.26 & 3.7 & 1.0      & 54 & 36 & .. & .. & .. \\
             &MM5 & 3.0& 40 & 0.28 & 3.8 & 1.0      & 64 & 46 & .. & .. & .. \\
             &MM6 & 3.0& 36 & 0.34 & 3.3 & 0.7      & 83 & 55 & .. & .. & .. \\ 
G050.06+00.06.. &MM1 & 4.8& 47 & 0.47 & 4.4 & 0.7      & 208 & 104 & 83 & 14.9(2.2) & 2.1(0.6)\\
             &MM2 & 4.8& 49 & 0.49 & 4.6 & 0.7      & 233 & 106 & 87 & 14.1(1.7) & 1.7(0.5)\\
G053.81-00.00.. &MM1  & 1.9& 74 & 0.14 & 6.9 & 3.6     & 28 & 36 & 28 & 12.4(1.4) & 2.1(0.5)  \\
\noalign{\smallskip}
\hline                  
\end{tabular}
\end{center}
{\bf Notes.} The first two columns give the HEC name and the millimeter clump
number, the following columns represent (in order of appearance): kinematic
distance, visual extinction based on $N_\mathrm{H_2}$ from the 1.2\,mm emission, size at FWHM, hydrogen column density, hydrogen volume density, mass, mass within 0.25\,pc diameter, virial mass, \amm\ rotational temperature, and \amm\ column density.\\
\end{table*}
\subsection{Extinction masses}

Since the color excess, or extinction, is a direct measure of the amount of
column density in a region, we could derive extinction masses.  
One magnitude of $A_V$ is related to a hydrogen column density, $
N_{\mathrm{H_2}}$, according to \citet{bohlin:1978} and \citet{frerking:1982} by
\begin{equation}
A_V = \frac{ N_{\mathrm{H_2}}}{0.94 \times 10^{21}}\hspace{0.5cm} [\mathrm{mag}].
\label{eq:ext}
\end{equation}
Multiplying by the mass of a hydrogen atom, $m_\mathrm{H}$, the mean
molecular weight, $\mu=2.33$, and the cloud surface, $A$, one arrives at the extinction mass: 
\begin{equation}
M_{\mathrm{ext}} = 3.7\times10^{26}\mu m_\mathrm{H}  <E(3.6\,\mu\mathrm{m}-4.5\,\mu\mathrm{m})> \big( \frac{A}{\mathrm{pc^2}}\big)\hspace{0.5cm} [\mathrm{M_\odot}]. \\
\end{equation}
The extinction mass is independent of temperature, but it still depends on
the distance via the cloud surface $A$. We derive extinction masses for the clouds from $\sim$30 to 6500\,$\mathrm{M_\odot}$ (Table \ref{ta:cloud}).

\subsection{Column densities and visual extinction}

As mm emission from cool clouds is usually optically thin, the column density and the mass of a
cloud are well sampled by the observed flux density.
The beam averaged column density is given by \citet{motte:2007}:
\begin{equation}
N_{\mathrm{H_2}} = \frac{F^{p}_{1.2\mathrm{mm}}}{\Omega m_\mathrm{H} \mu \kappa_{1.2\mathrm{mm}} B_{1.2\mathrm{mm}}(T_\mathrm{dust})}\,,
\end{equation}
where $F^{p}_{1.2\mathrm{mm}}$ is the peak flux, $\Omega$ the beam
solid angle, $\kappa_{1.2\mathrm{mm}}$ the dust opacity at 1.2\,mm per unit mass
  column density, assuming a gas-to-dust ratio of 100, $B_{1.2\mathrm{mm}}(T_{\mathrm{dust}})$ the (full) Planck function at the dust temperature, and $m_\mathrm{H}$ and $\mu$ as defined before. The column density depends strongly on the dust properties:
 the dust opacity, $\kappa_\lambda$, and the emission coefficient, $\beta$, are related as $\kappa_\lambda =
\kappa_{0}(\frac{\lambda}{\lambda_0})^{-\beta}$. We considered two different opacities; $\kappa_{\mathrm{1.2mm}} = 0.4\,\mathrm{cm^2g^{-1}}$
after \citet{hildebrand:1983} and $\kappa_{\mathrm{1.2mm}}=1.0\,\mathrm{cm^2g^{-1}}$ taken from
\citet{ossenkopf:1994}, Table 1 column 6, for dust grains with thin ice
mantles. The emission coefficient was kept at $\beta=2$, the advocated value for cold dust clumps \citep{hill:2006}. The dust opacities differ by a factor 2.5, meaning the $\kappa$ of \citet{hildebrand:1983} results in 2.5 times larger column densities than the opacity of \citet{ossenkopf:1994}. In this work we used the dust opacity of \citet{ossenkopf:1994}, $\kappa_{\mathrm{1.2mm}}=1.0\,\mathrm{cm^2g^{-1}}$.
For the dust temperature we assumed the rotational temperature derived from ammonia (see Section \ref{sec:temp}). Since the clumps have high densities, collisions will dominate over radiative processes and the temperature exchange between dust and gas will be efficient.
In sources without \amm\ detection we assumed a dust temperature of 16\,K, which was the average \amm\ rotational temperature. For clumps which have an embedded protostar the \amm\ rotational temperature, derived on a 40\arcsec\ scale, might underestimate the dust temperature leading to an overestimation of the derived masses and column densities.

We found clumps with column densities of the order of $10^{22}-10^{23}\,\mathrm{cm^{-2}}$ (Table \ref{ta:mass}). We derived corresponding peak visual
extinction values by applying Eq. \ref{eq:ext}. The peak fluxes, column densities and peak visual extinction of Table \ref{ta:mass} correspond to the positions listed in Table \ref{ta:flux}.
The clumps have peaks in $A_\mathrm{V}$ from 31 to 280\,mag,
with an average value of 75\,mag. These values are much higher than the mean visual extinction reached
by the extinction method for this selected sample, which is between 16 and 47\,mag.
This is expected because the peak values are larger than the mean and with the limited resolution of 54\arcsec\ higher extinction peaks (as found for example in IRDCs) are missed. Additionally, also the temperature might play a role, since the derivation of the $A_\mathrm{V}$ from the mm emission depends on temperature while the extinction method does not.



\subsection{Masses from 1.2 mm emission}

The clump mass can be derived from the 1.2\,mm emission by \citep{hildebrand:1983, motte:2007}:
\begin{equation}
M_{\mathrm{1.2mm}} = \frac{F^i_{\mathrm{1.2mm}} d^2}{\kappa_{\mathrm{1.2mm}} B_{\mathrm{1.2mm}}(T_{\mathrm{dust}})}\,,
\label{eq:mass}
\end{equation}
where $F^i_{\mathrm{1.2mm}}$ is the integrated flux density of the clump, $d$ the distance, and $\kappa_{\mathrm{1.2mm}}$, $B_{\mathrm{1.2mm}}$, and $T_{\mathrm{dust}}$ as defined before. The measured flux density can be
contaminated by free-free emission if an ionizing source is present. Since our sample contained very young objects, we can neglect such contamination.
The derived masses depend strongly on temperature and distance; a $10\%$ nearer distance decreases
the mass by $20\%$, a $10\%$ decrease in temperature increases the mass
by $17\%$. Nevertheless, the largest uncertainty in the mass derivation is caused by the dust properties, as described in the previous section.

The majority of the clumps have masses between 12 and 700\,$\mathrm{M_\odot}$ and are located at distances 
between 2 and 5\,kpc (see the squares in the left panel of Fig.~\ref{fig:distance-m-r} and Table \ref{ta:mass}).
We find no low-mass clumps ($M < 100\,\mathrm{M_\odot}$) further than 4\,kpc, which is an bias of our extinction method (discussed in Sect. \ref{bias}). Almost all high mass clumps ($M >
100\,\mathrm{M_\odot}$) are located at distances larger than 4\,kpc.

\begin{figure}[!htb]
\includegraphics[height=0.8\columnwidth, angle=-90]{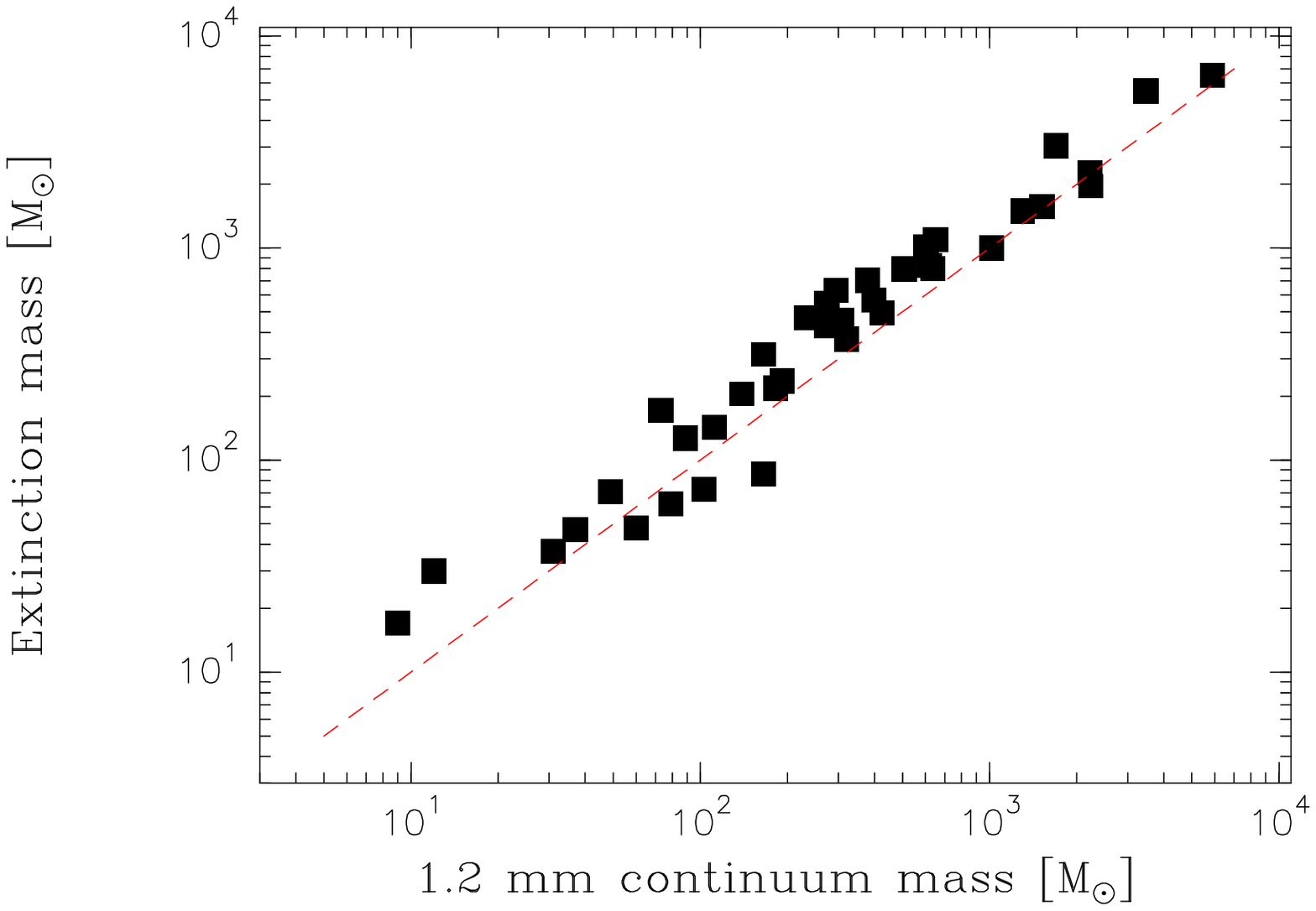}
\caption{\label{fig:av-bol}The cloud masses estimated from the color excess versus the masses
  estimated from the 1.2 mm continuum maps. The red dashed line is a help line indicating the trend in which both mass estimates are equal.}
\end{figure}

A second method was used as a comparison for the clump masses to check if the
source finding algorithm was biased to a source size (and hence clump mass).  
Based on the (near) kinematical distances, we defined
circles of 0.25\,pc diameter for each source. Then, we derived the integrated
flux, $F_{0.25\mathrm{pc}}$, and the mass, $M_{0.25\mathrm{pc}}$, for the region within this circle (see Tables \ref{ta:flux}, \ref{ta:mass} and \ref{ta:noclump}). The crosses in the left panel of Fig.~\ref{fig:distance-m-r} show $M_{0.25\mathrm{pc}}$ as a function of the distance. 
The clump masses within 0.25\,pc show a similar behavior as the clump masses determined by Gaussian fits.
\label{sect:mass-cloud}
On a larger scale, the cloud mass was derived from the 1.2\,mm emission according to Eq.~\ref{eq:mass} using the integrated flux down to 3$\sigma$. 
The cloud diameters are of order $\sim$90\arcsec , so after resampling the extinction
map, it was possible to compare the cloud masses derived by extinction with the masses from the 1.2\,mm emission.
We find the extinction masses to be larger by a factor $\sim1.3$ than the masses derived from
the dust continuum maps (see Fig.~\ref{fig:av-bol}). 
This is expected, since the bolometer
filters out large-scale structures by the sky noise subtraction and
chopping. The cloud masses derived from the 1.2\,mm emission and the
extinction are listed in Table \ref{ta:cloud}.

Finally, we estimated the volume-averaged gas density of the clumps, $n_\mathrm{H_2}$, following \citet{motte:2007}:
\begin{equation}
n_{\mathrm{H_2}} = \frac{M}{\frac{4}{3} \pi R^3 \mu m_\mathrm{H} }\,,
\end{equation}
where $M$ is the clump mass, $R$ the clump radius given by the geometrical mean of the
semi major and semi minor axis from the Gaussian fit. The average gas density
is $\sim2\times10^5\,\mathrm{cm^{-3}}$ and the individual results are given in Table \ref{ta:mass}.

\subsection{Virial masses}

\begin{figure}[!hptb]
\includegraphics[height=0.8\columnwidth, angle=-90]{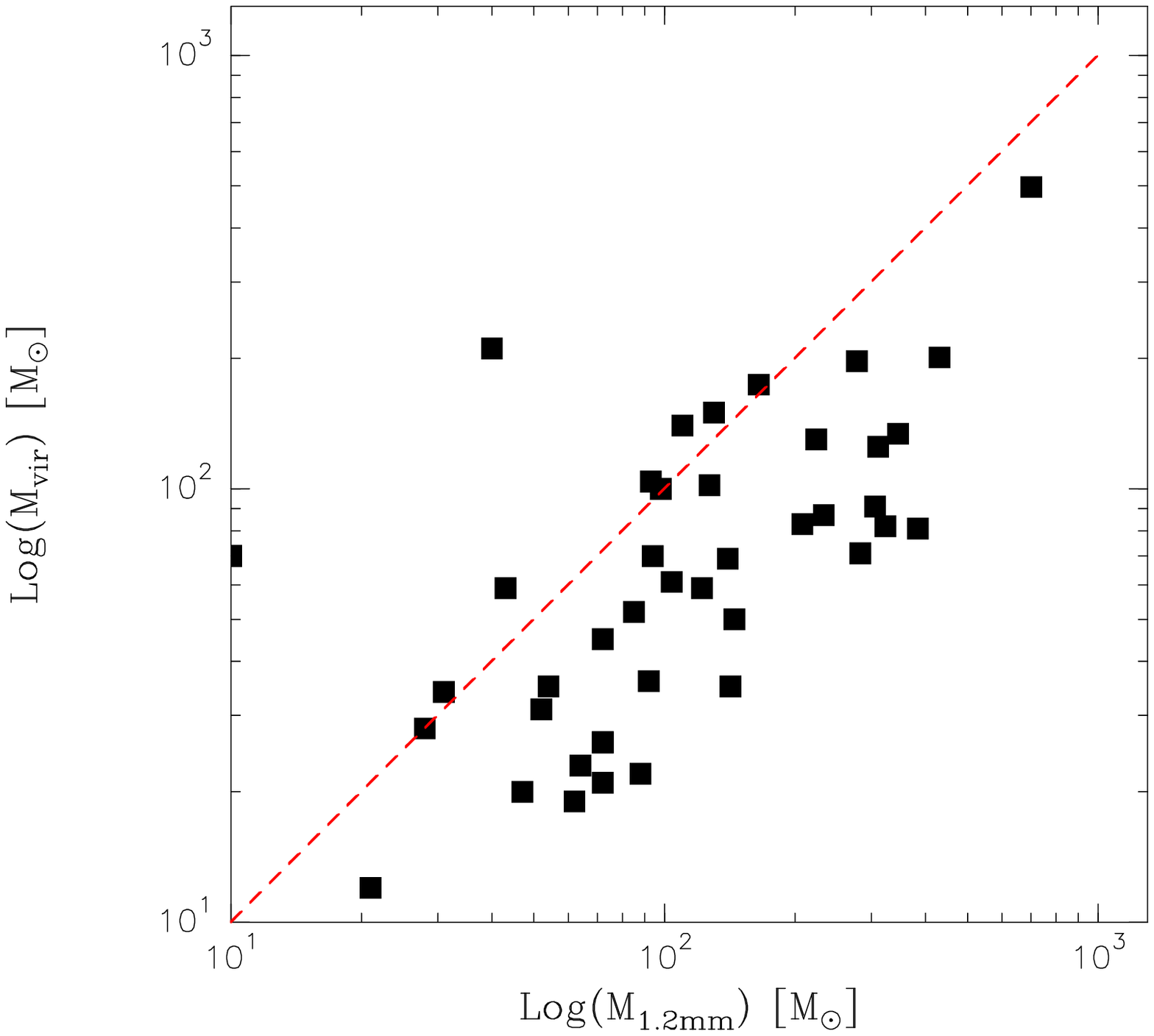}
\caption{\label{fig:mvir}Virial clump masses versus the 1.2 mm continuum clump masses. The red dashed line indicates the trend where the virial parameter is unity.} 
\end{figure} 

The thermal line widths are usually one order of magnitude smaller than the observed \amm\ line widths, which indicates that the observed line widths are dominated by turbulence, hence contain information of the average kinetic
energy within a clump. 

Given an optically thin line width and a clump radius, the
virial mass, $M_{\mathrm{vir}} = k_1\sigma^2R/G$, can be calculated \citep{macLaren:1988}. Here $\sigma$ is the three dimensional root-mean-square velocity, 
$R$ the clump radius and $k_1$ a density distribution constant. For a constant
density distribution $k_1 = 5/3$. After the conversion of $\sigma$ to the observable
FWHM line width $\Delta v$, $\sigma^2 = (3/8\ln2)\Delta v^2 $ \citep{rohlfs:2004}, the virial mass can be written as: 
\begin{equation}
M_{\mathrm{vir}} = 210 \times \big(\frac{\Delta v}{\mathrm{km\,s^{-1}}}\big)^2 \big(\frac{R}{\mathrm{pc}}\big) \hspace{0.5cm}[M_\odot].
\label{eq:mvir}
\end{equation}
The resultant virial masses, $M_{\mathrm{vir}}$, are listed for each clump
in Table \ref{ta:mass}.
The virial parameter is defined as $\alpha=\frac{M_{\mathrm{vir}}}{M}$ \citep{bertoldi:1992}. For $\alpha < 1$ , the clumps are dominated by gravity, however with $\alpha>>1$,  the clumps are confined by surface pressure and self-gravity is unimportant. 
Most clumps in high extinction clouds have clump masses larger than their virial masses (see Fig.~\ref{fig:mvir}), indicating that possibly most clumps are dominated by gravity and are collapsing.

\subsection{Temperatures}
\label{sec:temp}
The ammonia molecule, described thoroughly by \citet{ho:1983}, is often used as a molecular cloud thermometer \citep{danby:1988}. Its energy levels are parameterized by the $J$ and $K$ quantum numbers, measuring, respectively, the magnitude of the total angular momentum and its component along the symmetry axis. Each set of rotational transitions is arranged into so-called $K$-ladders, levels of fixed $K$-value. From the symmetry of the electric dipole moment of the molecule, all dipole transitions with nonzero $\Delta K$ are forbidden, meaning that the $K$-ladders are independent of each other.
The lowest transitions of each $K$-ladder are metastable, and can be excited
via collisions. Additionally, \amm\ also undergoes vibrational motion from the tunneling of the
nitrogen atom through the hydrogen plane, which splits the rotational energy levels into inversion doublets.

The inversion transitions are further split into five quadrupole hyperfine lines, which allow the calculation of the
optical depth (see the hyperfine lines in the \amm(1,1) transition in Fig.~\ref{fig:amm}). Typical values of the optical depth were between 1 and 3, showing that ammonia is optically thick in most cases.
With the optical depth known, the rotational
temperature follows from the ratio of the peak intensities of \amm(1,1) and \amm(2,2) lines after \citet{mangum:1992}:
\begin{equation}
T_{\mathrm{rot}} = -41.5 \big[ \ln \big[ -\frac{0.283 \Delta v_{22}}{\tau_\mathrm{main} \Delta
  v_{11} }\ln\big[ 1 - \frac{T_{22}}{T_{11}}(1 - e^{-\tau})\big]\big]   \big]^{-1}\hspace{0.5cm}[\mathrm{K}], 
\end{equation}
where $\tau_{\mathrm{main}}$ is the \amm(1,1) main group optical depth, $\Delta v_{11}$, $\Delta
v_{22}$, $T_{11}$, $T_{22}$ are the line widths and the peak
intensities of the \amm(1,1) and \amm(2,2) lines,
respectively. 

The line width ratio in this equation is debatable. Observations show that the \amm(1,1) and \amm(2,2) line widths are dominated by turbulence. 
If one assumes that both transitions arise from the same region, they experience the same turbulence, and the line widths are equal, $\Delta v_{22}=\Delta v_{11}$. In this case, the linewidth ratio drops out of the equation, which becomes the equation from \citet{ho:1983}. 
We observed slightly larger \amm(2,2) line widths for
several clumps, probably since this transition has a higher energy and is therefore more
sensitive to warmer and more turbulent regions. The \amm(2,2) line widths are therefore more sensitive to the peak temperature, determined by the turbulent outliers. We considered it reasonable to use $\Delta v_{22}$ =$\Delta v_{11}$, since it would give an `average' of the \amm(2,2) turbulence, and hence an `average' rotational temperature of that region. 
The two methods yield very similar results; the \citet{ho:1983} temperatures were slightly lower ($\Delta T_{\mathrm{rot}}
 \simeq 1$K) than when using the formula of \citet{mangum:1992}.

The clumps are on average cold $\sim$16\,K. The temperatures
 range from 10 to 25\,K. For such low temperatures, the kinetic temperature is well approximated by $T_{\mathrm{rot}}$ \citep{malcolm:1983,danby:1988}. 

We derived column densities of the \amm(1,1) line \citep[after][]{mangum:1992}:
\begin{equation}
N_{(1,1)} = 6.60 \times 10^{14} \Delta v_{11} \tau \frac{T_{\mathrm{rot}}}{\nu_{11}}\hspace{0.5cm}[\mathrm{cm}^{-2.}]
\end{equation}
From this we estimated the total ammonia column density \citep[following][]{li:2003}:
\begin{eqnarray}
N_{\mathrm{NH_3}} =  N_{(1,1)} \Big[1 +\frac{1}{3}\exp\big(\frac{23.1}{T_{\mathrm{rot}}}\big) +
\frac{5}{3}\exp\big(\frac{-41.2}{T_{\mathrm{rot}}}\big)\nonumber\\
+ \frac{14}{3}\exp\big(\frac{-99.4}{T_{\mathrm{rot}}}\big)\Big]\hspace{0.5cm}[\mathrm{cm}^{-2}].
\end{eqnarray}
The averaged ammonia column density for the clumps is $3.1\times10^{15}\,\mathrm{cm^{-2}}$.
The \amm\ rotational temperatures and column densities for all clumps are given in Table \ref{ta:mass} together with the other parameters derived in this Section.
\section {Discussion}

\subsection{High extinction structures}

Thanks to the high resolution Spitzer IRAC data, the extinction method allows to follow the mass distribution from Galactic size-scales down to single clouds. High extinction traces Galactic structure (see the Galactic distribution of the color excess in Fig.~\ref{fig:ll}), similar to the CO survey of
the Galactic plane \citep{dame:1987} and the dust continuum surveys, ATLASGAL at 850~$\mathrm{\mu m}$ \citep{schuller:2009} and BOLOCAM at 1.1\,mm \citep{rosolowsky:2009}. In addition, the distribution of 6.7 GHz Class II methanol masers traces high-mass star-forming regions and, thus, Galactic structure \citep{pestalozzi:2005}. All these surveys, except for the methanol masers, peak toward the
Galactic Center. The extinction maps miss the inner 1\degr\ around the Galactic Center, since the Spitzer IRAC data were not publicly available for this region at the time. 
However, a rising trend toward the Galactic Center was observed. The extinction distribution indicates that the column densities are higher towards the fourth quadrant, $0\degr> l >-60\degr$. We find no evidence for this from the CO distribution, nor from the methanol masers. Only the ATLASGAL survey \citep{schuller:2009} hints at a similar distribution, but needs to be extended in longitude range for giving conclusive evidence. 

The next eye-catching features in the extinction distribution were two bumps, peaking roughly around longitudes of $\sim40\degr$ and $-30\degr$ (see Fig.~\ref{fig:ll}), which were confirmed by the methanol maser distribution \citep{pestalozzi:2005}. 
We identified several spiral arms by comparing our results to the Galactic
models of \citet{vallee:2008}. In Fig.~\ref{fig:ll} each spiral arm traced by a maximum in the extinction distribution is marked. We can
trace the tangential point of the Sagittarius-Carina arm at $l\sim 50\degr$, the
tangential points of the Scutum-Crux arm  at $l\sim 35\degr$ and $l\sim
-45\degr$, the beginning of the Norma-Cygnus arm on the near end of the Galactic bar
$l\sim10\degr$, the beginning of the Sagittarius-Carina arm at the far end of
the Galactic bar $l\sim-5\degr$, the beginning of the Perseus arm
$l\sim-20\degr$, and the tangential point of the Norma-Cygnus arm at $l\sim-30\degr$.
Several of these features are also seen in the ATLASGAL survey, and to a lesser extent in the CO distribution. Hence, the total extinction distribution seems to agree with results of
previous studies, and tracks large scale mass structures.

The combination of submillimeter dust continuum emission and extinction maps extends the size scales down to clump sizes of fractions of parsecs. In Table \ref{ta:scale} the mean masses, radii, and volume densities for different size-scales of complexes, clouds and clumps are compared. Complexes are the low density ($A_\mathrm{V}>21\,\mathrm{mag}$ or $N_\mathrm{H_2}>2\times10^{22}\,\mathrm{cm^{-2}}$) regions surrounding the cloud. 
In general, the masses and sizes decrease towards the smaller scales, while
the volume density increases as expected. The cloud masses derived from the
extinction are higher than from the mm emission, as discussed in
Sect. \ref{sect:mass-cloud}. The complexes and the clouds are confined by
  the pressure of the surroudning medium rather than by gravity. This is based
  on their virial masses, where we used for the line width estimate the $^{13}$CO data from the Galactic Ring Survey (GRS)
  \citep{jackson:2006}. The $^{13}$CO line is a probe of the low density material
  and with the GRS resolution of 47\arcsec\ the line width is represenative for a cloud-scale average. Most of the
  clumps inside the clouds are, however, bound objects for which $M_\mathrm{vir}<M_\mathrm{1.2\,mm}$.

\begin{table}[!htb]
\begin{minipage}[t]{\columnwidth}
\renewcommand{\footnoterule}{}
\caption{\label{ta:scale}Mean properties of high extinction complexes, clouds and clumps}
\centering
\begin{tabular}{l|lrcr}
\hline\hline
Size scale & Method & Mass & Radius & Vol. den.\\
 & &($\mathrm{M_\odot})$& (pc) & (cm$^{-3}$)\\
\hline
 complex\footnote{The high extinction complexes should be treated as an
   indication since they have not so strictly defined boundaries as clouds and
 clumps.} & extinction & 4000 & 1.4~~ &  6$\times 10^3$\\
 clouds  & extinction & 910 & 0.7~~ & 1$\times 10^4$\\
 clouds  & 1.2 mm emission \footnote{smoothed maps by 20\arcsec\ Gaussian} & 700 & 0.7~~ & 9$\times 10^3$\\
 clumps  & 1.2 mm emission \footnote{unsmoothed maps} & 130 & 0.15 & 2$\times 10^5$ \\ 
\hline
\end{tabular}
\end{minipage}
\end{table}

\subsection{Evolutionary sequence}

Our observations suggest different classes of clouds and in the following we discuss their possible relation to different evolutionary stages.
Clouds, which either have no clumps or clumps of which the peak in the mm emission is less than twice the mean emission in the cloud, were defined as diffuse clouds -- this definition is based on the cloud morphology and therefore different from the ``classical diffuse clouds'' defined by $A_\mathrm{V}<1\,$mag \citep{snow:2006}. 
Clouds which contain clumps with a higher contrast, above twice the mean cloud emission, were considered to be peaked clouds. If the peak flux was above thrice the mean emission and there were two or more clumps the cloud was classified as a multiply peaked cloud. The ratio of the clump peak emission to the mean emission of the cloud is given in Table \ref{ta:cloud} together with the classification. The mean physical properties such as temperature, masses, column densities and line widths are put together in Table \ref{ta:fase} for the three classes. 

{\it Diffuse clouds} are the most likely candidates to form, or harbor, starless clumps, which are expected to be cold, more extended and more massive. The clump, or cloud, consists of low column density material and contains no massive compact object. Indeed, for very diffuse clouds, the source find algorithm yielded few clumps. On the smoothed maps, the algorithm returned values which were close to cloud size-scales without much substructure. This supports the idea that these diffuse clouds represent the earliest stage in which few condensations have formed and gravitational collapse has not yet started. More evidence for the young nature of diffuse clouds was found in the 24~$\mathrm{\mu}$m MIPS data: generally the (smoothed) dust emission followed the
24~$\mathrm{\mu}$m-dark regions and only a few 24~$\mathrm{\mu}$m sources of
$\sim$100$\,\mathrm{MJy~sr^{-1}}$, located within 20\arcsec\ of mm peak, were found
toward the clouds. In two cases 24\,$\mu$m source $>100\,\mathrm{MJy~sr^{-1}}$
were found the edge of the cloud, at $\sim$1\arcmin\ from the mm peak. 
 Examples are G012.73--00.58, G013.28--00.34, and G034.85+00.43 (see Fig.~\ref{fig:mips}).

As the first clumps become more compact and their mm peak flux rises above twice the mean emission of cloud, in the {\it peaked cloud} stage, the clump is accreting material from a lower density reservoir or envelope surrounding it, and consequently the temperature and turbulence rise. Since star formation is a process which is dynamic and includes feedback and
triggering, already in this early stage a cloud can contain more than one clump, though possibly the other peaks barely stick out above the mean mm emission of the cloud. The peaked clouds are generally dark at 24~$\mathrm{\mu}$m,
except for a bright 24~$\mathrm{\mu}$m source located within
20\arcsec\ of the mm peak (see
G013.91--00.51, G016.93+00.24, and G053.81--00.00 in Fig.~\ref{fig:mips}). The
peak flux of the 24\,$\mu$m sources is on average
$\sim250\,\mathrm{Mjy~sr^{-1}}$, which is higher than what was found for the diffuse clouds.

After more accretion from the reservoir, the clump will have a much higher mm peak flux compared to the initial stage, reaching at least trice the mean emission of the cloud. The clump mass will have decreased toward the
more evolved stage, since not all of the mass reservoir will not be accreted
onto the clump, as the star formation efficiency is apparently less then 100\%. 
The column density, temperature and turbulence will increase even further than in the peaked cloud stage. While in the peaked stage, there were few clouds which had multiple clumps, in the more evolved stage one expects to find more clouds with multiple clumps. Most of these clumps should be well above the mean emission of the cloud. 
In these {\it multiply peaked clouds}, 
the clumps will be in different cloud stages of evolution depending on the
initial conditions of the clump. Also not all clumps will form high-mass
stars, so they will have different properties.
And indeed, several clumps in the multiply peaked clouds show bright, $\gtrsim 500\,\mathrm{Mjy~sr^{-1}}$,
24~$\mathrm{\mu}$m emission, indicating a protostar, within 10\arcsec\ or less of the mm peak, while
other clumps are infrared dark. Such clouds with clumps in various phases are not completely infrared dark, since the dust is locally heated from
the already hot HMPO or UCH{\sc ii} region. Examples for multiply peaked clumps
are shown in Fig.~\ref{fig:mips}:
G014.63--00.57, with the very bright component MM1 and the cloud
G018.26--00.24 with five clumps. 

Additionally, one can check the evolutionary stage by searching for signs of
star formation such as water masers, Class II methanol masers, and shocks. Table  \ref{ta:fase} also contains entries for such star formation tracers.  Water masers are thought to be caused by outflows during (low and high-mass) star formation \citep{menten:1996}. The Class II methanol masers are uniquely associated with high-mass star formation, and are usually found prior or coexistent with an H {\sc ii } region \citep{menten:1991, ellingsen:2006, pestalozzi:2007}. While the water masers are found towards both the peaked and the multiply peaked clouds, the methanol masers are only found toward the latter class.
Another indication for star formation comes from gas excited by shocks, such as H$_2$($v$=0--0), $S$(9,10,11) or CO($v$=1--0) gas. The IRAC 4.5\,$\mu$m band contains both these transitions, and can therefore be used to search for ongoing star formation. The close association of the extended 4.5\,$\mu$m objects with Class II methanol masers, found by \citet{cyganowski:2008}, suggests that they might trace ongoing {\em massive} star formation. We find two multiple peaked clouds in our sample that contain such an extended 4.5\,$\mu$m source, of which one also contains Class II methanol masers. 

The peaked and multiply peaked clouds show active star formation, while the
diffuse clouds do not. Possibly, these clouds might not form stars or will not
have column densities high enough to form massive stars. However, some, for example G013.28--00.34, seem to be forming clumps which might evolve into active star-forming clumps. Table \ref{ta:fase} shows that the clump temperatures are very similar, especially in the case of peaked and the multiply peaked clouds; the diffuse clouds stand out  with their temperature of $13.5\pm1.5$\,K against the average temperature of all clouds $16\pm1.8$\,K. As star formation produces stars in a large range of masses, so cluster-forming regions or clumps within one cloud are expected to have different column densities and temperatures. Given the range of column densities for the clumps $3-30\times10^{22}\,\mathrm{cm^{-2}}$ not all clumps might evolve into clusters with massive stars, even if mass accretion is still continued from the low density material surrounding the clump.

\begin{table}[!h]
\begin{minipage}[t]{\columnwidth}
\caption{\label{ta:fase} Mean physical properties of the clumps and
    star formation indicators in diffuse, peaked
    and multiply peaked clouds.}
\centering
\renewcommand{\footnoterule}{}
\begin{tabular}{l|ccc}
\hline
\hline
  &Diffuse & Peaked & Multiply peaked\\
\noalign{\smallskip}
\hline
\noalign{\smallskip}
Mass ($\mathrm{M_\odot}$) &{\it 230}\footnote{As most diffuse clouds contained no clumps, we derived the sizes from the smoothed bolometer maps of the diffuse clouds. These are therefore not properties of the clumps in diffuse clouds, but properties of the diffuse clouds themselves.} &  185   & 150\\
Size (pc) &{\it 0.52}$^a$ & 0.36 & 0.30\\
$N_\mathrm{N_{H_2}}$ ($10^{22}$cm$^{-2}$) &{\it 4.5}$^a$&5.5&8.5\\
$T_\mathrm{rot}$ (K)& 13.5 & 15.7 & 17.5\\ 
$\Delta v$ (km\,s$^{-1}$)& 1.2 & 1.4 & 1.6 \\
IRDC corr.\footnote{Catalogs of \citet{simon:2006a} and \citet{simon:2006b}.} (within 2\arcmin) & 83\% & 50\% & 70\%\\
24$\mu$m\footnote{$>100\,\mathrm{MJy~sr^{-1}}$} (within 1\arcmin)& 17\% &  83\% & 100\%\\
Ext. 4.5$\mu$m \footnote{Extended 4.5$\mu$m sources from the catalog of \citet{cyganowski:2008}}  (within 1\arcmin)& none & none & 29\%\\
Masers\footnote{H$_2$O: \citet{jaffe:1981} and this work, CH$_3$OH: \citet{szymczak:2000}} (within 1\arcmin) &none & $\mathrm{H_2O}$ 25\%& $\mathrm{H_2O}$ 57\%\\
&&& $\mathrm{CH_3OH}$ 29\%\\
\noalign{\smallskip}
\hline
\end{tabular}
\end{minipage}
\end{table}

\subsection{Comments on individual sources}

We performed a systematic search using the SIMBAD Astronomical Database of our
cloud to find signs of star formation such as maser emission, H{\sc ii}
regions and IRAS sources. We discuss here several diffuse, peaked and multiply peaked high extinction clouds, which
are representative examples of typical clouds in our sample. Additionally, several unique and interesting clouds are mentioned. 

\subsubsection{ G013.28--00.34}
The cloud G013.28--00.34, with its one centrally located clump MM1, is a good example of a cloud in a very early phase of
protocluster formation. This clump has a low column density
$3\times10^{22}\,\mathrm{cm^{-2}}$ and contains less than 5\% of
the total cloud mass. Likely, this clump did not yet start to collapse, since from the virial mass we know that the clump is confined by gas pressure. Additionally, the clump
contains no 24\,$\mu$m emission, which is a further evidence for stage prior to star formation. The moderate \amm\ line widths of $1.6\,\mathrm{km~s^{-1}}$ indicate a source of turbulence or large scale motions, however, without an \amm\ map of the cloud it is impossible to isolate the source for the turbulence.

\subsubsection{G013.91--00.51}
In the cloud G013.91--00.51 the 1.2\,mm emission follows an infrared dark
patch. The one clump in this cloud which is three times above the mean emission is located nearby two infrared
emission peaks and can be possibly triggered by the star formation in these
infrared bright sites. The clump is very massive, 145\,$\mathrm{M_\odot}$, and
cold. The virial mass is 50\,$\mathrm{M_\odot}$ suggesting a collapse.

\subsubsection{  G014.63--00.57}
Cloud G014.63--00.57 is nested around two bright infrared sources. The 1.2\,mm peak fluxes are high,
especially for clumps nearby the infrared sources. The brightest clump,
MM1, contains a compact bright infrared source, IRAS 18164--1631, with a flux density at 25\,$\mu$m $F_{\mathrm{25\mu m}}=23$\,Jy \citep{helou:1988} and has wide \amm\ line widths of  $1.8\,\mathrm{km\,s^{-1}}$ indicating large turbulence. Additionally, water masers were detected towards this clump \citep{jaffe:1981} and \citet{cyganowski:2008} showed that there is extended 4.5\,$\mu$m emission -- a further indication for outflows.  
All evidence points to a state of ongoing star formation for the MM1 clump. The other clumps of G014.63--00.57 are still infrared dark.
The MM1 and MM2 clumps have the theoretically required column density for massive star formation (see Sect. \ref{sect:theory}).

\subsubsection{ G016.93+00.24}
In G016.93+00.24 the 1.2\,mm emission follows the infrared dark filament. In the
west end of the filament, the 1.2\,mm peak MM1, is a few arc seconds offset
from a infrared bright emission peak. The \amm\ line widths are relatively
narrow, $0.9\,\mathrm{km~s^{-1}}$, and the temperature is low, $14\,$K, which suggests that the clump is in a
very early phase with little turbulence.

\subsubsection{ G017.19+00.81}

\begin{figure}
\includegraphics[height=\columnwidth, angle=-90]{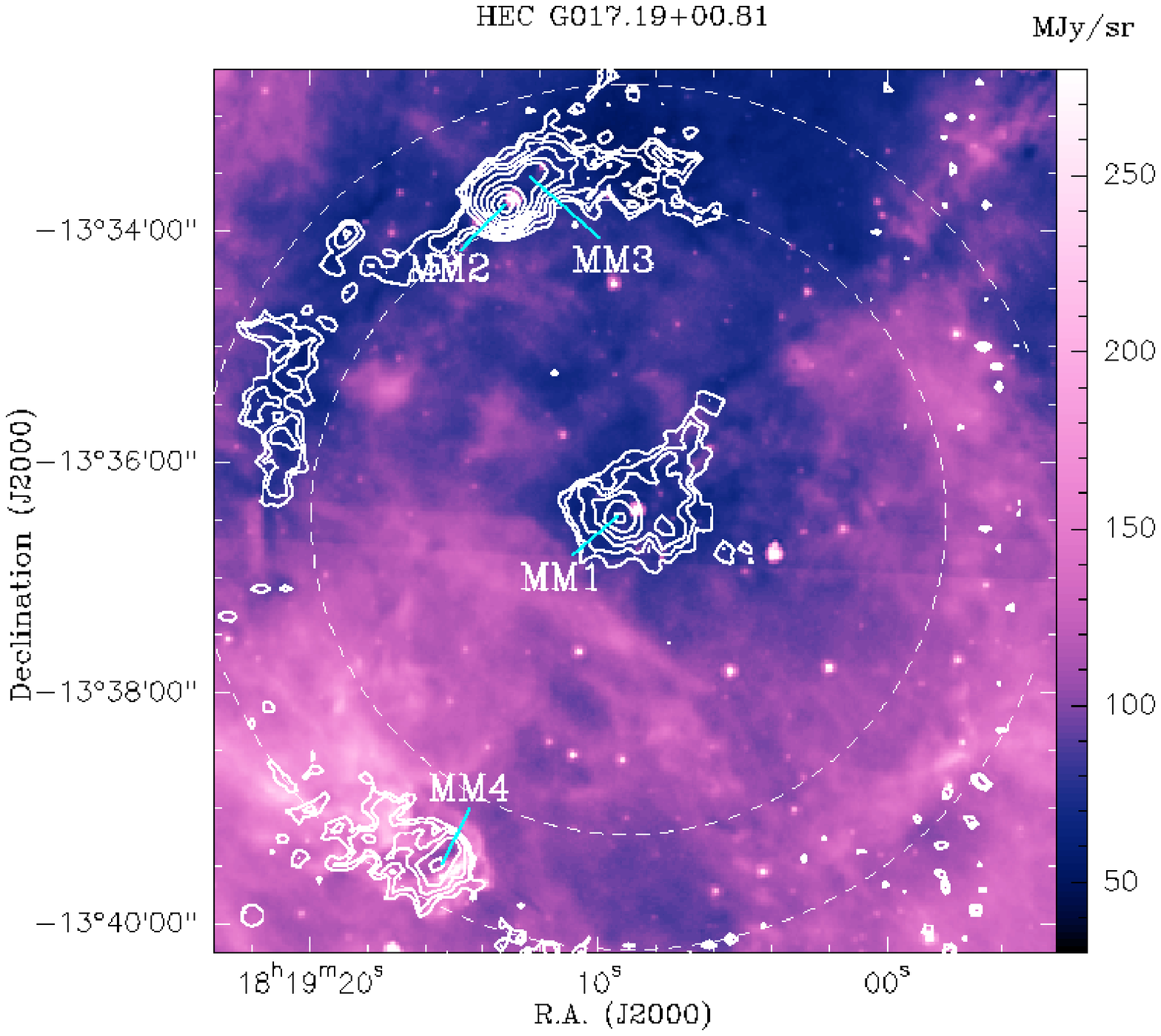}
\caption{A Spitzer GLIMPSE 8.0 $\mu$m image of cloud G017.19+00.81 overlaid with
  the 1.2\,mm emission in white contours starting at $2\sigma$
  ($\sigma=0.015$\,Jy~beam$^{-1}$) and increasing by $\sqrt{2}$. All clumps are at
    the same kinematic distance of $2.3\pm0.2$\,kpc. The dashed
  circles guide the eye to the arc-like structure, which is infrared dark
in the north-east and infrared bright in the south west. }
\label{fig:g17-glimpse}
\end{figure}

The multiple peaked cloud G017.19+00.81 harbors several clumps, which are
  all at a kinematic distance of $2.3\pm0.2$\,kpc, arranged in an
infrared dark arc (Figs.~\ref{fig:mips} and \ref{fig:g17-glimpse}). Such a morphology might point to
triggered star formation. The location of this cloud, on the edge of the bright H{\sc
  ii} region M16, the Eagle nebula, suggests that we are observing new massive
star-forming clumps in an early stage of evolution, whose creation might be linked to
the already ongoing star formation in this region. G017.19+00.81 differs from the classical
triggered star formation scenario described by \citet{zavagno:2005}. First,
the absence of an H{\sc ii} region in the center of the arc; no radio emission
was detected either in the 20\,cm MAGPIS survey \citep{helfand:2006}, the 21\,cm VLA survey \citep{condon:1998}, or in the 6\,cm Parkes survey \citep{haynes:1979} -- the object was not covered by the  5\,cm VLA survey \citep{becker:1994}.
Second, an arc triggered by an H{\sc ii} region is expected to be infrared
bright, whereas G017.19+00.81's arc is partially infrared dark. 

The clumps in the cloud have similar temperatures and sizes. The upper two
left clumps, MM2 and MM3, represent most of the mass in this cloud, whereas
the southern MM4, shows the largest amount of turbulence. Toward the MM2
clump we detected water masers, within a beam of 40\arcsec. It is coincident
with a bright 24\,$\mu$m source indicating a protostar. MM2 seems to be an evolved clump, where star formation is just beginning. Additionally, the column density of this clump theoretically allows high-mass star formation. 
The MM4 clump is located nearby an H{\sc ii}
region, which might explain the wide line width of $2.8\,\mathrm{km~s^{-1}}$ and high degree of
turbulence. Nearby (1\arcmin) there is an infrared-bright source, IRAS 18164--1340, which is only detected at long mid-infrared wavelengths of 100\,$\mu$m \citep[$F_{\mathrm{100\mu m}}= 770$\,Jy,][]{helou:1988}
We performed a more detailed study of this region; a molecular line survey
with IRAM 30m and an interferometer study with the SMA (Rygl et al.\,, in preparation).
 
\subsubsection{G018.26--00.24}

G018.26--00.24 is a cloud with five 1.2\,mm peaks located northward of an infrared bright region with several H{\sc ii} regions and water maser detections.
There are several infrared sources present in the cloud, however, the most mm
emission peaks are infrared dark. MM3 is an exception, here the infrared peak
is just a few arc seconds from the mm peak. MM1 has the largest separation from
the infrared bright region. This clump has the highest column density
and the narrowest line width of the five clumps in this cloud. The overall cloud shape suggests that the gas and dust was collected by a driving force south-west of the cloud, possibly the infrared bright source at ($\alpha=18^{\mathrm{h}}25^{\mathrm{m}}02\rlap{$.$}\,^{\mathrm{s}}8,\delta=-13\degr09\arcmin30\arcsec$, J2000).

\subsubsection{  G022.06+00.21}

G022.06+00.21 contains one clump with a bright mm peak, MM1, and second clump with a weaker peak, MM2.
MM1 is coincident with IRAS 18278--0936 \citep[$F_{\mathrm{25\mu m}}=12$\,Jy,][]{helou:1988}, has
6.7\,GHz methanol masers \citep{szymczak:2000}, extended 4.5\,$\mu$m
emission \citep{cyganowski:2008}, and has a sufficient column density for
high-mass star formation. Additionally, we detected water masers
towards MM1, within a beam of 40\arcsec. MM1 is a relatively warm clump,
$25\,\mathrm{K}$, with ongoing star formation suggested by the masers and outflow tracers. MM2 is more quiescent and
coincident with a weak infrared source.

\subsubsection{G024.37--00.15}
G024.37--00.15 contains two clumps, one infrared bright, MM1, and one infrared dark, MM2. MM1 coincides with the bright IRAS source 18337--0743 \citep[$F_{\mathrm{25\mu m}}=47$\,Jy,][]{helou:1988}, and has a wide linewidth $2.1~\mathrm{km~s^{-1}}$. The MM2 core is in an earlier state of star formation, exhibiting SiO emission \citep{beuther:2007b} and water masers.

\subsubsection{G024.61--00.33}
In G024.61--00.33, the strongest mm emission peak, MM1, is coincident with IRAS
18346--0734 \citep[$F_{\mathrm{25\mu m}}=25$\,Jy,][]{helou:1988} and nearby, separated by $\sim30\arcsec$, \citet{szymczak:2000} detected 6.7 GHz methanol masers. The second clump, MM2, is infrared dark.

\subsubsection{G024.94--00.15}
The mm emission follows closely the infrared dark filament in cloud G024.94--00.15. The clumps, MM1 and MM2, are located near weak infrared sources. We found water maser emission toward MM1, within a beam of 40\arcsec, indicating early phases of star formation.
 
\subsubsection{G030.90+00.00 A, B, C, and D}
G030.90+00.00 consists of several high extinction clouds at very different distances (Fig.~\ref{fig:mips}).
It is located at the tangential point of the Scutum-Crux spiral arm ($V_{\mathrm{LSR}}\sim105$\,km\,s$^{-1}$), where it overlaps with the Sagittarius-Carina ($V_{\mathrm{LSR}}\sim35$ and $\sim75$\,km\,s$^{-1}$) \citep{vallee:2008}. The Scutum-Crux arm connects at this longitude of 30\degr\ with the Galactic bar, a region with highly shocked gas and explosive star formation \citep{garzon:1997}, featuring additionally the mini-starburst region W\,43. We found one clump, MM2, which, considering the kinematic distance at 7.2 kpc, is located on the end of the bar.

\subsubsection{G034.77--00.81}
G034.77--00.81 is a cloud with very diffuse 1.2\,mm emission located in a infrared
dark part of the sky. By smoothing with a 20\arcsec~Gaussian the extended
emission became more evident. We detected weak \amm(1,1) emission toward the
center of the extended emission. The ammonia emission indicates that there is
a dense region in this cloud, however, it has too low column density or is too extended to be detected by MAMBO-2. Such clouds as   G034.77--00.81 could be the most early stages of within our sample.

\subsubsection{G035.49--00.30 A and B}
Cloud G035.49--00.30 A contains one clump, MM1, coincident with a bright infrared
source. In the nearby G035.49--00.30 B we found five clumps located in an elongated infrared dark cloud with several
weak infrared sources. We detected water maser emission towards MM1 and MM3,
within a beam of 40\arcsec. MM1 is the most evolved clump, with ongoing star
formation. MM3 should be in a earlier stage of star formation, since the peak of
the mm emission is infrared dark.


\subsection{Comparison with infrared dark clouds}
\label{bias}

Recent studies have shown that IRDCs generally have a filamentary shape and a high column density of $\sim$$10^{23}$\,cm$^{-2}$ \citep{carey:1998,carey:2000,rathborne:2006}. IRDCs, which are very compact or have large aspect ratios, might not be detected in the extinction maps because of the limited resolution. The extincion method is more sensitive to the more extended and therefore on average lower column density material than generally found in IRDCs. However, the follow-up study of the 1.2\,mm emission maps with MAMBO is sensitive enough to reach the typical IRDC column densities, allowing a comparison between the properties of the high extinction clouds and the IRDCs.

We find that almost 70\% of the high extinction clouds have also IRDC
characteristics using the Catalog of IRDC candidates from \citet{simon:2006a}
and the Spitzer IRDC catalog of \citet{peretto:2009}.
This overlap caused that we incidentally studied the same source as \citet{rathborne:2006}, namely G035.49--00.30 \citep[listed as MSXDC G035.39--00.33 in the paper by][]{rathborne:2006}, allowing a direct comparison. Given the size of our bolometer map, we identified the same clumps within this cloud. Taking into account different analysis methods, we find similar fluxes and masses as  \citet{rathborne:2006}.

\begin{figure}[!htb]
\centering
\includegraphics[height=0.8\columnwidth, angle=-90]{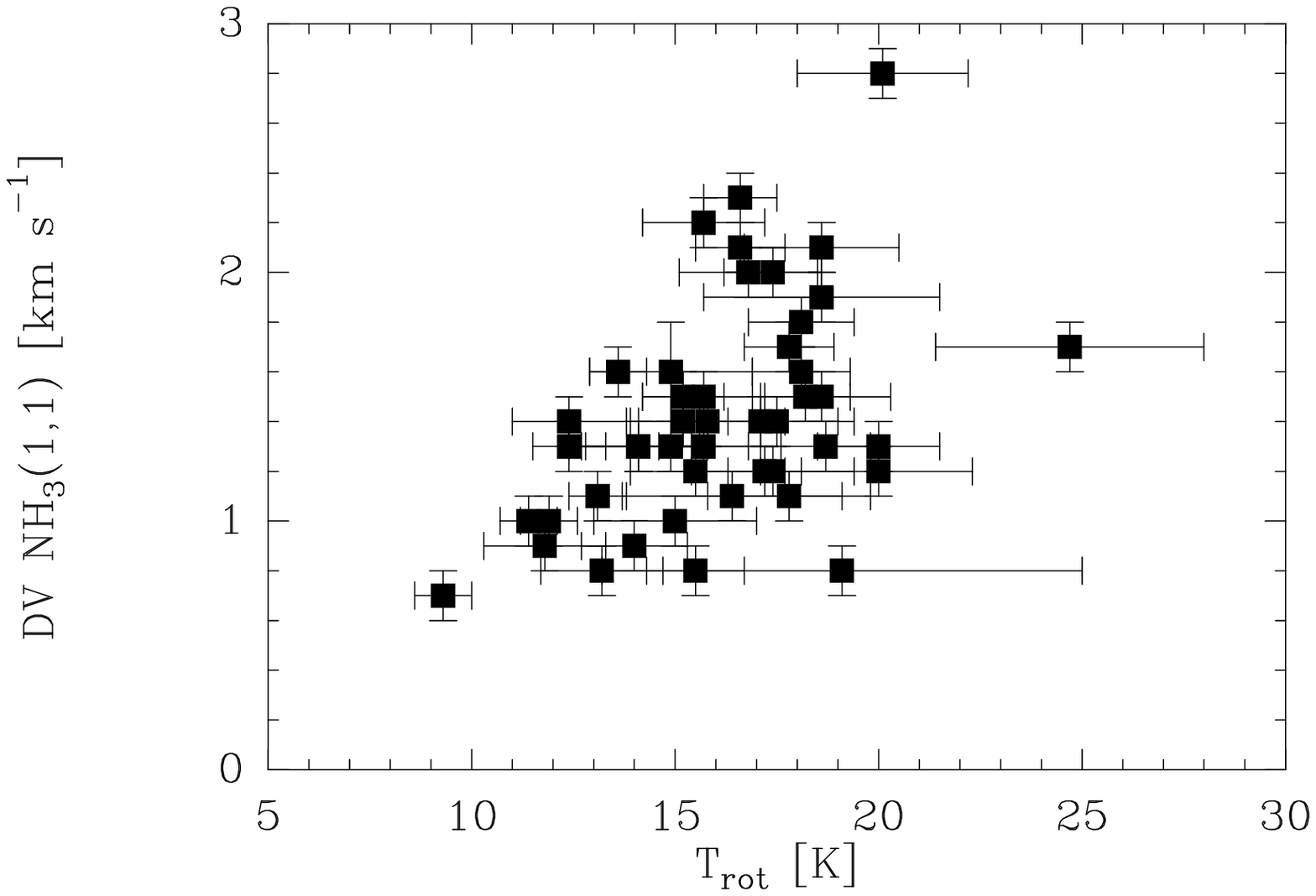}
\caption{\label{fig:dv-t}The \amm(1,1) line width as a function of the \amm\ rotational
  temperature. Linear regression finds a weak trend, correlation coefficient of 0.46, of increasing line width with temperature.}
\end{figure}

\begin{figure}[!htb]
\centering
\includegraphics[height=0.8\columnwidth, angle=-90]{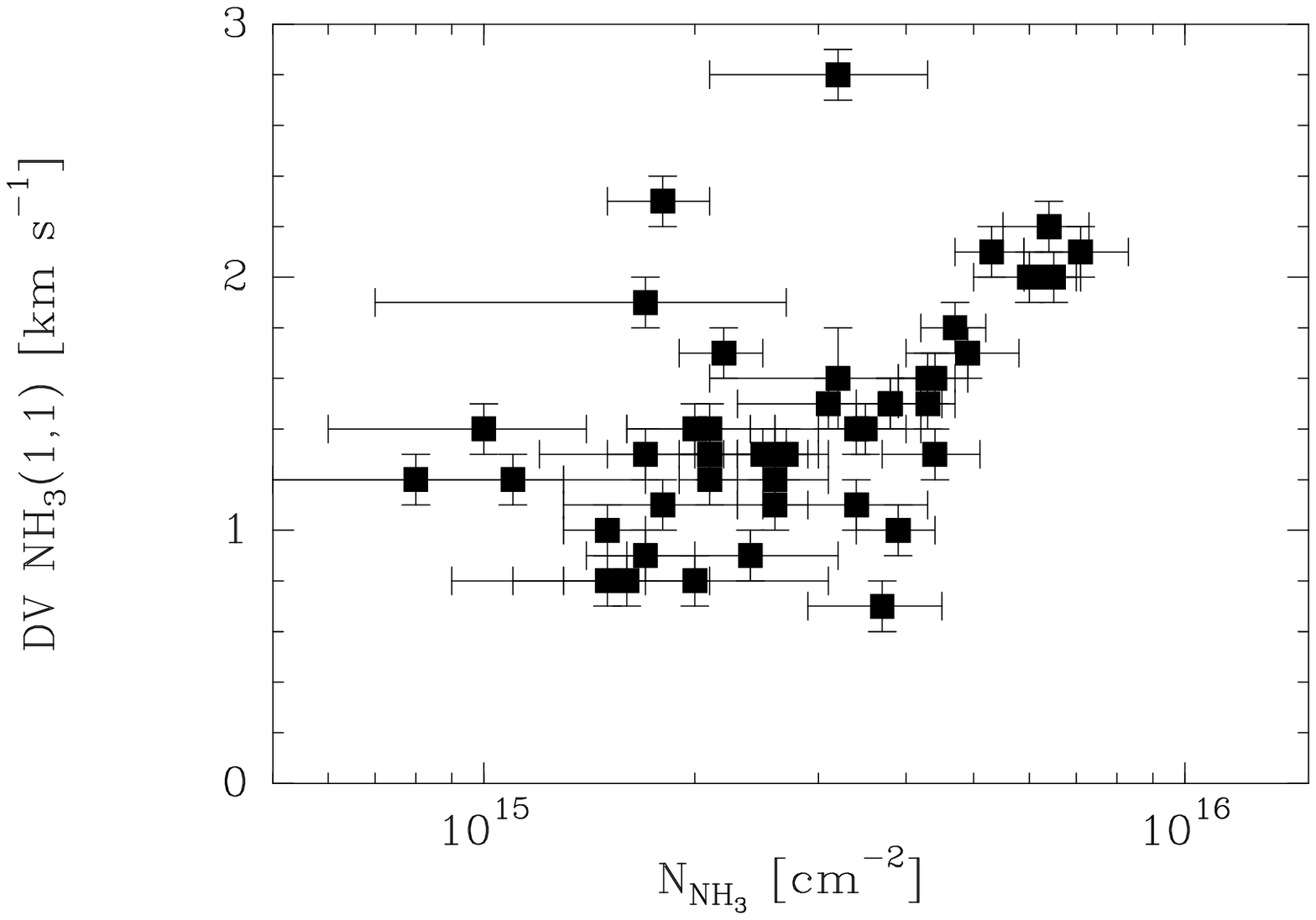}
\caption{\label{fig:dv-n}The \amm(1,1) line width as a function of the \amm\ column density. In
  general, larger line widths are found for higher \amm\ column densities with a correlation coefficient of 0.53.}
\end{figure}

In general, the temperature and the level of turbulence (line withs) of the clumps in high extinction clouds was similar to values found for IRDCs \citep{pillai:2006}. 
Figures \ref{fig:dv-t} and \ref{fig:dv-n} show that the turbulence tends to increase with
ammonia column density and temperature, which is similar for IRDCs \citep{pillai:2006}. The masses and volume averaged densities of the clumps in our sample are similar to the averaged values for clumps in IRDCs \citep[referred to as ``cores'' in][]{rathborne:2006}. 
However, at large distances the high extinction method is biased toward more massive clumps. Comparing with the IRDCs sample of \citet{rathborne:2006} our sample is missing low-mass clumps ($M_{\mathrm{clump}}<100\,\mathrm{M_\odot}$) at large distances ($d>4\,$kpc), see the left panel of Fig.~\ref{fig:distance-m-r}. However, our goal, to search for high-mass molecular clouds, is not affected.

Although, the majority of the high extinction clouds is infrared dark,
there are a number of infrared bright sources. Especially amongst the multiply
peaked clouds, we find several cases of already ongoing star formation via
detections of H{\sc ii} regions, water and Class II methanol masers, and
extended 4.5\,$\mu$m emission. Since we did not require the high extinction clouds to be infrared dark, the method is also sensitive to the evolved stages of star formation. In addition, the color excess selection is sensitive to embedded young stars with a color excess of their own.


\subsection{Comparison with more evolved objects}

Later stages of massive star formation are the HMPOs and the UCH{\sc ii}
regions. Since these objects are usually still embedded in their natal cloud, they
emit most strongly in the millimeter to infrared. Additionally, UCH{\sc ii}
regions emit free-free emission, which is detectable at radio frequencies. 
Most red and bright IRAS sources are such embedded
stages of evolved star (or cluster) formation \citep{wood:1989}.
Ammonia observations of HMPOs and UCH{\sc ii} regions show that they are usually warm \citep[$T\geq22$\,K,
][]{sridharan:2002, churchwell:1990} and very turbulent (line widths of 2--3\,km\,s$^{-1}$). Water and Class II methanol masers are observed
towards these regions \citep{walsh:1998}, indicating outflows and infrared
radiation, respectively. 

The high extinction clumps have lower temperatures and narrower line widths. 
Many of the more evolved clumps in high extinction clouds have
their peak of mm emission nearby bright infrared sources. 
In several cases the two emission peaks even coincide. For three such cases, G014.63--00.57
MM1, G017.19+00.81 MM2, and G022.06+00.21 MM1, we find slightly higher rotational temperatures, respectively 18, 19 and 25\,K, and detections of water masers reminiscent of outflows in the early phase of star formation.

The clump masses in the high extinction clouds are almost one order magnitude lower than the masses of HMPOs \citep{beuther:2002}. 
Likely, by removing the median emission before performing the source extraction, the low column density mass reservoir of the clump is subtracted resulting in less massive clumps than in the HMPO studies. Also the HMPOs are on average larger with an average FWHM of 0.5\,pc, than the sizes of our clumps ($\sim$0.3\,pc). 

In most of our clumps the ammonia column densities were of an order of magnitude higher than those for HMPOs. 
As a clump forms protostellar objects, the temperature and line widths
increase, while the clump will start losing its envelope, which will decrease
the ammonia column density on clump size-scales. Similar behavior of temperature, line width and ammonia column density is also observed for
IRDCs \citep{pillai:2006}.

\subsection{Comparison with theoretical models}
\label{sect:theory}
There are two competing models to explain the formation of massive stars, core accretion \citep{mckee:2003,krumholz:2005,krumholz:2009} and competitive accretion \citep{klessen:1998,bonnell:2001,bonnell:2002,bonnell:2006}. They differ primarily in how the mass is collected which will ultimately make the massive star. In the first model, core accretion, a massive star forms in a massive core, hence the mass reservoir of the core determines the mass of the star. In the second theory, a clump of gas will fragment into many low-mass objects containing masses around the thermal Jeans mass. The accretion will continue in this N-body system, where the overall system potential funnels gas down to the center of the potential, to be accreted by the massive stars forming there. 
A powerful way to distinguish between these scenarios are better
determinations of the massive cores within the clumps. For the core accretion
model these cores should be distributed similar to a stellar mass
function. The degree of mass segregation of the cores in a clump and the limit of fragmentation will test the competitive accretion model. Such massive core studies require a resolution much higher than this study; they have to resolve the clumps into cores. Several studies on fragmentation were recently carried out using the Plateau du Bure or the SMA interferometers \citep{beuther:2007a,rathborne:2007,rathborne:2008,zhang:2009,swift:2009}. They show that the 1.2 mm clumps fragment into smaller cores, but cannot tell at what level the fragmentation halts. With the advent of ALMA, the limit of fragmentation will finally come within observational reach.

Simulations of core accretion models by \citet{krumholz:2008} put forward a
minium column density of 1\,g\,cm$^{-2}$ for massive star formation. For
reference, this corresponds to 4750\,$\mathrm{M_\odot\,pc^{-2}}$ or $N_{\mathrm{H_2}} = 2.7\times10^{23}\,\mathrm{cm^{-2}}$. This limit would imply that only one clump from the high extinction clouds can form massive stars, namely G014.63--00.57 MM1. 
We note here that \citet{krumholz:2008} use a $\kappa_{\mathrm{1.2mm}}\sim
0.4\,\mathrm{cm^{2}\,g^{-1}}$, while we used $\kappa_{\mathrm{1.2mm}}\sim
1.0\,\mathrm{cm^{2}\,g^{-1}}$. To compare this limit with our measured column
densities we would have to scale our results by a factor 2.5, and in this case four clumps would theoretically be able to form massive stars.
Of these four clumps only one, G014.63--00.57 MM2, is infrared dark. The other three clumps are infrared bright, indicating that they already commenced forming stars. It suggests that the theoretical column density limit is too high for the very initial stages of star cluster forming clumps. 

\section{Summary}

We made infrared extinction maps of the first and fourth Galactic quadrant, compiling a catalog of compact high extinction features. We studied 25 high extinction clouds in detail, using IRAM 30m bolometer observations and pointed ammonia observations with the Effelsberg Telescope. The main results of the paper can be summarized as:

\begin{enumerate}
\item{Using the average color excess of the Spitzer $3.6-4.5$\,$\mu$m IRAC bands, we are able to trace high density structures across the Galactic plane. The extinction selected clouds are found on distances between 1 and 7\,kpc. Most of them are concentrated in the range of $1-4$\,kpc, and trace the same regions of the Galactic plane as IRDCs.}
\item{The extinction method is more sensitive to large scale lower column density clouds than the bolometer observations of 1.2 mm emission, which resulted in a few non detections of clouds in the mm emission despite a mean visual extinction of $\sim$30 magnitudes from the extinction map. In general, there is a good correlation between the cloud masses derived from the extinction maps and from the 1.2\,mm emission.}
\item{From the 1.2 mm emission, we have found clumps with column densities of $3-30\times10^{22}\,\mathrm{cm^{-2}}$ and masses of $12-400$\,$M_\odot$. Evidently, not all clumps found by high extinction will be able to form massive stars. It is expected for cluster-forming regions to produce low and high-mass stars, thus to have a wide range in masses.}
\item{High extinction clouds contain a wide range of evolutionary stages. 70\% of the high extinction clouds are associated with infrared dark clouds, however several clouds show more evolved stages of ongoing (massive) star formation. Three different classes of clouds are proposed:
\begin{itemize}
\item{Diffuse clouds have no signs of masers, H{\sc ii} regions or extended
    4.5\,$\mu$m objects, which are general signposts of star
    formation. The 1.2 mm emission shows no mm peaks above twice the mean
    emission of the cloud. These clouds are cold ($T\sim13.5\,\mathrm{K}$) and line widths far above the thermal value mark the importance of turbulence. The observations suggest that the diffuse clouds are in an early phase, where the accretion of gas and dust into clumps might be on the verge of beginning or might never reach the necessary column density to form stars. }
\item{When the clumps manage to accrete more matter so that the mm emission exceeds
    twice the mean cloud emission they enter the next (peaked) stage. During the peaked stage already some clouds show more than one clump, which indicates that star formation can start in different regions in the same cloud at different times.}
\item{Multiply peaked clouds show many signs of ongoing star formation, such
    as detections of masers, H {\sc ii} regions or extended 4.5\,$\mu$m
    objects. Several clumps are infrared bright and show generally slightly higher temperatures and turbulence than diffuse or peaked clouds. Clumps within a cloud are not necessary in the same state, and will likely not all form massive stars. We find four clumps that satisfy the theoretical column density requirement for high-mass star formation. }
\end{itemize}
}

\end{enumerate}

\begin{acknowledgements}
      We thank the anonymous referee, whose comments and suggestions resulted
      in a much improved paper. We also thank Malcolm Walmsley for the helpful
      discussion. This work is based on observations with the 100m telescope of the
      MPIfR (Max-Planck-Institut f\"ur Radioastronomie) at Effelsberg and the
      IRAM 30m telescope at Pico Veleta. We are grateful to the staff of both
      observatories for their support. We thank Marion Wienen and
      Lies Verheyen for their help with the Effelsberg observations. This
      research has made use of the NASA/ IPAC Infrared Science Archive, which
      is operated by the Jet Propulsion Laboratory, California Institute of
      Technology, under contract with the National Aeronautics and Space
      Administration and the SIMBAD database, operated at CDS, Strasbourg,
      France. This publication makes use of molecular line data from the Boston University-FCRAO Galactic Ring Survey (GRS). The GRS is a joint project of Boston University and Five College Radio Astronomy Observatory, funded by the National Science Foundation under grants AST-9800334, AST-0098562, \& AST-0100793.
      KLJR was supported for this research through a stipend from the
      International Max Planck Research School (IMPRS) for Astronomy and Astrophysics at the Universities of Bonn and Cologne.
\end{acknowledgements}

\bibliographystyle{aa}
\bibliography{13510bib}
  










\end{document}